\documentclass[aps,amsmath,amssymb,onecolumn]{revtex4}
\usepackage[dvips]{graphicx,color}
\usepackage[T1]{fontenc}
\usepackage{times}
\usepackage{mathrsfs}
\usepackage{amsmath}
\usepackage{graphicx}
\usepackage{epsfig}
\usepackage{dcolumn}
\usepackage{bm}
\definecolor{r}{rgb}{1.0,0,0}

\begin{document}
\title{Parameterisation of lateral density and arrival time distributions of Cherenkov photons in EASs as functions of independent shower parameters for different primaries}
\author{P. Hazarika$^1$, G.S. Das$^1$, U. D. Goswami$^1$}
\affiliation{$1.$ Department of Physics, Dibrugarh University, Dibrugarh 786 004, Assam, India} 
\begin{abstract}
The simulation of Cherenkov photon's lateral density and arrival time 
distributions in Extensive Air Showers (EASs) was performed with the CORSIKA 
code in the energy range: 100 GeV to 100 TeV. On the basis of this simulation 
we obtained a set of approximating functions for the primary $\gamma$-ray 
photons, protons and iron nuclei incident at zenith angles from 0$^\circ$ to 
40$^\circ$ over different altitudes of observation. Such a parameterisation is 
important for the primary particle identification, for the reconstruction of 
the shower observables and hence for a more efficient disentanglement of the 
$\gamma$-ray showers from the hadronic showers. From our parameterisation 
analysis, we have found that even though the geometry of the lateral density 
($\rho_{ch}$) and the arrival time ($t_{ch}$) distributions is different for 
different primaries at a particular energy ($E$), at a particular incident 
angle ($\theta$) and at a particular altitude of observation ($H$) up to a 
given distance from the showe core ($R$), the distributions follow the same 
mathematical functions $\rho(E,R,\theta,H) = a E^{b}\exp[-\{c R + (\theta /d)^{2}-f H\}]$ and $t(E,R,\theta,H) = l E^{-m}\exp(n/R^{p})({\theta}^q+s)(u {H}^2+v)$ respectively but with different values of function parameters.
\end{abstract}
\maketitle
\section{Introduction}
The major unsolved problems in the study of Cosmic Rays (CRs) are related to 
the understanding of their origin, propagation and acceleration mechanisms 
\cite{Nakamura, Gaisser, Blasi, Bhattacharjee}. CRs are made up of particles 
of high, ultra-high, 
and extremely high energy of extra terrestrial origin, which constantly 
impinge the Earth's atmosphere. The primary CRs extend over twelve decades of 
energy accompanied by a corresponding decline in the intensity \cite{Nakamura, 
Gaisser, Blasi, Bhattacharjee}. 
It is regarded that the bulk of the CRs originate from within the galaxy. Since
the CRs are mainly charged particles, they are deflected in the intragalactic 
magnetic fields and hence they reach us isotropically loosing the direction(s) 
of their source(s). However, the celestial sources that emit CRs also emit 
$\gamma$-rays. The $\gamma$-rays being neutral, do not loose their direction. 
Thus the detection of $\gamma$-rays can help us in estimating the locations of 
such astrophysical sources.

The Atmospheric Cherenkov Technique (ACT) is a technique which is extensively 
used to detect $\gamma$-rays with the help of ground based telescopes within 
the energy range of some hundred GeV to few TeV \cite{Rene, Holder}. This 
technique is based on the registration of the very brief flash of Cherenkov 
radiation emitted by the relativistic charged particles present in the 
EASs initiated by the primary $\gamma$-rays in the atmosphere 
\cite{Hoffman, Weekes, Acharya}. As this method is indirect, so for the proper 
analysis of data of the Cherenkov detectors and also to effectively reject the 
huge CRs background from the signal, a detailed Monte Carlo simulation study 
of Cherenkov photons have to be carried out. This will help us in proper 
estimation of the energies and arrival directions of the primary $\gamma$-rays.
It needs to be mentioned here that, we have already conducted a detailed study 
on the arrival time and the density distributions as well as on the angular 
distributions of Cherenkov photons initiated by the $\gamma$-rays and the 
hadronic primaries, incident at various zenith angles with a wide range of 
energies, particularly at high altitude of observation. We have also carried 
out hadronic interaction model dependent studies \cite{Hazarika, Das}. However, 
the lateral density and arrival time distributions of Cherenkov photons depend 
not only on the energy and type of the primary particle, but also on the 
distance from the shower axis, the direction of the shower axis and the 
altitude of 
observation level. Hence a complete model which depends on the distance ($R$) 
from the shower axis, primary energy ($E$), zenith angle ($\theta$) and 
altitude of observation ($H$) is developed in this work and compared to the 
predictions
 of Monte Carlo simulations. The parameterisations can help us to approximate the simulated Cherenkov photon's density and arrival time distributions. A complete analytical description of this kind will help us for primary energy reconstruction and to properly differentiate $\gamma$-rays from the hadronic showers.

The outline of the paper is as follows. In Section II, the detailed simulation 
process is discussed. In Section III, an analytical expression is derived for 
the density distributions of Cherenkov photons as a function of energy of the 
primary ($E$), distance from the shower core ($R$), zenith angle ($\theta$) 
and altitude of observation level ($H$). The results are compared and analysed 
to a detailed CORSIKA simulation \cite{Knapp}. Section IV describes a similar 
approach to the parametrisation of arrival time of Cherenkov photons. Finally,
in Section V we have summarized the results of our work.
\section{Monte Carlo Simulations}
For our study, the CR simulation code CORSIKA version 6.990 with the hadronic 
interaction models QGSJET 01C for high energy interactions and GHEISHA 2002d 
for low energy interactions has been used for the simulations of EASs 
generated by $\gamma$, proton and iron primaries \cite{Knapp, Kalmykov, 
Fesefeldt}. EGS4 code  \cite{Nelson} is used for the simulation of the
electromagnetic component of the EAS. The choice of our model combination is 
based on the fact that the density and arrival time distributions of the 
Cherenkov photons are almost independent of hadronic interaction 
models \cite{Hazarika}. Using this combination of hadronic interaction models 
we have generated the EASs for the vertically incident monoenergetic 
$\gamma$-ray, proton and iron primaries as well as for those inclined at 
zenith angles 10$^{\circ}$, 20$^{\circ}$, 30$^{\circ}$ and 40$^{\circ}$ for 
energies ranging from 100 GeV to 100 TeV. These energies lay within the typical
range of ACT energy and are selected for the different primaries on the basis 
of their equivalent number of Cherenkov yields \cite{Hazarika}. In order to 
consider the dependence of Cherenkov photon distribution on the altitude of 
observation level, we have also generated showers at 500 m, 1075 m 
(Pachmarhi observation level, longitude: 78$^\circ$ 26$^\prime$ E, latitude: 
22$^\circ$ 28$^\prime$ N), 2000 m, 3000 m and 4270 m (Hanle observation level, 
longitude: 78$^\circ$ 57$^\prime$ 51$^{\prime\prime}$ E, latitude: 32$^\circ$ 
46$^\prime$ 46$^{\prime\prime}$ N) \cite{Britto, Majumdar}. The number of 
showers generated  at different energies, zenith angles and the altitudes of 
observation level for the $\gamma$-ray, proton and iron primaries is listed in 
the Table \ref{tab1}.
\begin{table}[ht]
\caption{Total number of showers generated at different energies, zenith angles and the altitudes of observation level for the $\gamma$-ray, proton and iron primaries}  \label{tab1}
\begin{center}
\begin{tabular}{ccc}\hline
Primary particle & ~~~~Energy & ~~~~Number of Showers \\\hline

$\gamma$-photon& ~~~~~~~100 GeV & ~~~~~~10000\\
& ~~~~~~~500 GeV & ~~~~~~~~5000\\
& ~~~~~~~~~~~1 TeV & ~~~~~~~~2000\\
& ~~~~~~~~~~~2 TeV & ~~~~~~~~1000\\\\
Proton&~~~~~~250 GeV&~~~~~~10000\\
& ~~~~~~~~~~1 TeV & ~~~~~~~~5000\\
& ~~~~~~~~~~2 TeV & ~~~~~~~~2000\\
& ~~~~~~~~~~5 TeV & ~~~~~~~~~~800\\\\
Iron&~~~~~~~~~~1 TeV&~~~~~~~~8000\\
& ~~~~~~~~~~5 TeV & ~~~~~~~~4000\\
& ~~~~~~~~10 TeV & ~~~~~~~~2000\\
& ~~~~~~~~50 TeV & ~~~~~~~~1000\\
& ~~~~~~100 TeV&~~~~~~~~~~600\\ \hline
\end{tabular}
\end{center}  
\end{table}
For detecting a TeV EAS with a large zenith angle, a very wide area detector 
array is required. So we have taken the detector geometry as a horizontal flat 
detector array, with 25 detectors in the E-W direction with a separation of 
25 m and 25 detectors in the N-S direction with a separation of 20 m. Each 
detector is considered to have an area 9 m$^2$. The cores of the EASs are 
considered to be coincident with the centre of the detector array. In case of 
the longitudinal distribution of Cherenkov photons, photons are counted only 
in the step where they are emitted. The emission angle of the Cherenkov photons is chosen as wavelength independent. The Cherenkov radiation wavelength range 
is taken as 200-650 nm. The threshold energies (in GeV) are chosen for hadrons, muons, electrons and photons as 3.0, 3.0, 0.003, 0.003 respectively. The 
position and time (with respect to the first interaction) of each photon 
hitting the detector on the observation level are recorded. To reduce the size 
of the data file, the variable bunch size option of Cherenkov photon is set 
to "5". The multiple scattering length for e$^-$ and e$^+$ is decided by the 
parameter STEPFC in EGS code which has been set to 0.1 here \cite{Nelson}. The 
US standard atmosphere parameterised by Linsley has been used here \cite{US}.
The choice of this atmospheric model will not affect our results, as we have
already shown in one of our previous works \cite{Das} that the Cherenkov 
photon density and arrival time distributions are almost independent of 
atmospheric models.    

In order to test our analytical model, the lateral density of the Cherenkov photons is obtained by counting the number of photons incident on each detector per shower. To obtain the arrival time of a Cherenkov photon over a detector, the time taken by the photon to reach the detector with respect to the first photon of the shower hitting the array is calculated. Since there are several photons hitting each detector per shower, so average of their arrival times is calculated for each detector. Moreover, the variation of Cherenkov photon density and arrival time with respect to core distance is found by calculating their average values for the specified number of showers in order to cancel out the effects of shower to shower fluctuations that may be present. As mentioned earlier, the general characteristics and features of Cherenkov photon density and arrival time distributions have already been discussed and part of the results shown here as a 
completeness of this work, have already been presented in \cite{Hazarika, Das}. 
In the following sub headings we will present the results of their 
parameterisation.
\section{Parameterisation of Cherenkov photon's density}
\subsection{As a function of radial distance ($R$)}
\begin{figure*}[hbt]
\centerline{\hspace{0.7cm}\includegraphics[width=5.5cm, height=4.3cm]{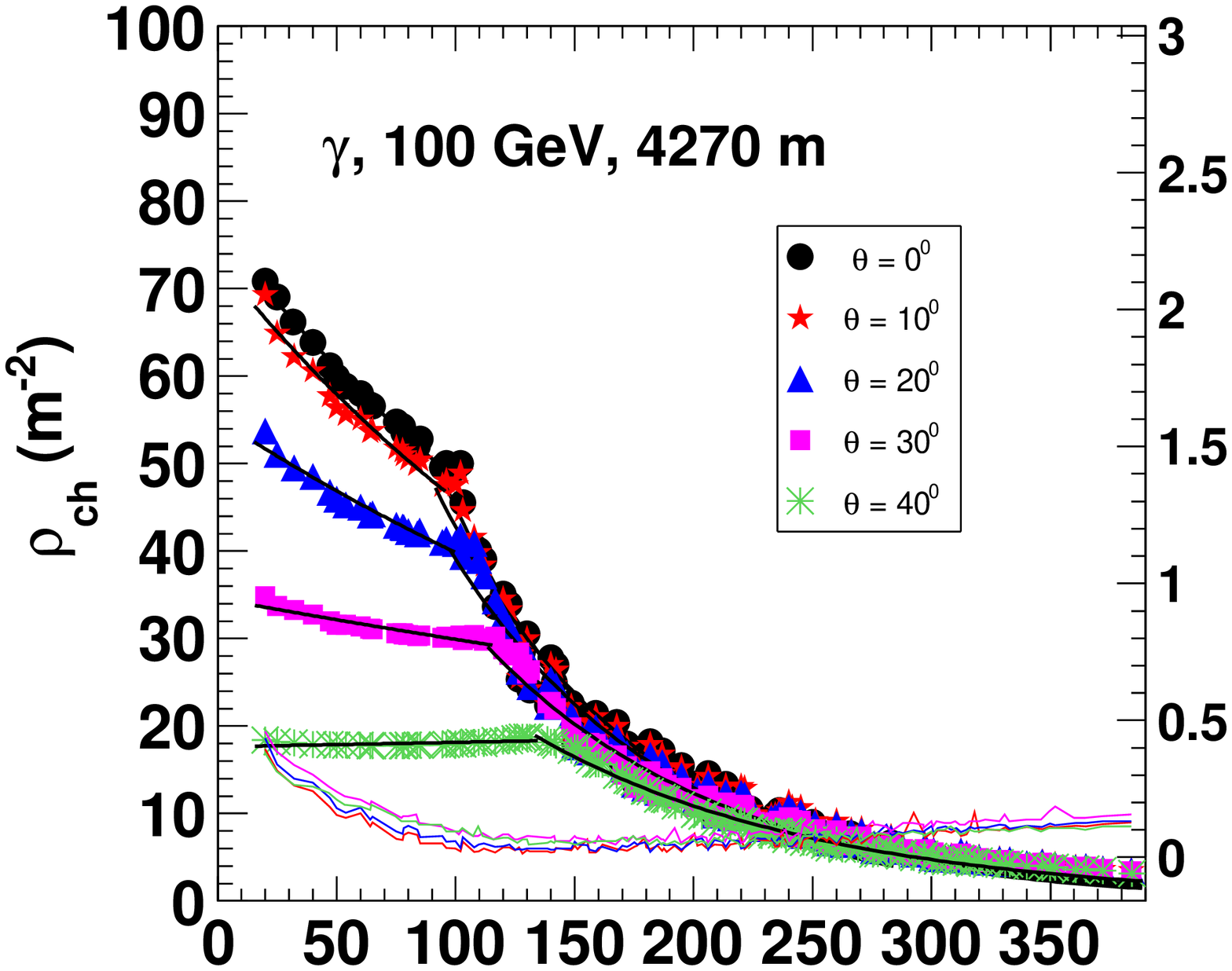} \hspace{0.2cm}
\includegraphics[width=5.2cm, height=4.3cm]{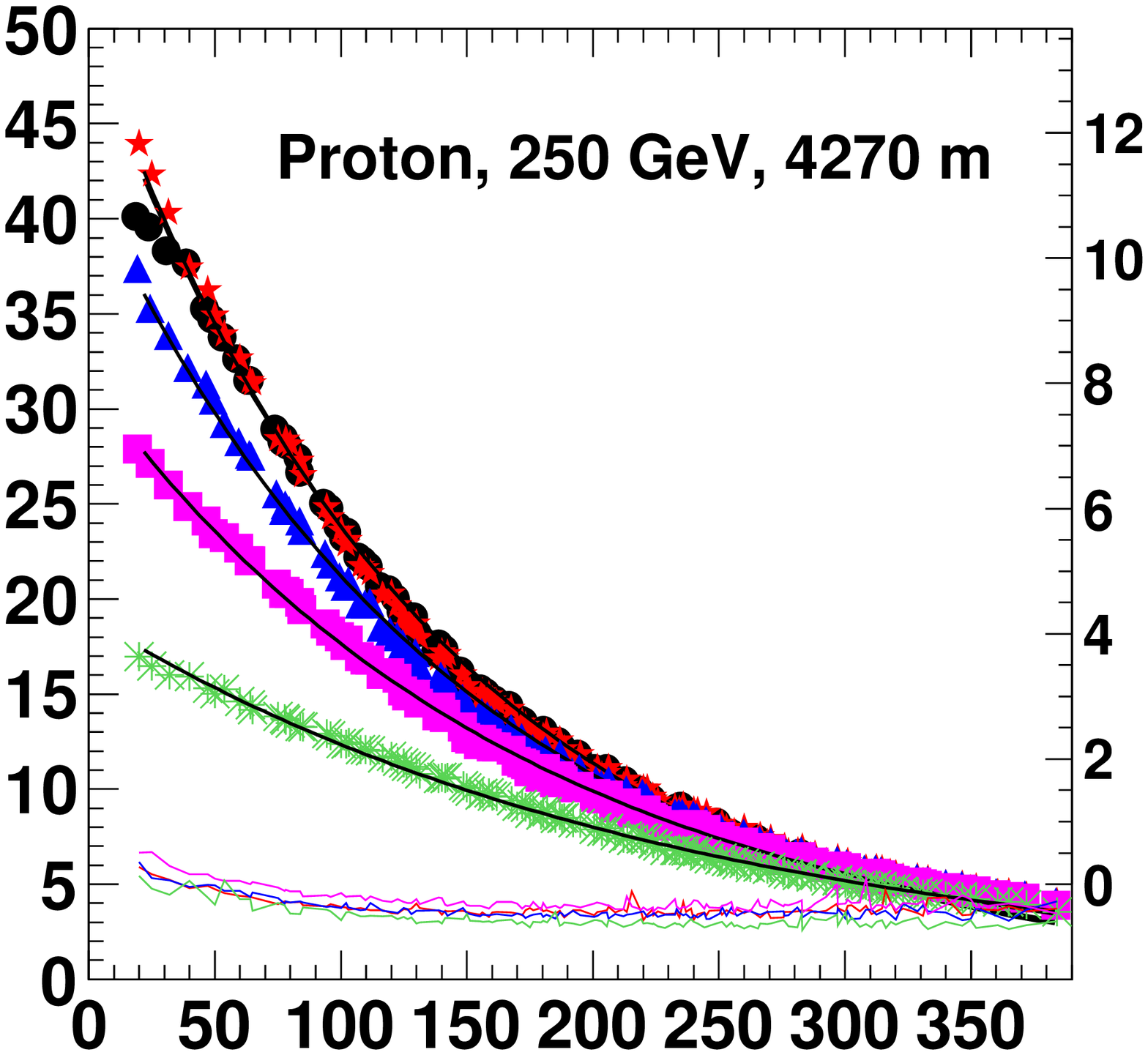}
\includegraphics[width=5.7cm, height=4.3cm]{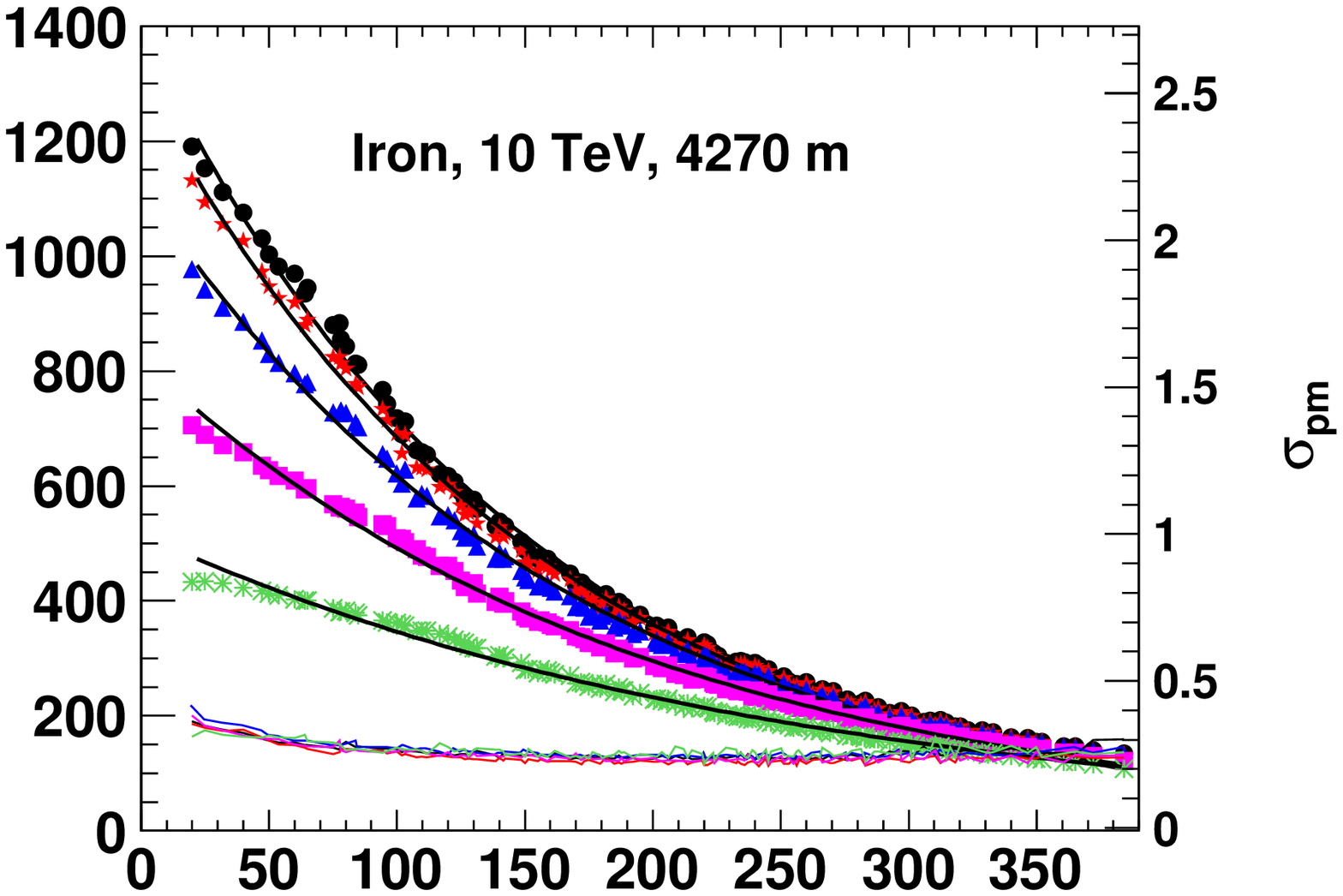}}
\centerline{\hspace{0.2cm}\includegraphics[width=5.25cm, height=4.3cm]{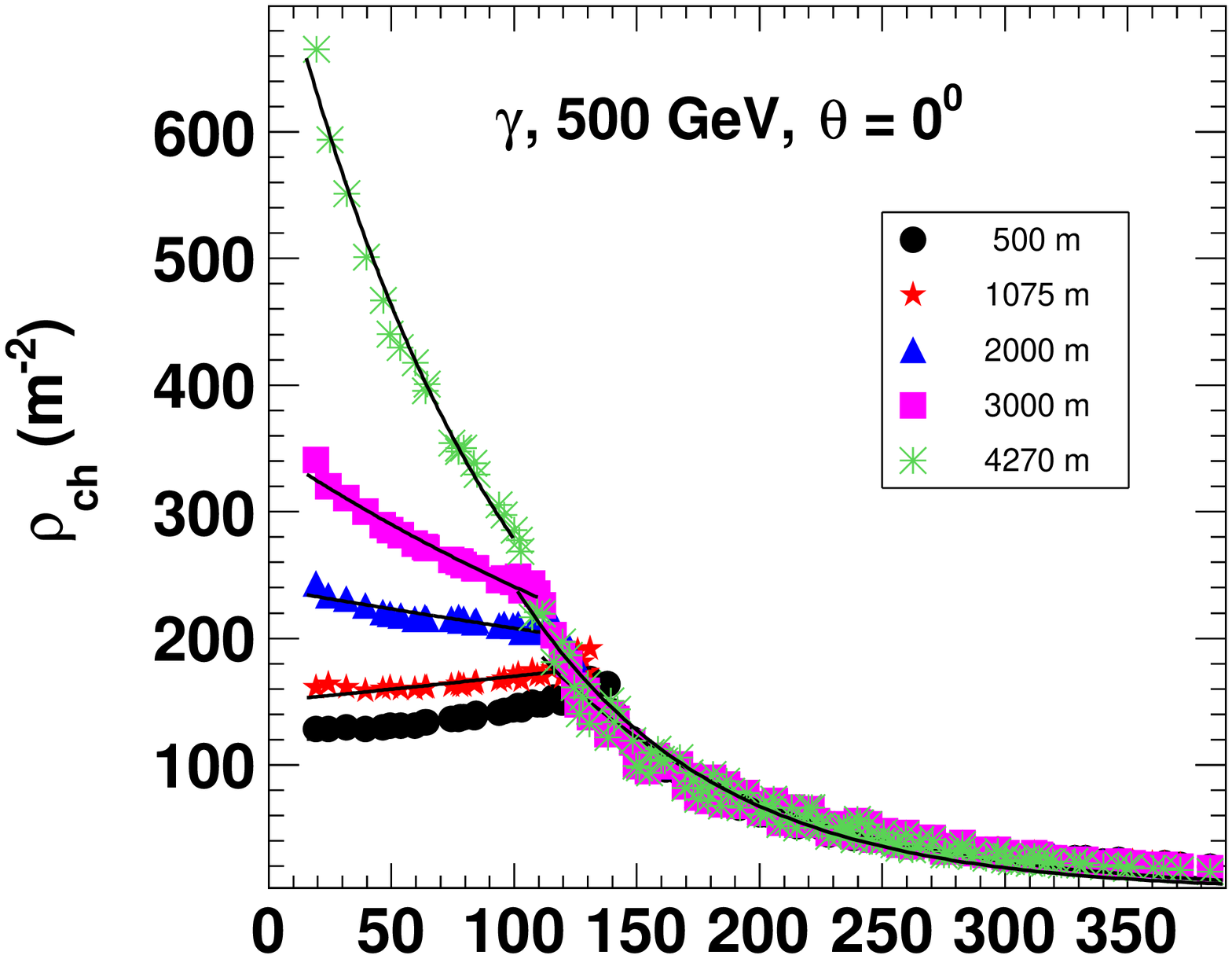}
\includegraphics[width=5.25cm, height=4.3cm]{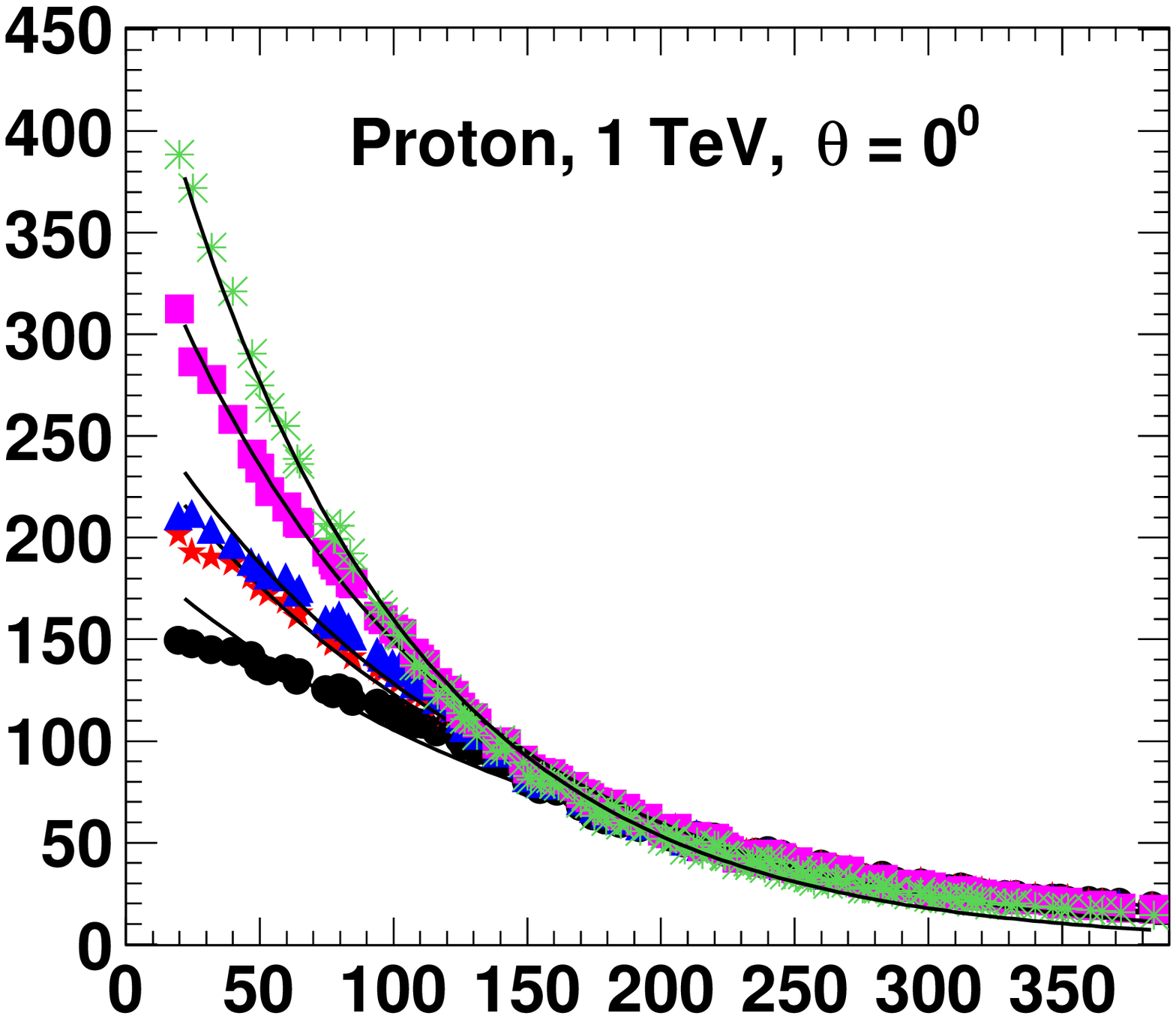}
\includegraphics[width=5.4cm, height=4.3cm]{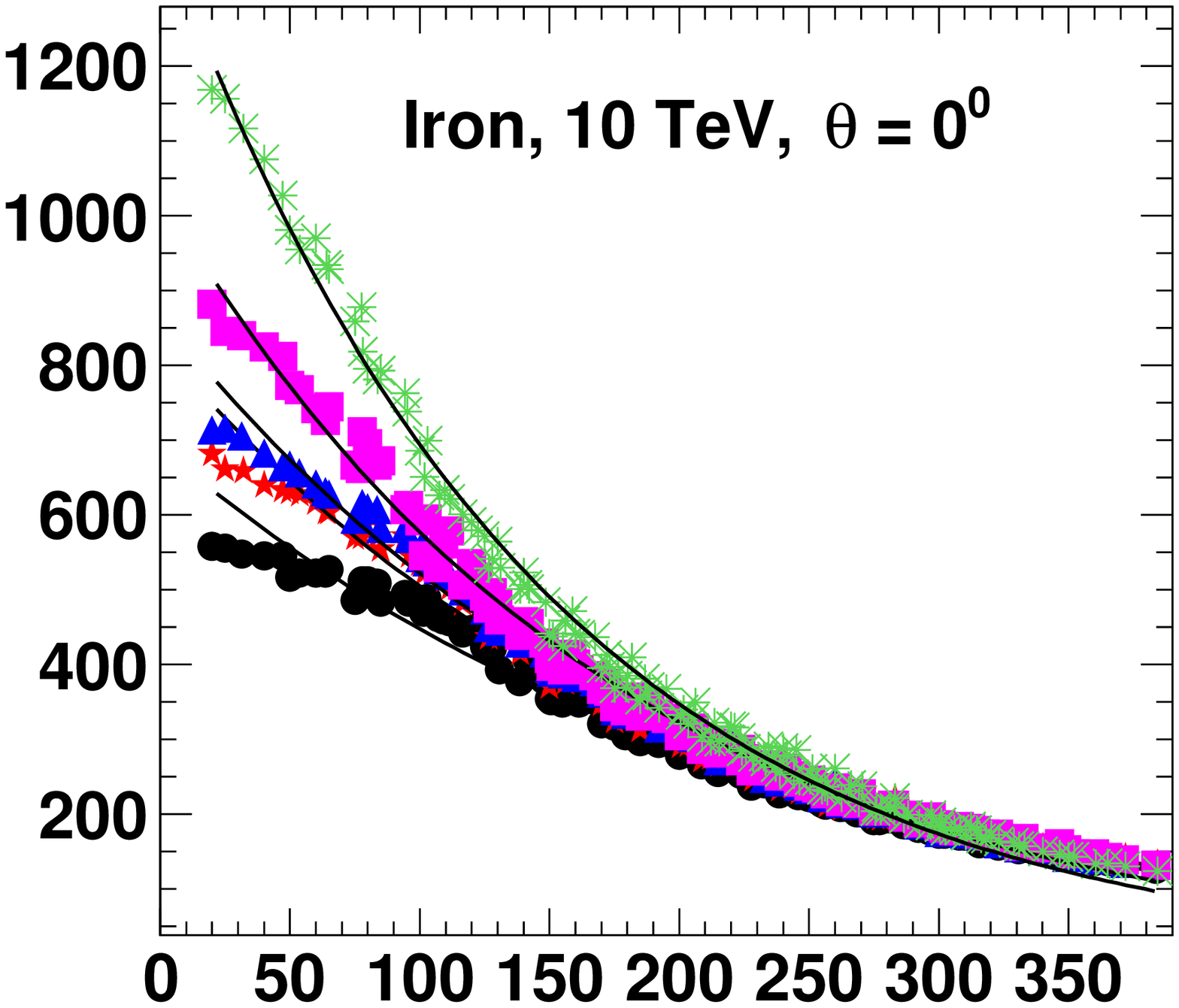}}
\centerline{\includegraphics[width=5.5cm, height=4.3cm]{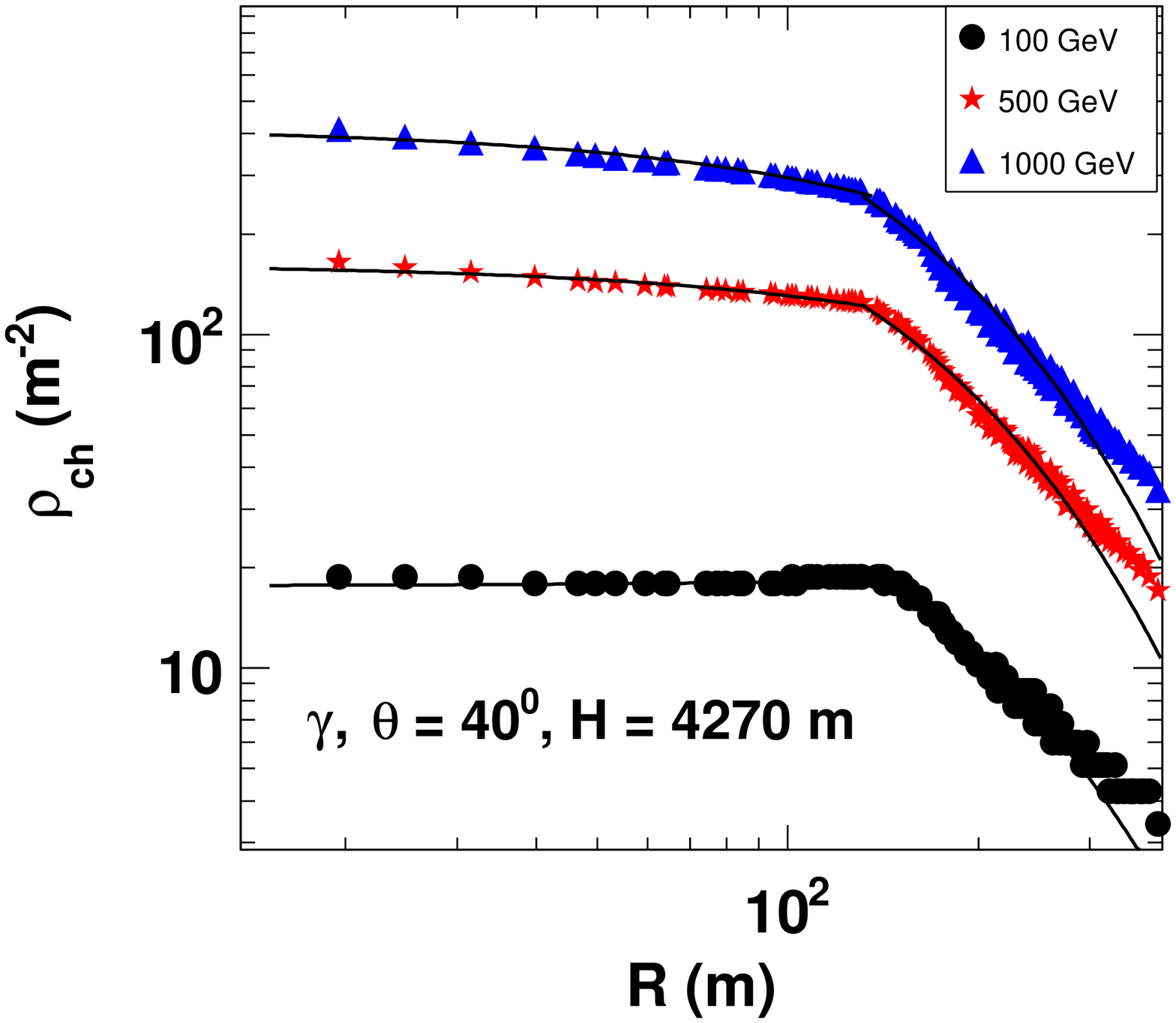}
\includegraphics[width=5.3cm, height=4.3cm]{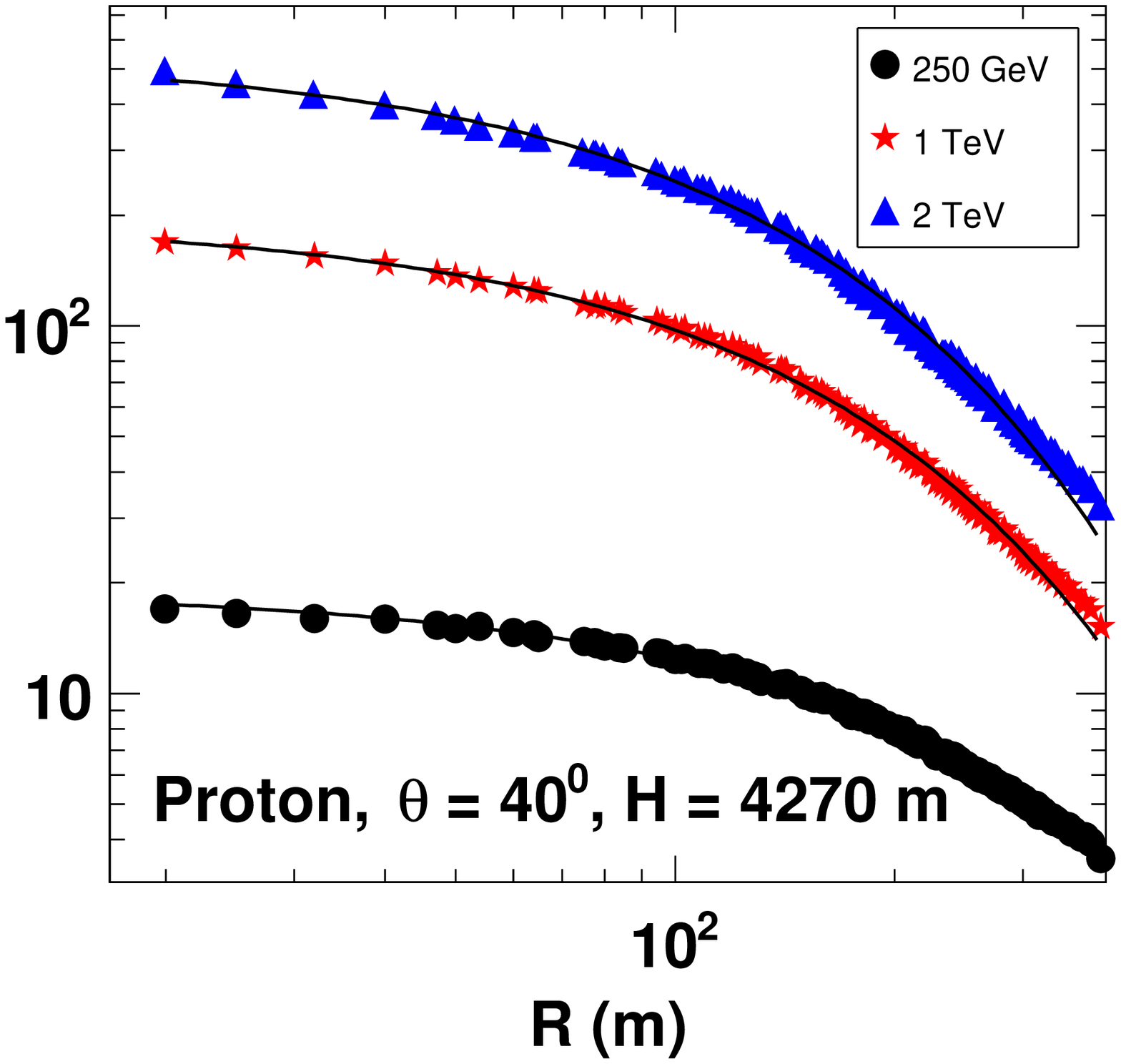}
\includegraphics[width=5.2cm, height=4.3cm]{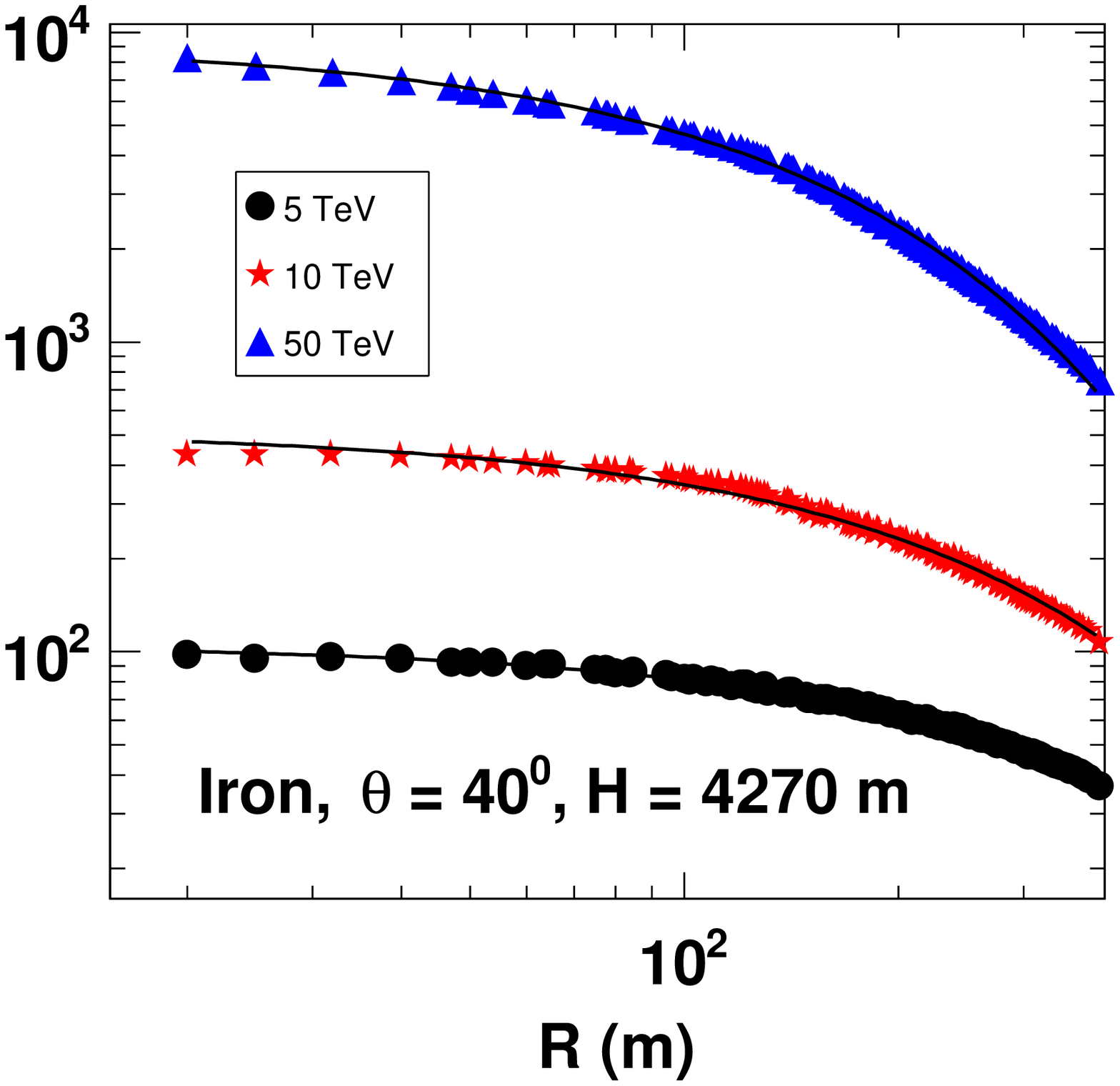}}
\caption{Average Cherenkov photon density ($\rho_{ch}$) for $\gamma$-ray, 
proton and iron primaries is plotted as a function of distance from the shower 
core ($R$). The plots in the upper most panel show these variations for zenith 
angles 
$0^{\circ}$, $10^{\circ}$, $20^{\circ}$, $30^{\circ}$ and $40^{\circ}$ keeping 
$E$ and $H$ fixed. The r.m.s values per mean ($\sigma_{pm}$) of the Cherenkov 
photon densities with respect to the distance from the shower core of different primaries is also shown. The plots in the middle panel show the $\rho_{ch}$ 
variations for 
different values of $H$ keeping $\theta$ and $E$ fixed. Finally, the plots in 
the bottom 
panel do the same for different energy of the primary keeping $\theta$ and 
$H$ fixed at a particular value. The solid lines in the respective plots show 
the result of our parameterisation (\ref{eq1}). The fits are within the limit 
of statistical error ($< \pm 10\%$). Same function is used to fit the plots on 
both sides of 
the hump (wherever necessary) but with different function parameters.}
\label{fig1}
\end{figure*}
It is seen that for all primary particles, energies and zenith angles, the lateral density of Cherenkov photons follow a negative exponential function as given by the equation
\begin{equation}
\rho(R) = a_{0} \exp(-b R),
\label{eq1}
\end{equation}
where $\rho(R)$ is the position dependent density function of Cherenkov photons, $a_{0}$ is the coefficient, $b$ is the slope of the function, $R$ is the distance from the shower core. $a_{0}$ and $b$ have different values for different 
primaries \cite{Hazarika}. This parameterisation Eq.(\ref{eq1}) of Cherenkov 
photon density as a function of core distance ($R$) is applied to the simulated
 data as shown in Fig.\ref{fig1}. Fig.\ref{fig1} shows three different cases 
of the variation of average density of Cherenkov photons ($\rho_{ch}$) as a 
function of distance from the shower core ($R$) of $\gamma$-ray, proton and 
iron primaries for different energies, zenith angles and altitudes of 
observation (for details check the caption of Fig.\ref{fig1}). The best fit 
functions, represented by the Eq.(\ref{eq1}) are shown by the solid lines in 
the plots. For these fittings we used the $\chi^{2}$-minimization method 
available in the ROOT software (the technique is applied to all the other 
parameterisations presented in the paper unless mentioned otherwise) 
\cite{root}. From Fig.\ref{fig1} it is clear that the results of our 
parameterisation is in good agreement with the simulated CORSIKA data for all 
the three primaries except at very small ($<$ 50 m) and very large core 
distances. This may be due to the lesser number of Cherenkov photons produced 
at very near and at very large distances from the shower core. In addition to 
this, for $\gamma$-ray due to the presence of the significant characteristic 
hump, the fitting is not good over the position of hump. So for the 
$\gamma$-ray primary, the fit is made at two segments, one before the position 
of hump and other after the position of hump with different function parameters
 and this method is applied to any plot wherever is required \cite{Das}. Again,
 since the hump becomes smaller as energy of the primary and the altitude of observation increases, the parameterisation becomes better even for the $\gamma$-ray primary. With increasing zenith angle ($\theta$), the distance of the hump from the core increases with increasing prominence \cite{Das}. Due to which at large zenith angle the parameterisation is not good for the $\gamma$-ray primary. However this is not the case for proton and iron primaries. For all the three primaries the parameterisation is at better agreement with the simulated data at higher altitudes of observation. As an example, the Table \ref{tab2} shows values
of the fitted parameters of the parameterisation Eq.(\ref{eq1}) to the 
$\rho_{ch}$ 
distributions as a function of $R$ for $\gamma$-ray, proton and iron primary at 
100 GeV, 250 GeV and 10 TeV energies respectively and at $H$ = 4270 m and 
$\theta$ = 30$^{0}$.
\begin{table}[ht]
\caption{Values of the fitted parameters of Eq.(\ref{eq1}) to the $\rho_{ch}$ distributions as a function of $R$ for $\gamma$-ray, proton and iron primary at 100 GeV, 250 GeV and 10 TeV energies respectively and at $H$ = 4270 m and $\theta$ = 30$^{0}$.} \label{tab2}
\begin{center}
\begin{tabular}{ccc}\hline
Primary & ~~$a_{0}$ & ~~~$b$ \\\hline\\[-7pt]

$\gamma$& ~~~5.069 $\pm$ 0.093 & ~~~1.5448 $\pm$ 0.0057\\[2pt]
Proton& ~~~4.683 $\pm$ 0.012 & ~~~1.2221 $\pm$ 0.0007\\[2pt]
Iron& ~~~3.415 $\pm$ 0.009 & ~~~1.6747 $\pm$ 0.0002\\\hline
\end{tabular}
\end{center}
\end{table}
\subsection{As a function of energy ($E$)}
\begin{figure*}[hbt]
\centerline
\centerline{\hspace{-0.1cm}\includegraphics[width=5.55cm, height=4.45cm]{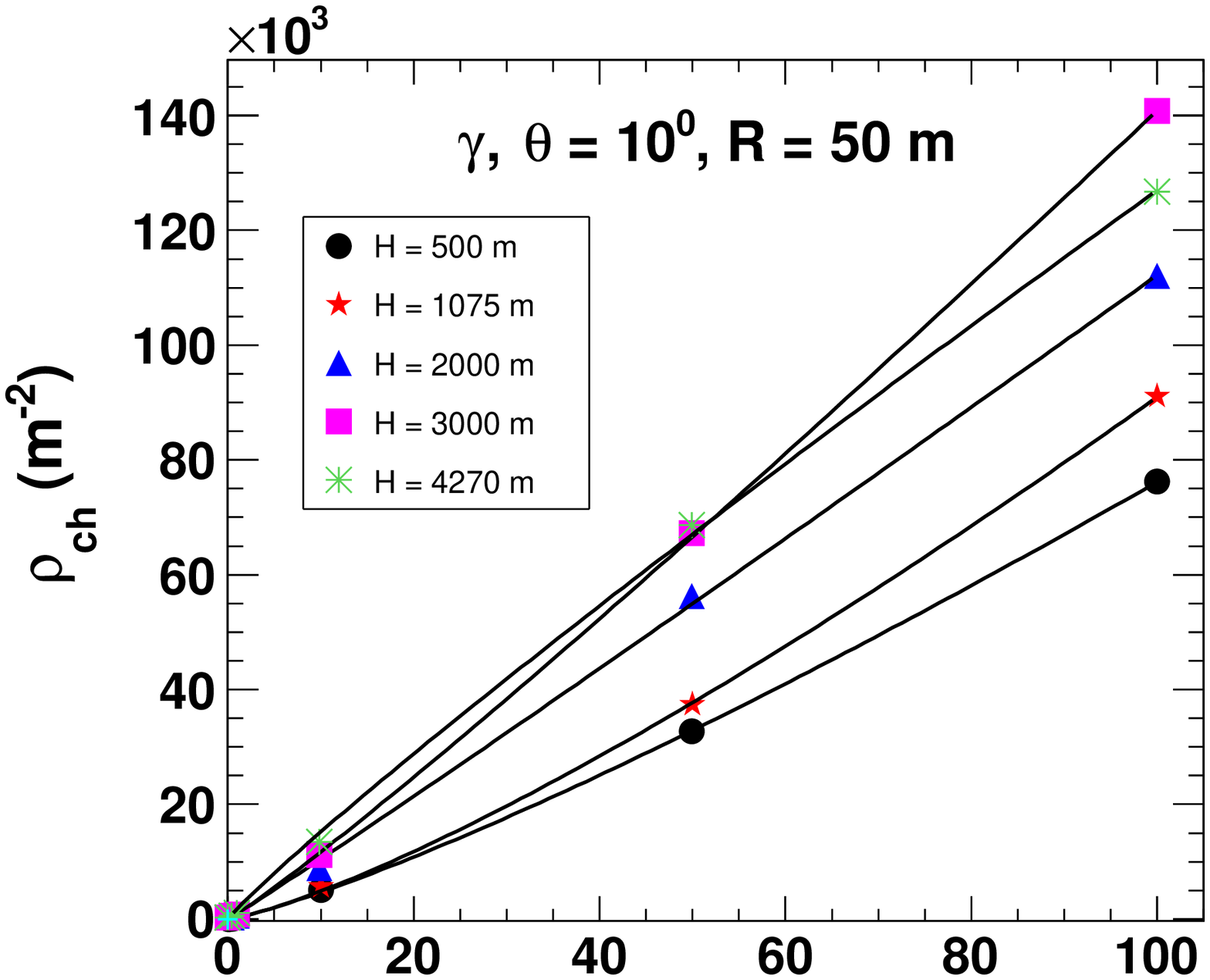} \hspace{0.1cm}
\includegraphics[width=5.10cm, height=4.35cm]{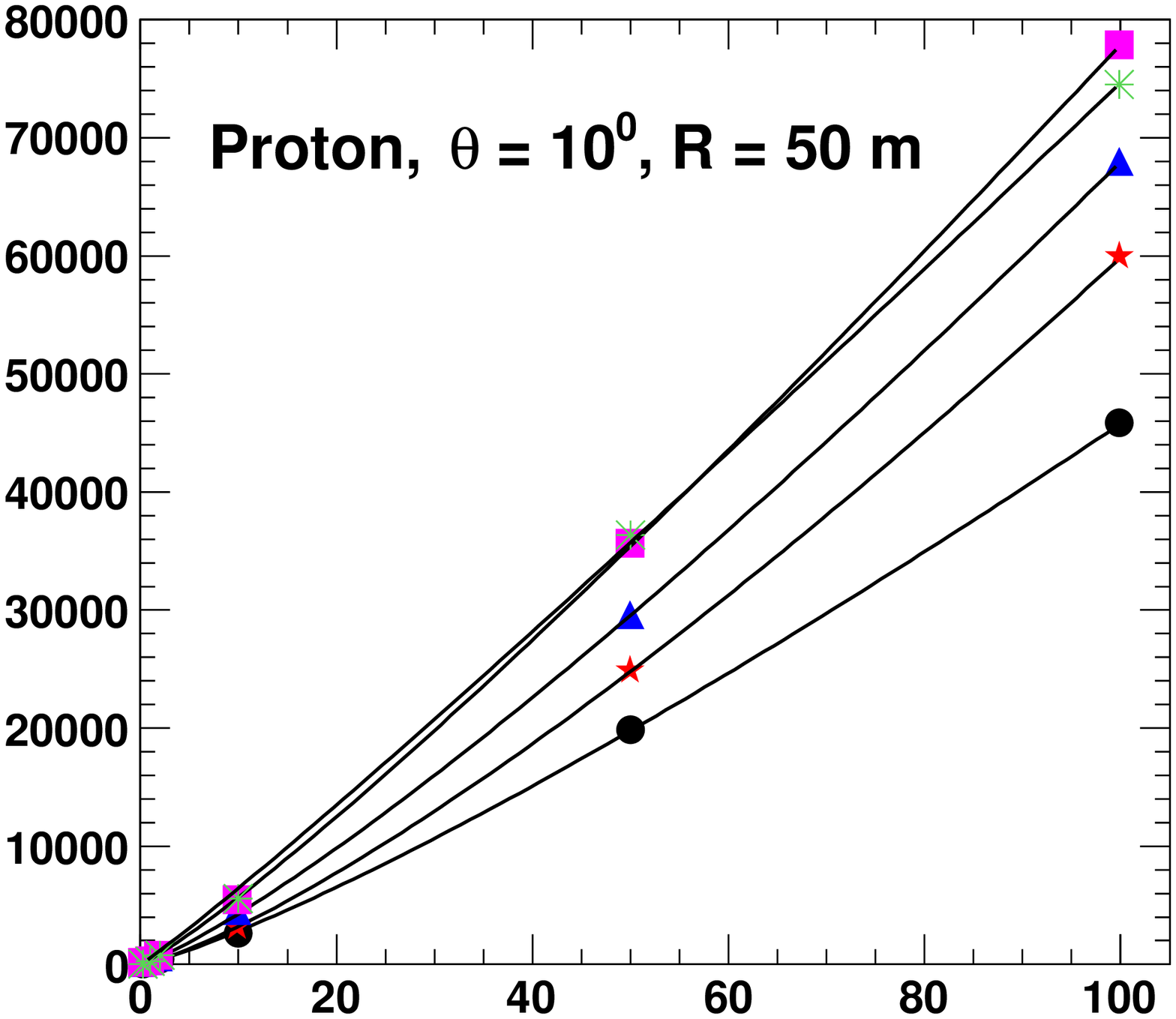}
\includegraphics[width=5.35cm, height=4.23cm]{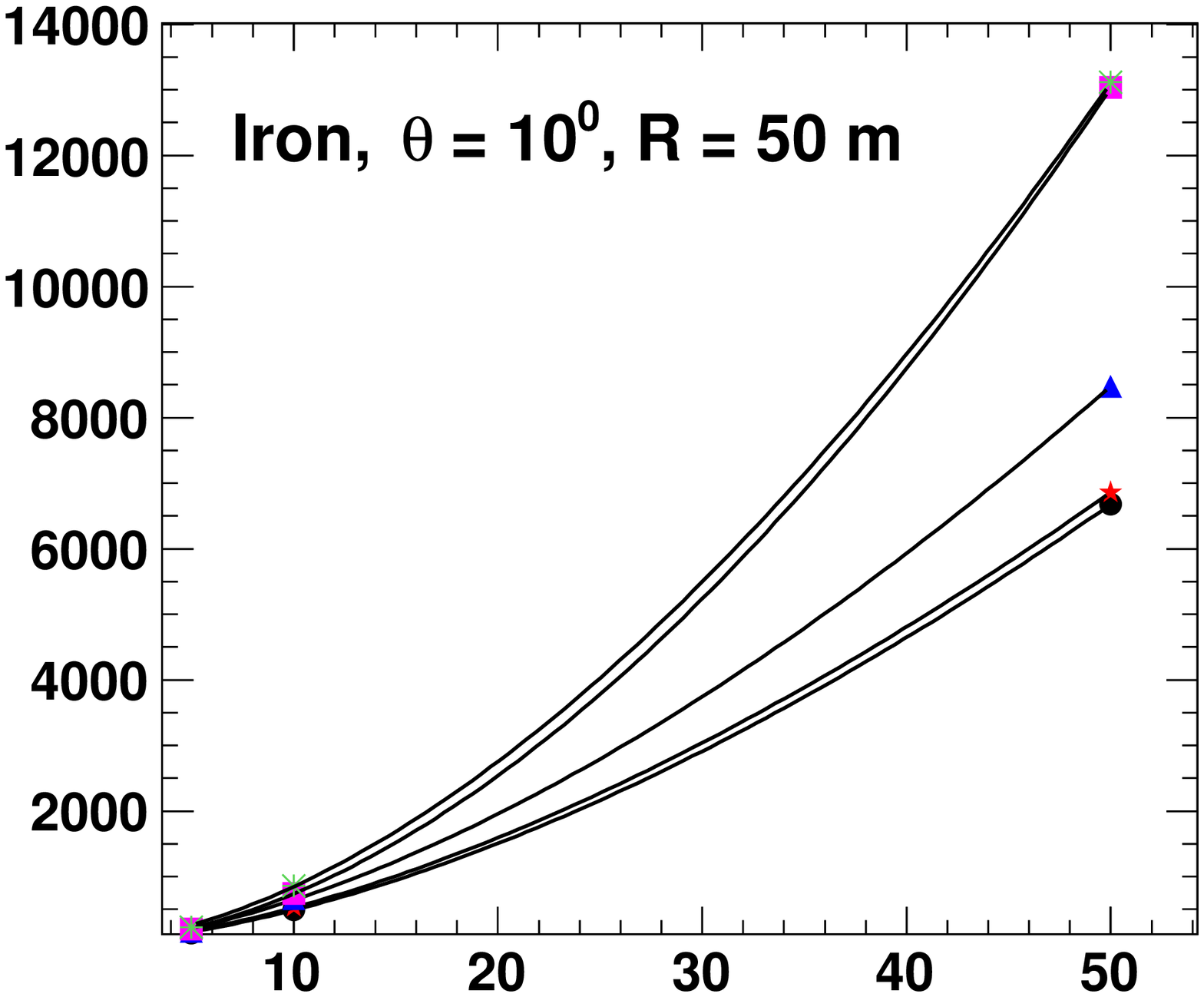}}
\centerline{\includegraphics[width=5.35cm, height=4.5cm]{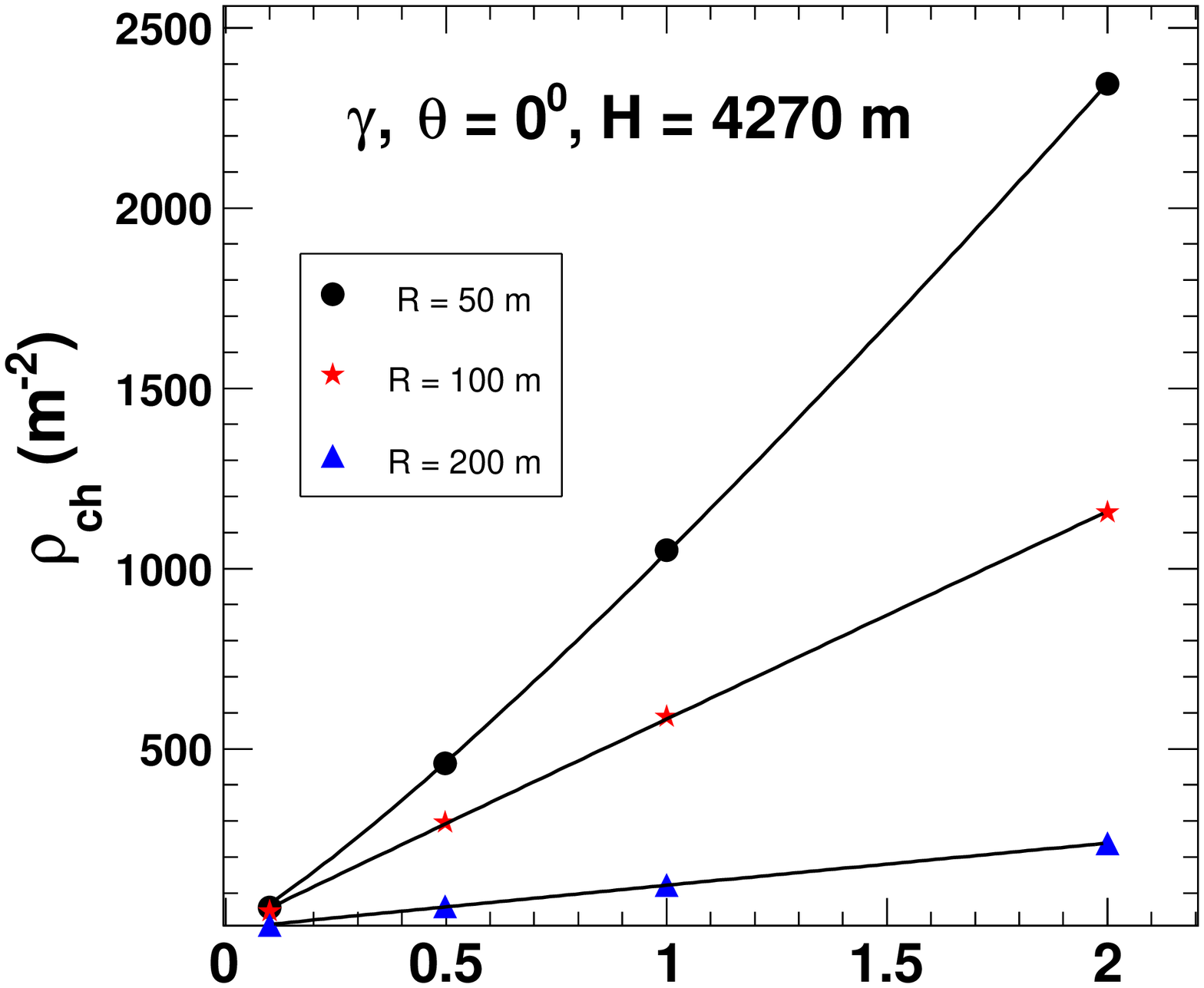}
\includegraphics[width=5.35cm, height=4.5cm]{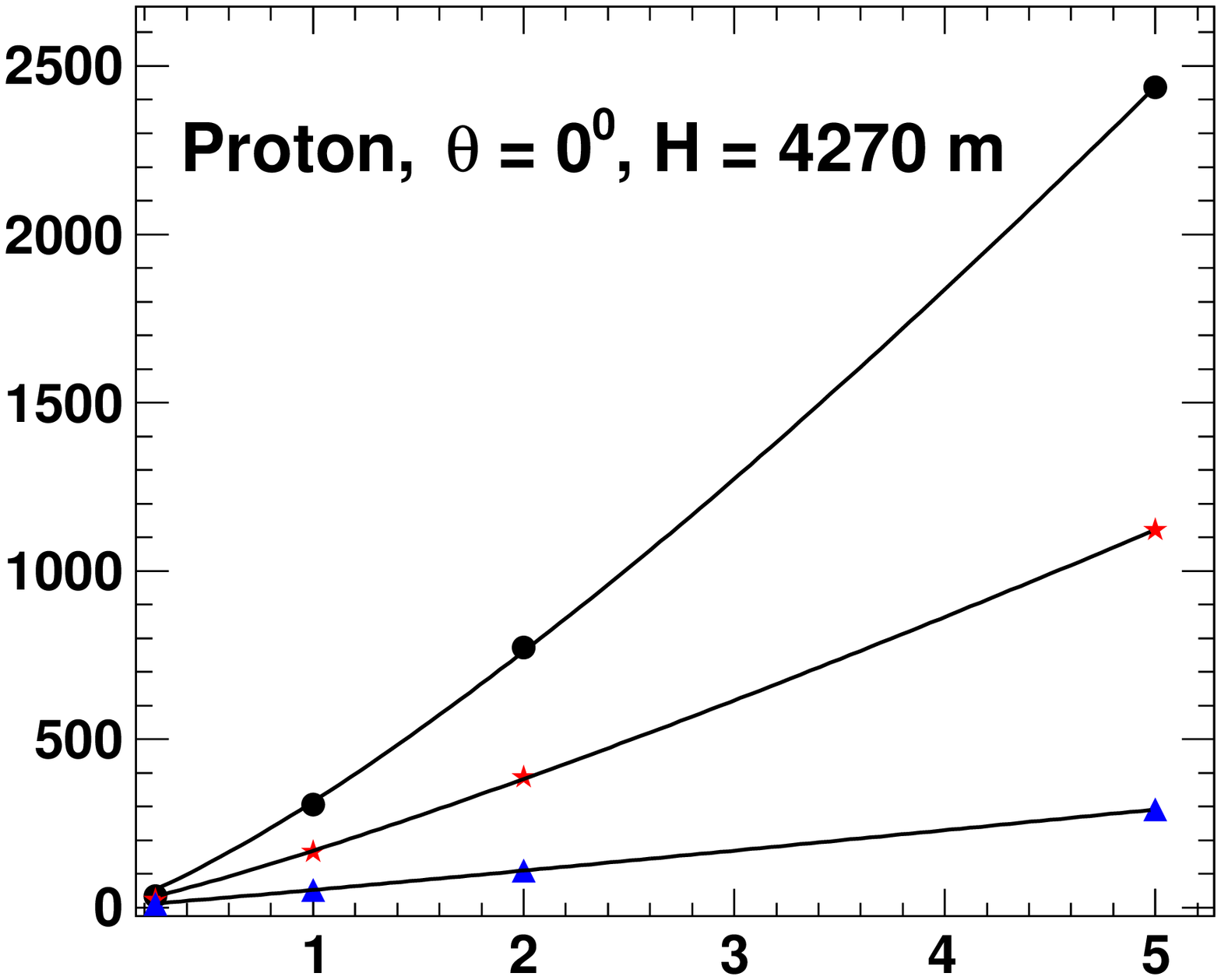}
\includegraphics[width=5.35cm, height=4.5cm]{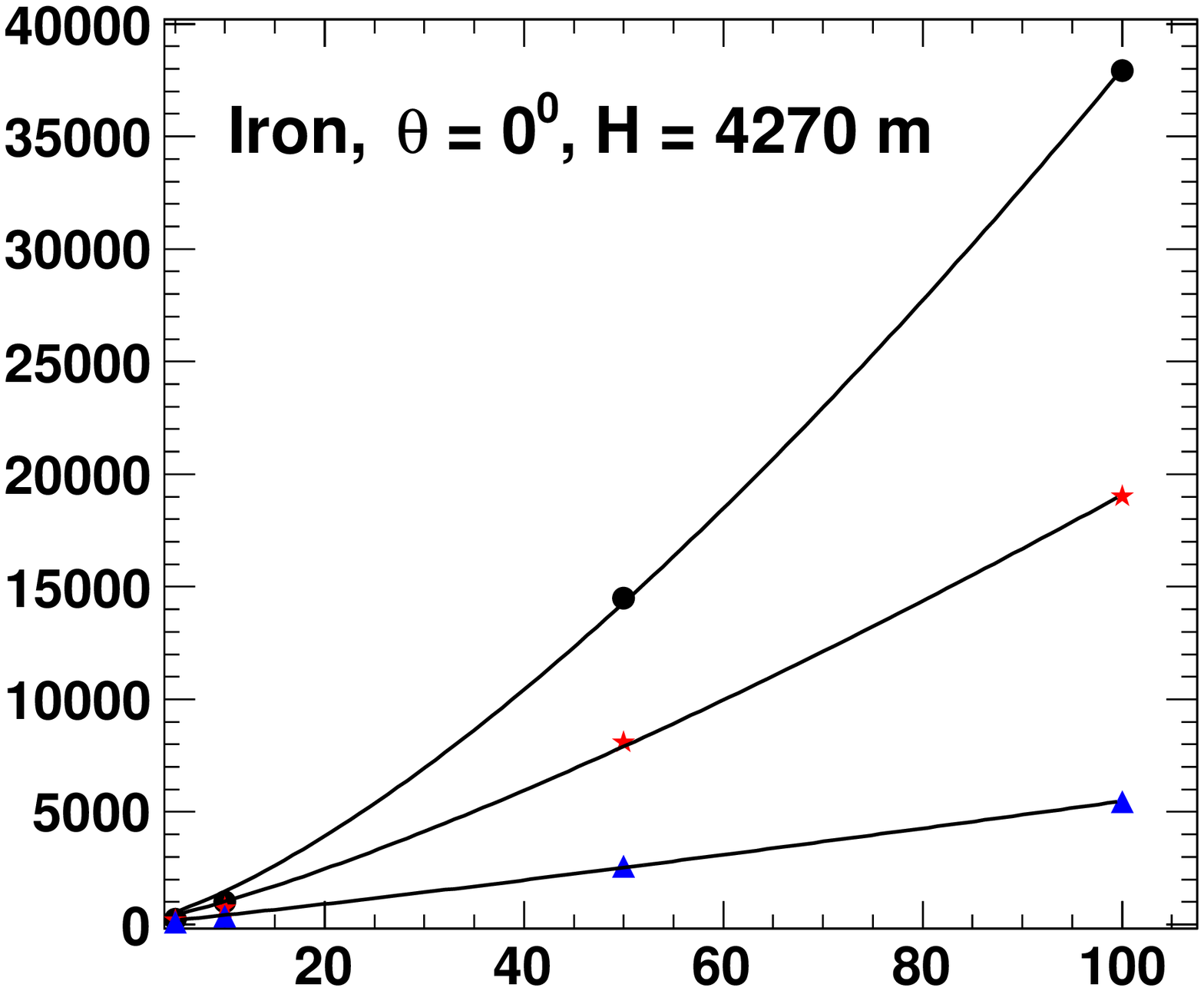}}
\centerline{\includegraphics[width=5.35cm, height=4.5cm]{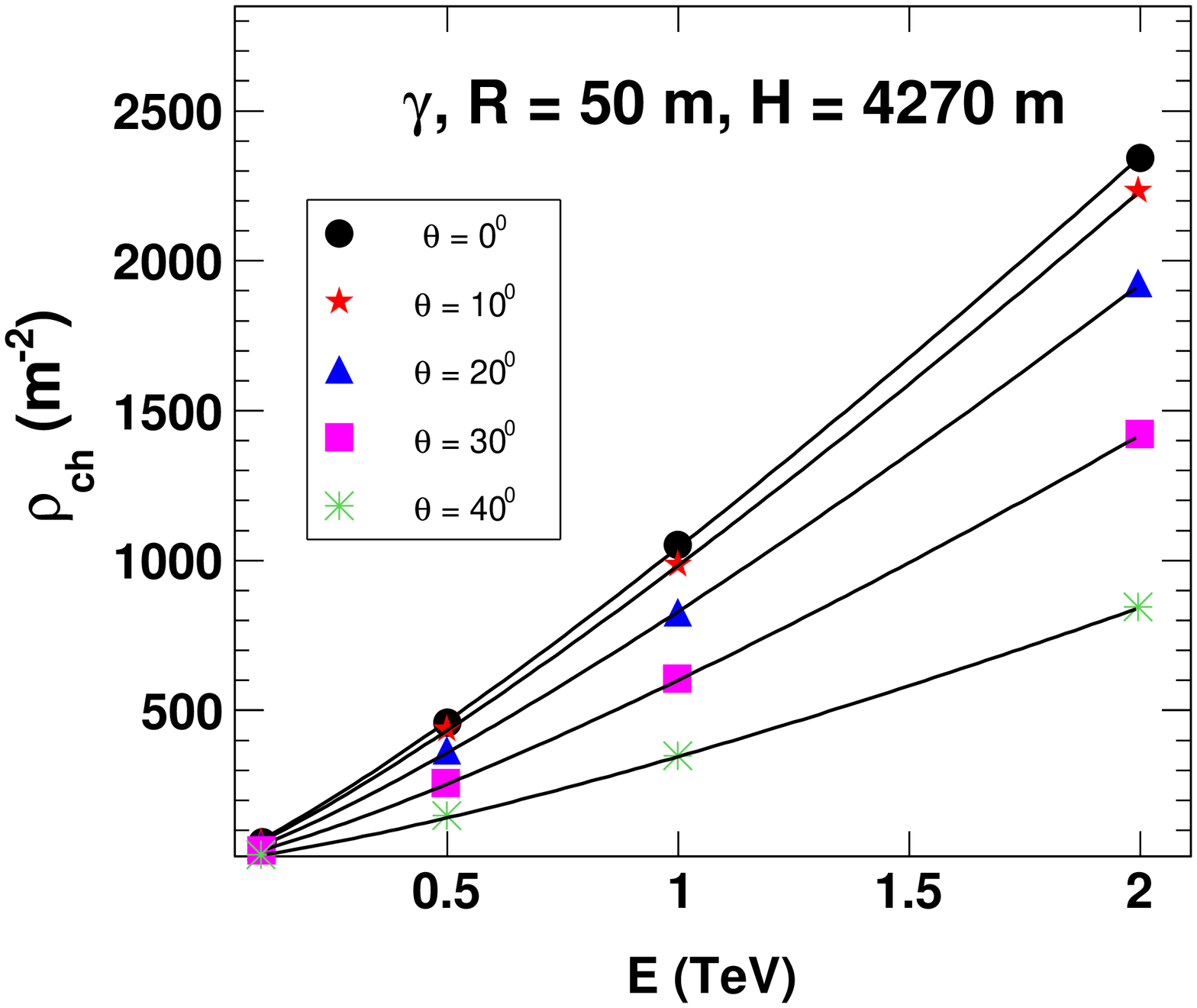}
\includegraphics[width=5.35cm, height=4.5cm]{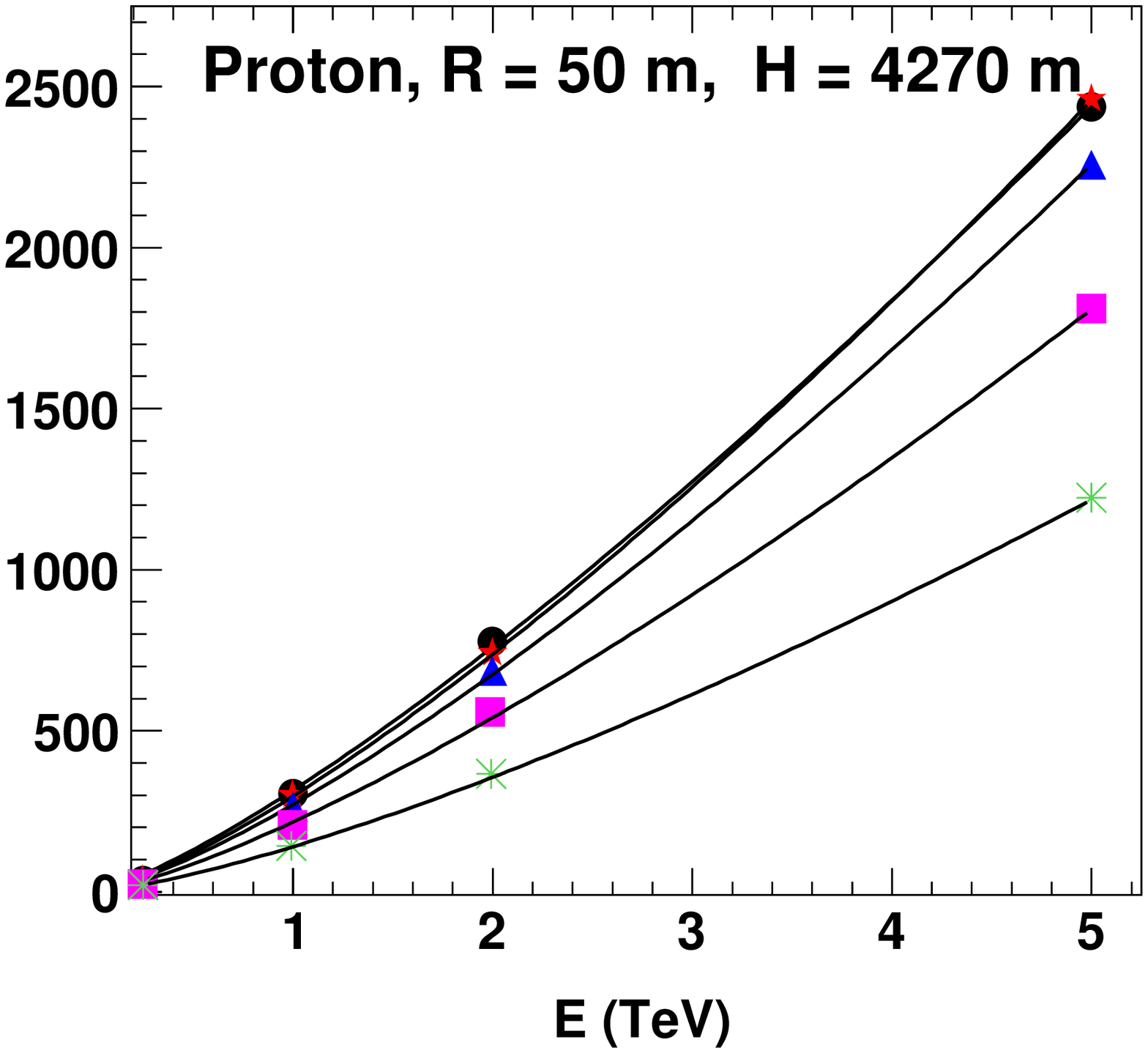}
\includegraphics[width=5.35cm, height=4.5cm]{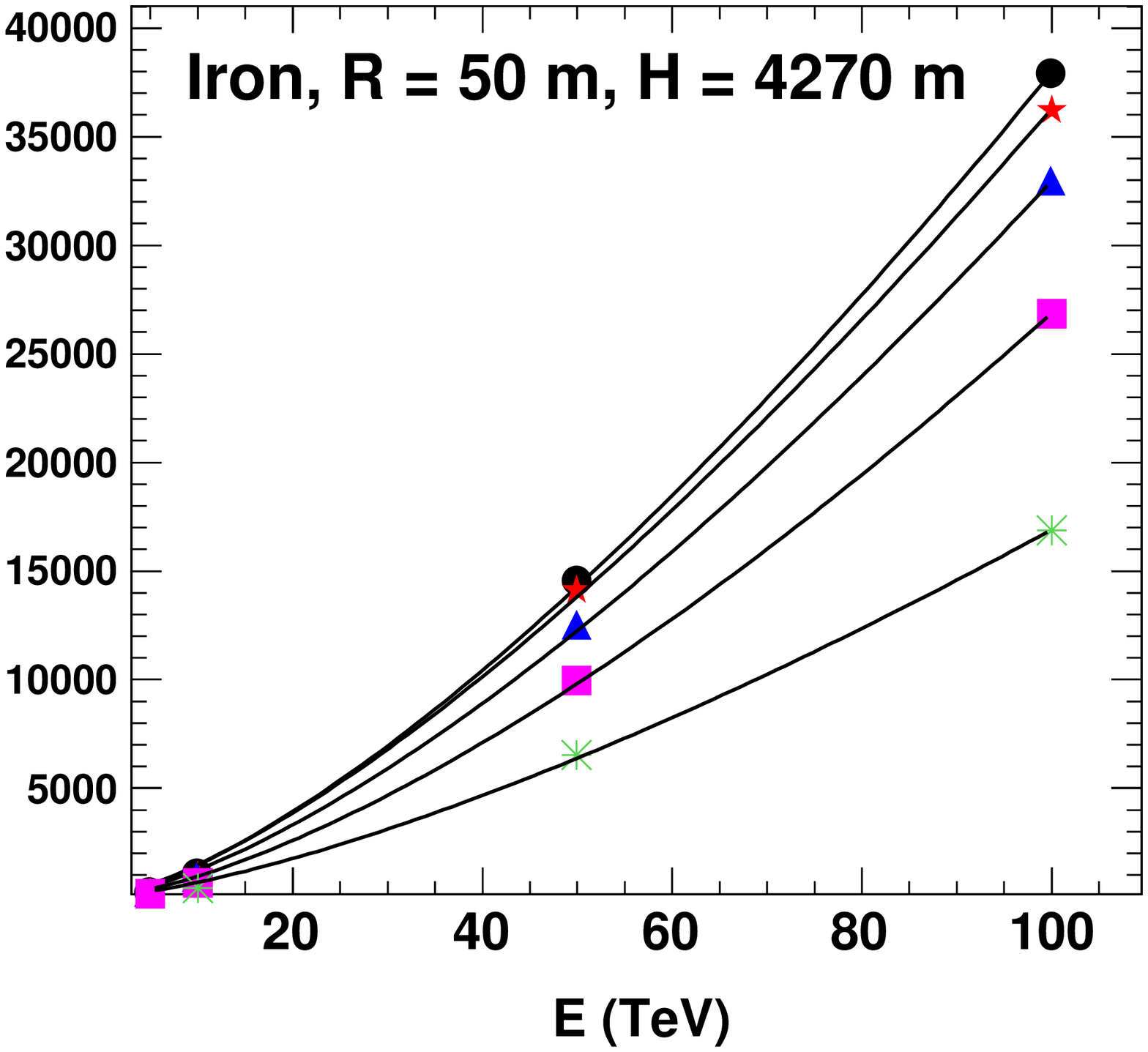}}
\caption{Average Cherenkov photon density ($\rho_{ch}$) for $\gamma$-ray, 
proton and iron primaries is plotted as a function of energy ($E$). The  plots 
in the upper 
most panel show these variations for different altitudes of observation $(H)$ 
keeping $\theta$ and $R$ fixed. The plots in the middle panel show the 
$\rho_{ch}$ 
variations for different values of $R$ keeping $\theta$ and $H$ fixed. Finally, 
the plots in the bottom panel do the same for different values of $\theta$ 
keeping $R$ and 
$H$ fixed at a particular value. The solid lines in the respective plots show 
the result of our parameterisation (\ref{eq2}). The fits are within the limit 
of statistical error ($< \pm 10\%$).}
\label{fig2}
\end{figure*}
The total number of Cherenkov photons produced in a EAS is proportional to 
some power of the primary energy i.e. the density increases with energy for 
all primaries \cite{Nerling, Al-Rubaiee, Das}. Hence we can parameterise the 
dependence of Cherenkov photon density on energy of the primary as
\begin{equation}
\rho(E) = a_{1} E^{c},
\label{eq2}
\end{equation}
where $\rho(E)$ is the energy dependent density function of Cherenkov photons, 
$a_{1}$ and $c$ are parameters of the function, $E$ is the energy of the 
primary. $a_{1}$ and $c$ have different values for different primaries. Our 
parameterisation Eq.(\ref{eq2}) as compared to the CORSIKA simulation results 
are shown in Fig.\ref{fig2}. $\gamma$-ray produces only EM showers whereas 
proton and iron primaries produce hadronic showers along with the EM showers. 
Also different heights of the shower maximum for the three primaries results 
in an almost linear and faster rate of variation of $\rho_{ch}$ with energy of 
the $\gamma$-ray primary, whereas a comparatively slower and non linear 
increase for the proton and iron primaries \cite{Das}. This experimental fact 
is verified by our parameterisation Eq.(\ref{eq2}). The proposed 
parameterisation Eq.(\ref{eq2}) gives a good overall description of the 
simulated data for all values of the core distance ($R$), altitude of 
observations ($H$) and primary energies ($E$) considered in our paper as 
clearly seen in Fig.\ref{fig2}. Due to some technical issues and space 
constrains we could not choose a large range of energies. However, we have 
tried to extend our study upto 100 TeV for the iron primary. Table \ref{tab3}
shows the values of the fitted parameters of the Eq.(\ref{eq2}) to the 
$\rho_{ch}$ distributions as a function of $E$ for $\gamma$-ray, proton and 
iron primary at $H$ = 4270 m, $\theta$ = 10$^{0}$ and $R$ = 50 m, to get
an idea about the parameters in the Eq.(\ref{eq2}).
\begin{table}[ht]
\caption{Values of the fitted parameters of the Eq.(\ref{eq2}) to the $\rho_{ch}$ distributions as a function of $E$ for $\gamma$-ray, proton and iron primary at $H$ = 4270 m, $\theta$ = 10$^{0}$ and $R$ = 50 m.} \label{tab3}
\begin{center}
\begin{tabular}{ccc}\hline
Primary & ~~$a_{1}$ & ~~~$c$ \\\hline\\[-7pt]

$\gamma$& ~~11.1469 $\pm$ 2.4964 & ~~~0.9220 $\pm$ 0.0349\\[2pt]
Proton& ~~~~0.0448 $\pm$ 0.0084 & ~~~1.0638 $\pm$ 0.0309\\[2pt]
Iron& ~~~~0.0014 $\pm$ 0.0001 & ~~~1.7027 $\pm$ 0.0002\\\hline
\end{tabular}
\end{center}
\end{table}
\subsection{As a function of zenith angle ($\theta$)}
\begin{figure*}[ht]
\centerline
\centerline{\includegraphics[width=5.5cm, height=4.5cm]{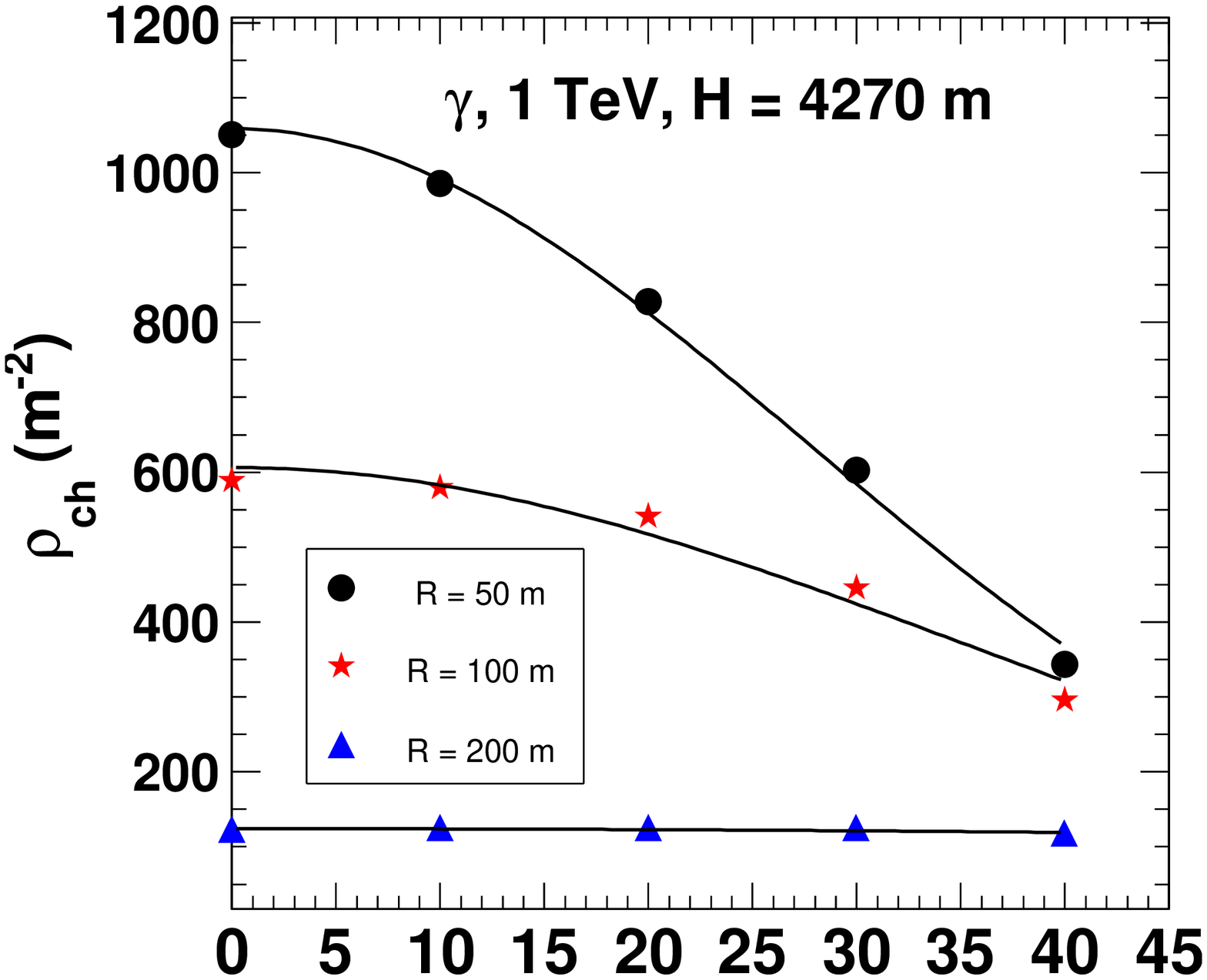}
\includegraphics[width=5.35cm, height=4.5cm]{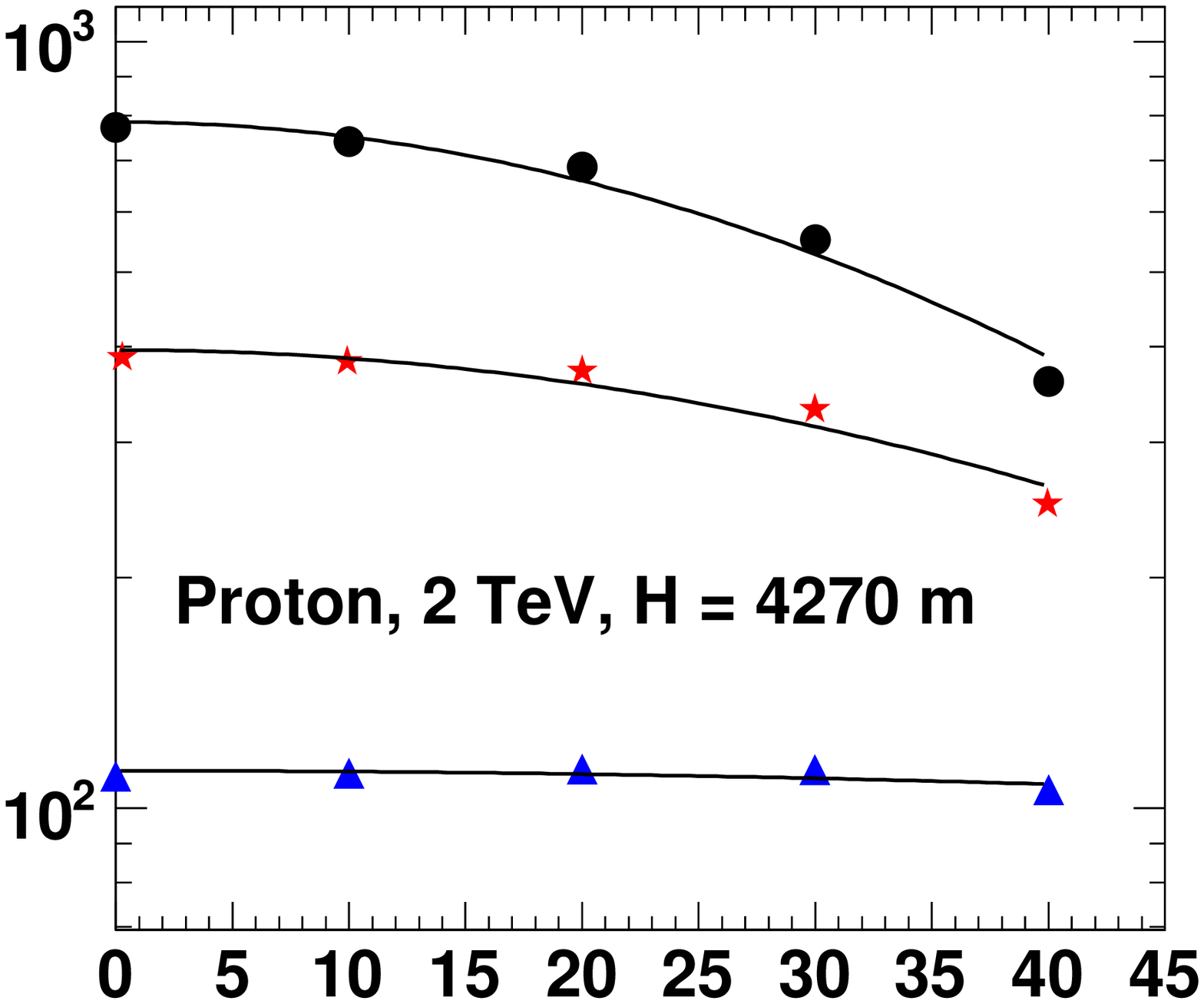}
\includegraphics[width=5.45cm, height=4.5cm]{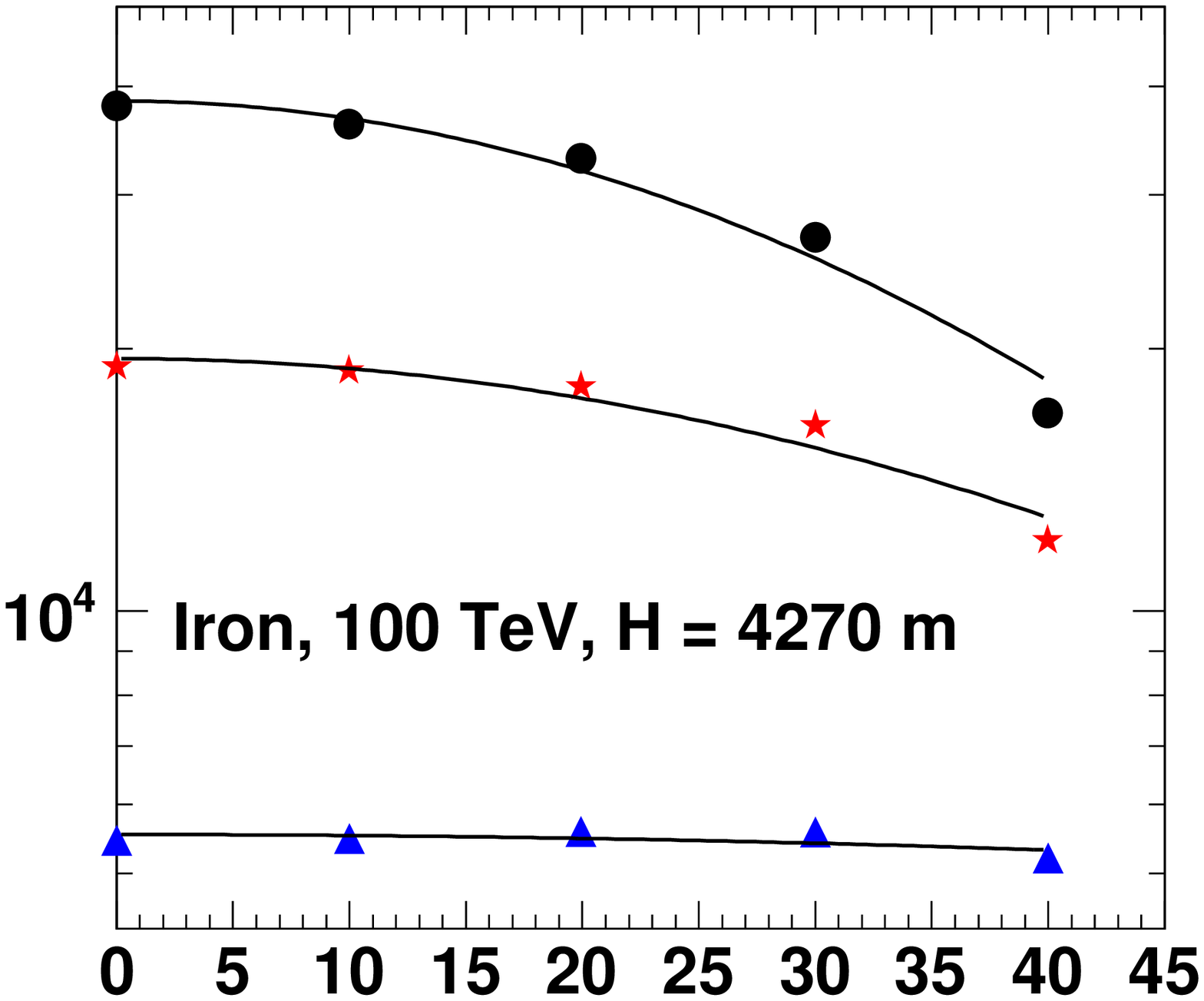}}
\centerline{\hspace{-0.5cm} \includegraphics[width=5.35cm, height=4.5cm]{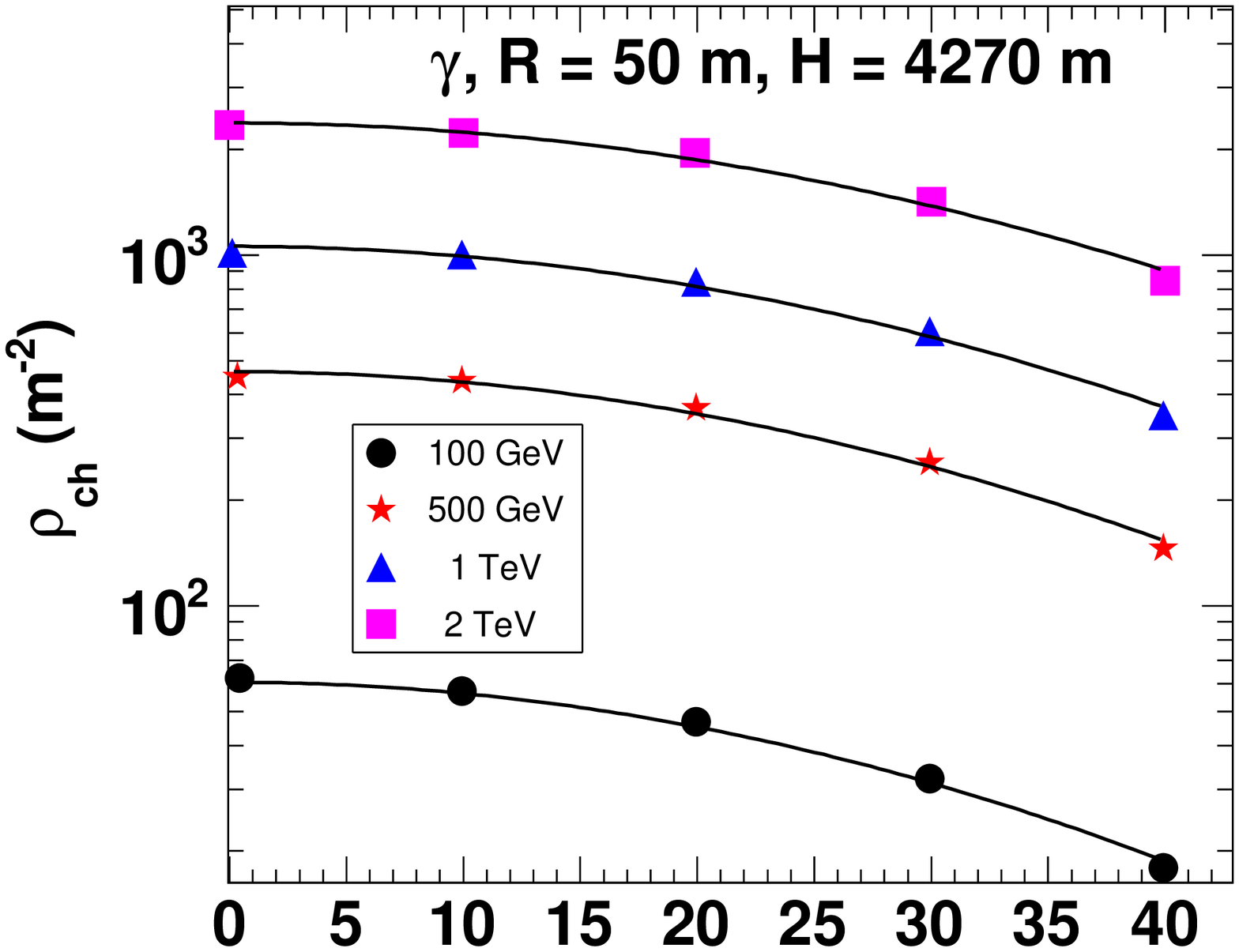}
\includegraphics[width=5.35cm, height=4.5cm]{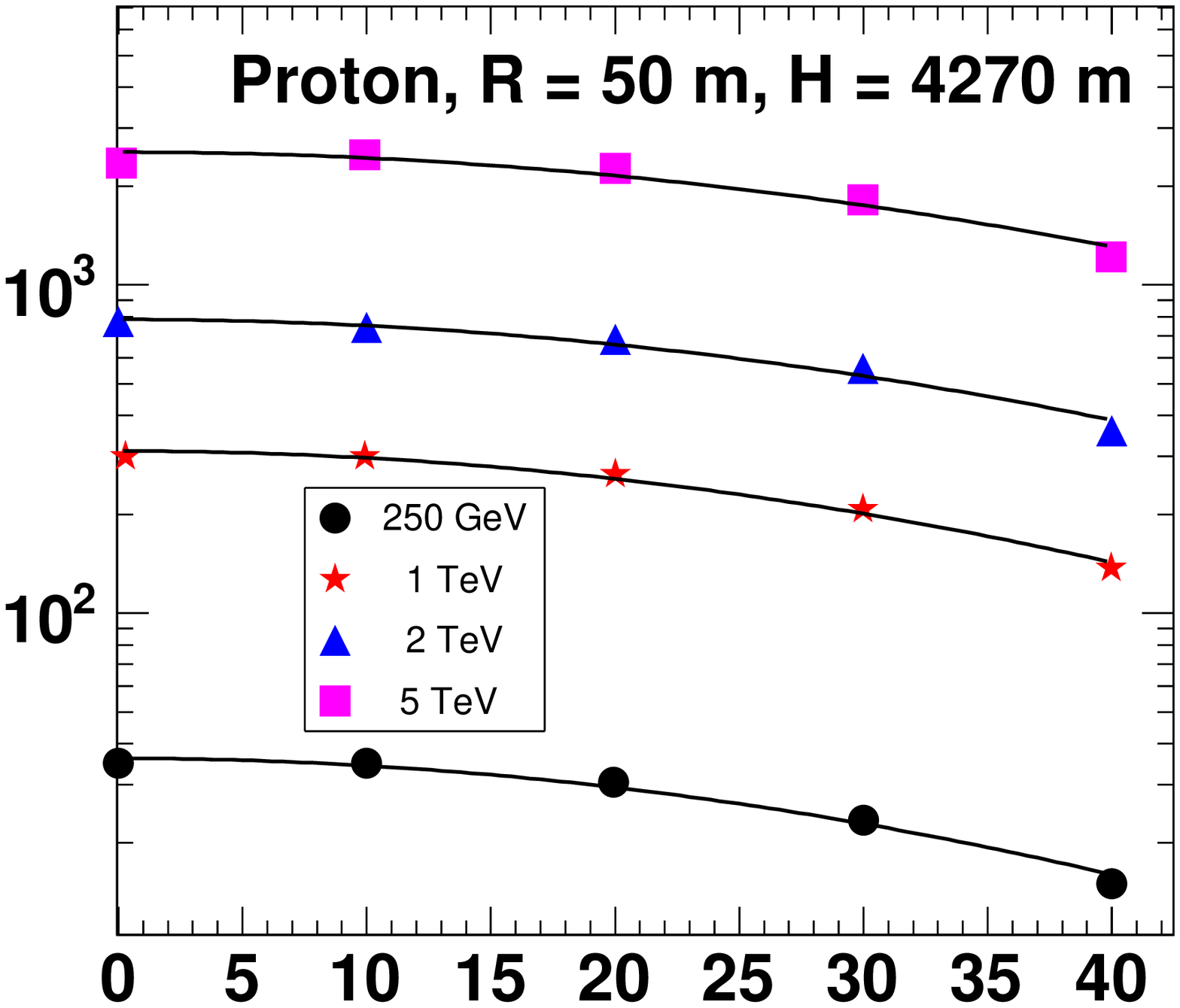}
\includegraphics[width=5.35cm, height=4.5cm]{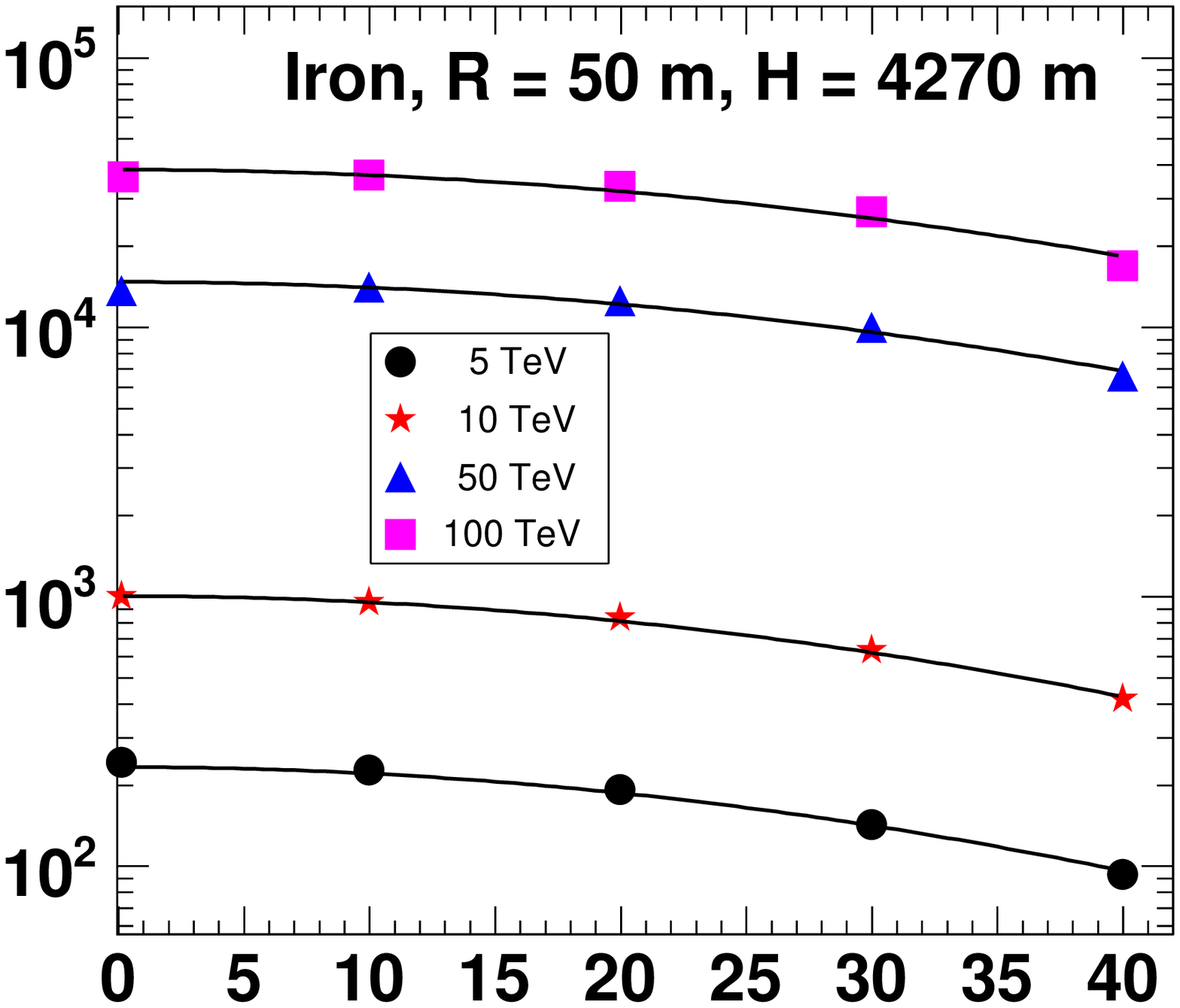}}
\centerline{\includegraphics[width=5.5cm, height=4.5cm]{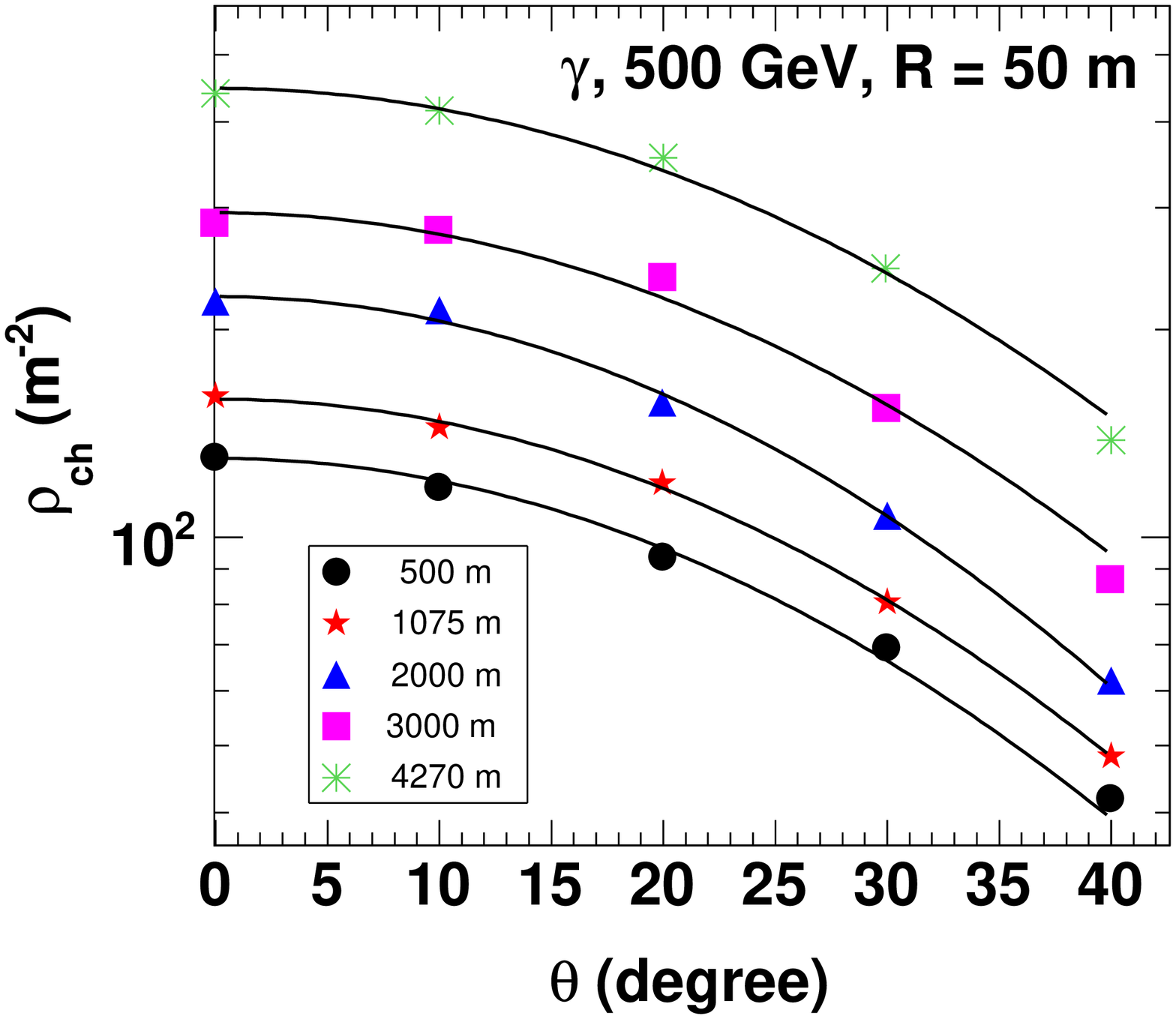}
\includegraphics[width=5.5cm, height=4.5cm]{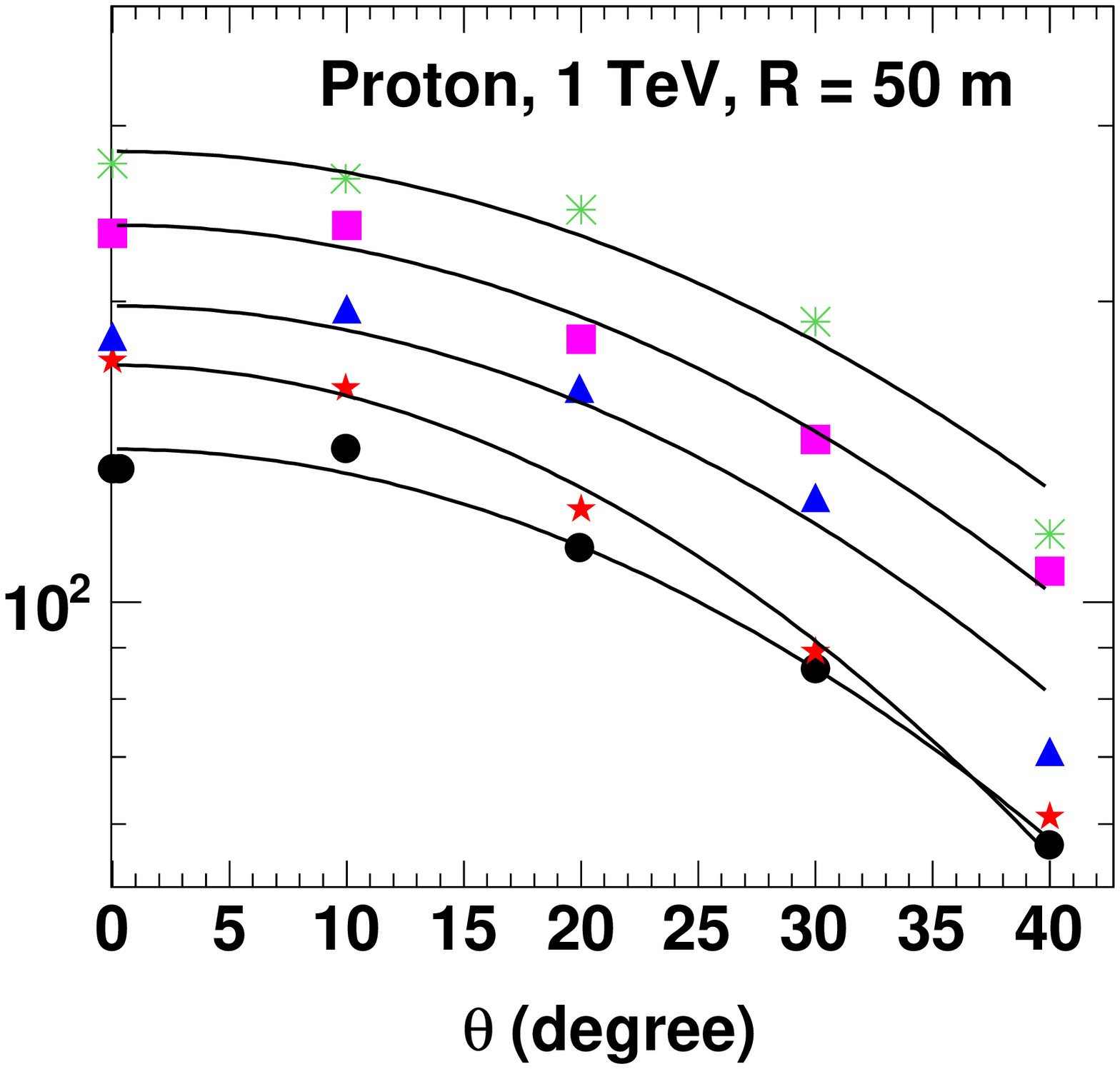}
\includegraphics[width=5.5cm, height=4.5cm]{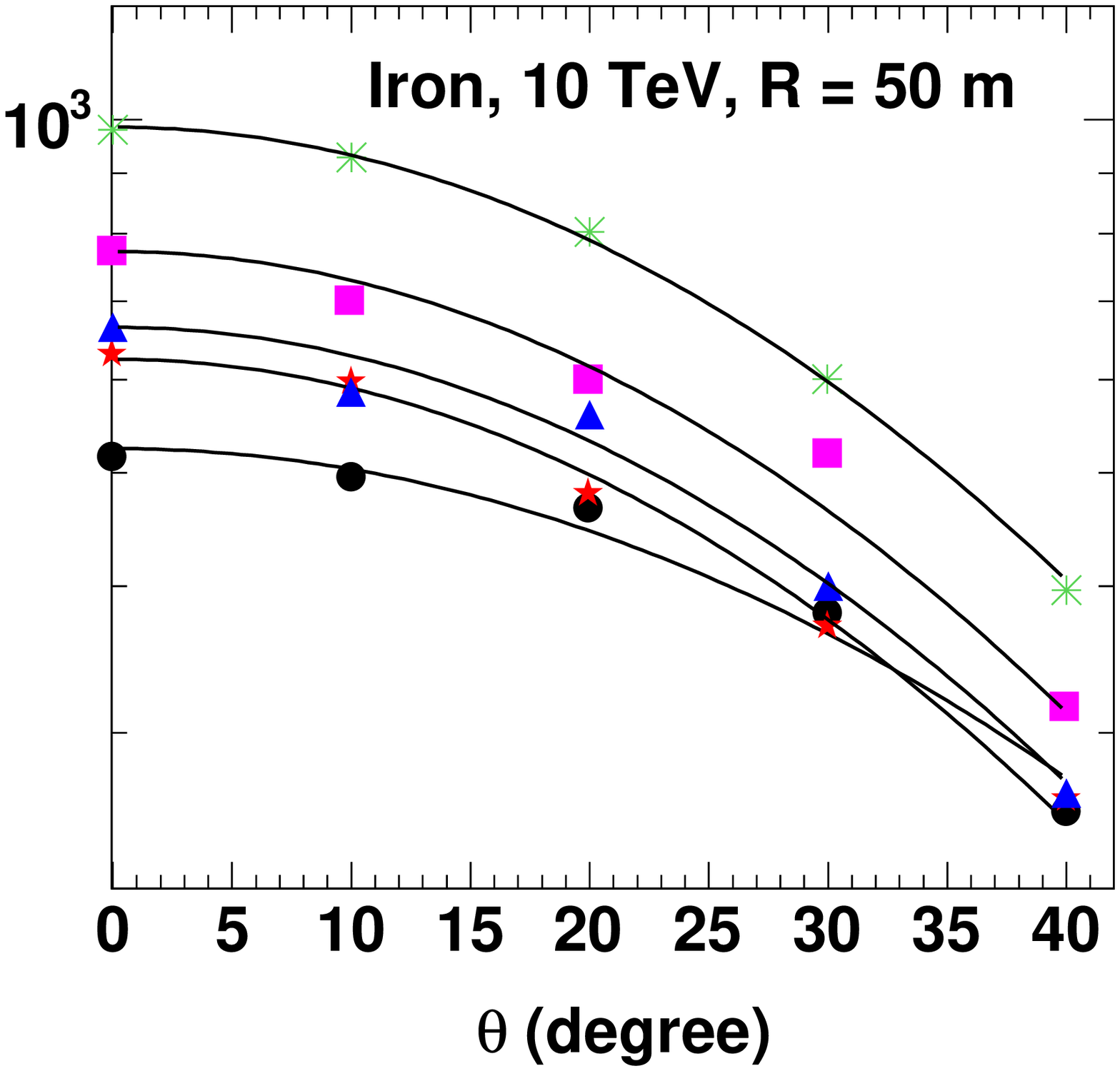}}
\caption{Average Cherenkov photon density ($\rho_{ch}$) for $\gamma$-ray, proton and iron primaries is plotted as a function of zenith angle ($\theta$). The 
plots in the upper most panel show these variations for different core 
distances $(R)$ keeping $E$ and $H$ fixed. The plots in the middle panel show 
the $\rho_{ch}$ variations for different values of $E$ keeping $H$ and $R$ 
fixed. Finally, the plots in the bottom 
panel do the same for different values of $H$ keeping $R$ and $E$ fixed at a particular value. The solid lines in the respective plots shows the result of our parameterisation (\ref{eq3}). The fits are within the limit of statistical 
error ($< \pm 10\%$).}
\label{fig3}
\end{figure*}

The study of the variation of average density of Cherenkov photons with zenith angle shows that for increase in the zenith angle, the density decreases gradually near the shower core, but remains almost constant far away from the core \cite{Das, Abdulsttar}. The general characteristics of the variation of $\rho_{ch}$ with $\theta$ can be parameterised as
\begin{equation}
\rho(\theta) = a_{2} \exp(-(\theta /d)^{2}),
\label{eq3}
\end{equation}
where $\rho(\theta)$ is the zenith angle dependent density function of 
Cherenkov photons, $a_{2}$ and $d$ are parameters of the function, $\theta$ is 
the zenith angle of the incident primary. $a_{2}$ and $d$ have different values 
for different primaries. The results of our parameterisation Eq.(\ref{eq3}) 
slightly vary from the simulated data for high energies at large core distances for all the incident primaries as seen in Fig.\ref{fig3}. This is may be because at large core distances, the variation of density with zenith angle is negligible as only high energetic charged particles could reach at larger distances from the core over the observation level. Again, with increasing zenith angle, depending on the energy of the primary most of the low energy charged particles gets absorbed as the shower now has to cross an additional slant depth. As such the density of Cherenkov photons decreases with increasing ($\theta$). So the quality of our parameterisation is limited by the low statistics. To give an 
imprerssion of the parameters of the Eq.(\ref{eq3}), we have shown in the Table
\ref{tab4} the values of the fitted parameters of this equation to the 
$\rho_{ch}$ distributions as a function of $\theta$ for $\gamma$-ray, proton 
and iron primary at 1 TeV, 2 TeV and 100 TeV energies respectively and at $H$ = 4270 m and $R$ = 100 m.  
\begin{table}
\caption{Values of the fitted parameters of the  Eq.(\ref{eq3}) to the $\rho_{ch}$ distributions as a function of $\theta$ for $\gamma$-ray, proton and iron primary at 1 TeV, 2 TeV and 100 TeV energies respectively and at $H$ = 4270 m and $R$ = 100 m.} \label{tab4}
\begin{center}
\begin{tabular}{ccc}\hline
Primary & ~~$a_{2}$ & ~~~$d$ \\\hline\\[-7pt]

$\gamma$& ~~~0.7581 $\pm$ 0.0088 & ~~~1.6179 $\pm$ 0.0016\\[2pt]
Proton& ~~~0.1553 $\pm$ 0.0026 & ~~~1.6241 $\pm$ 0.0021\\[2pt]
Iron& ~~~0.0032 $\pm$ 0.0001 & ~~~1.7467 $\pm$ 0.0003\\\hline
\end{tabular}
\end{center}
\end{table}
\subsection{As a function of altitude of observation ($H$)}
\begin{figure*}[hbt]
\centerline
\centerline{\includegraphics[width=5.5cm, height=4.5cm]{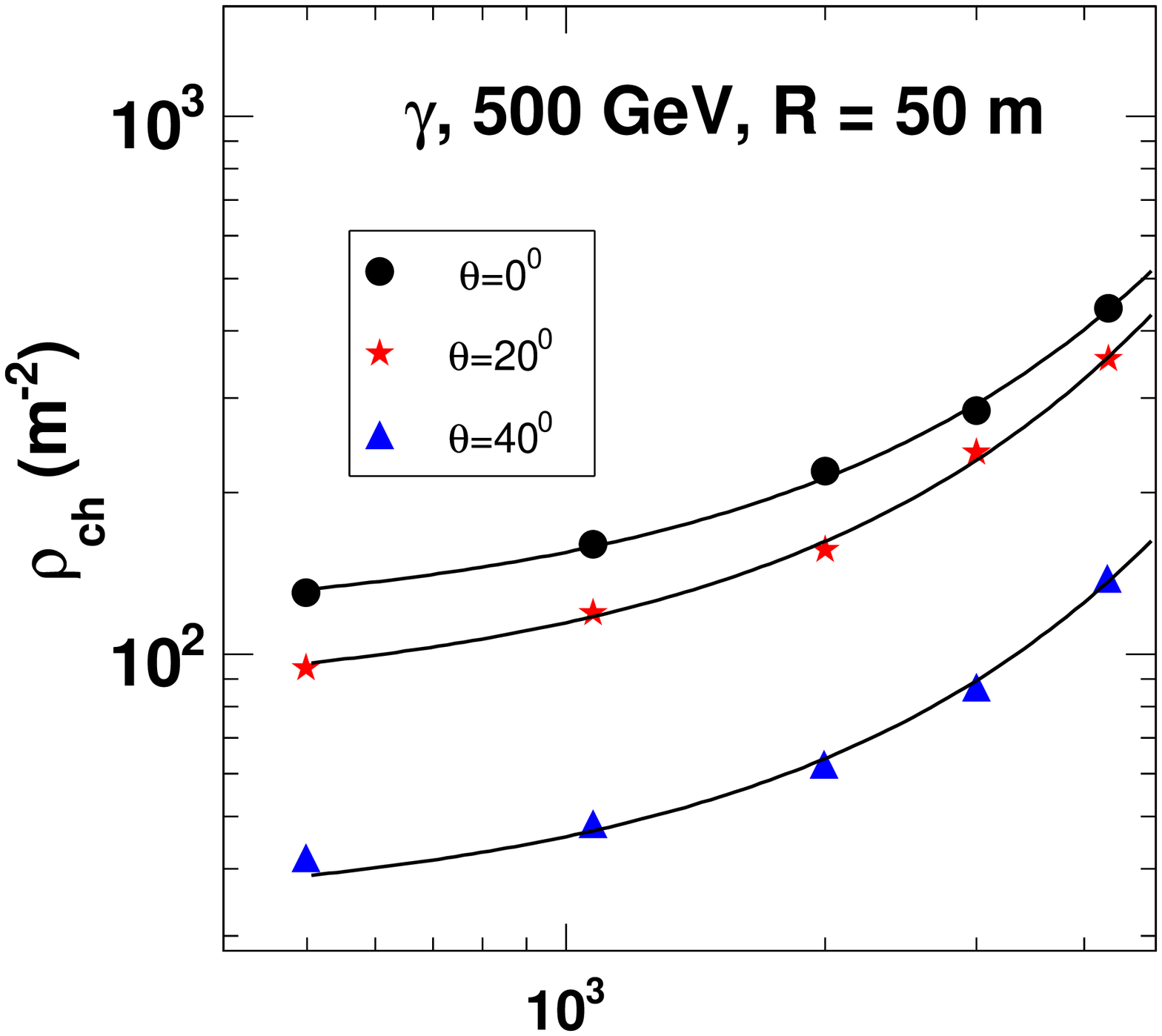}
\includegraphics[width=5.5cm, height=4.5cm]{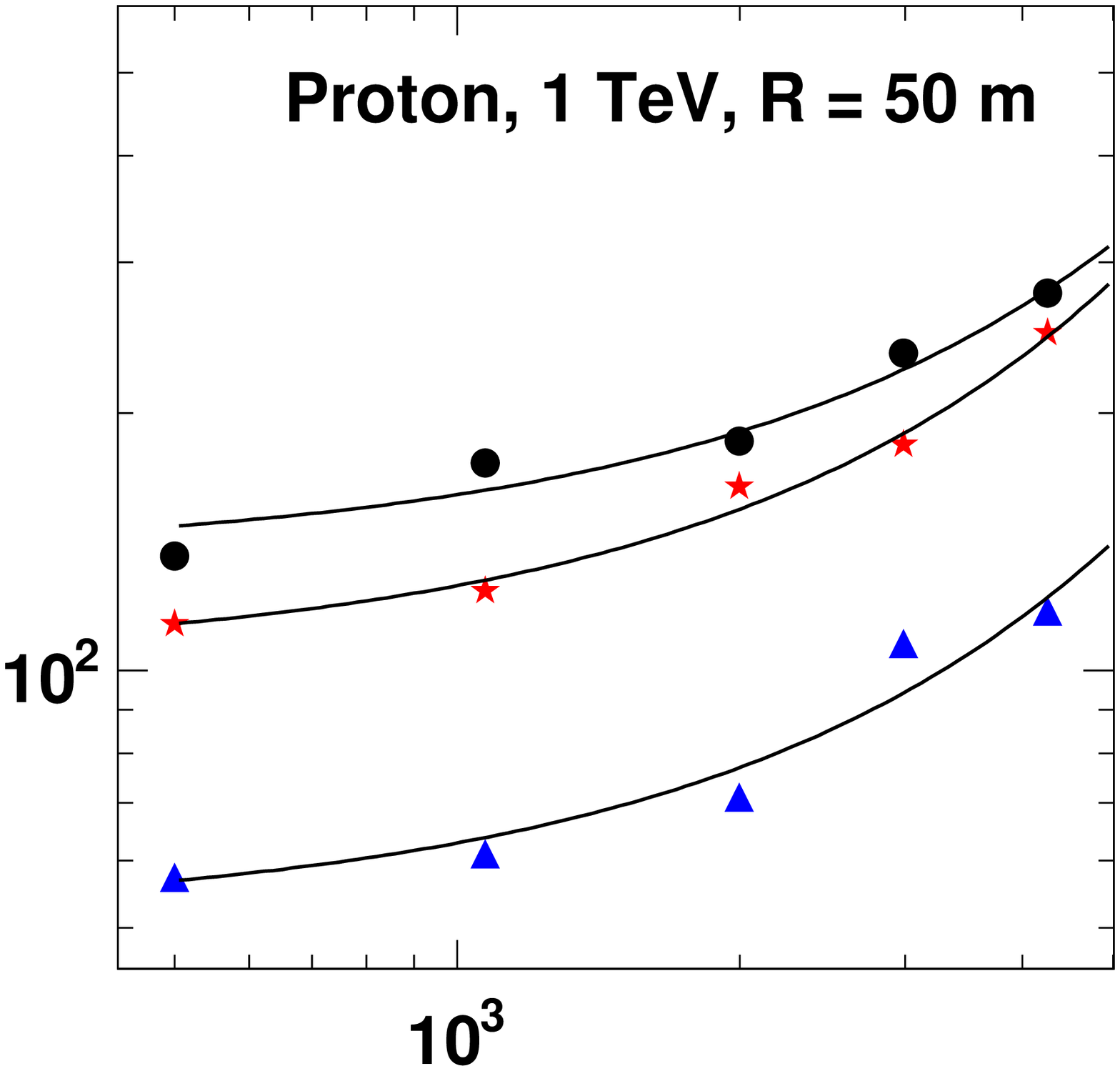}
\includegraphics[width=5.5cm, height=4.5cm]{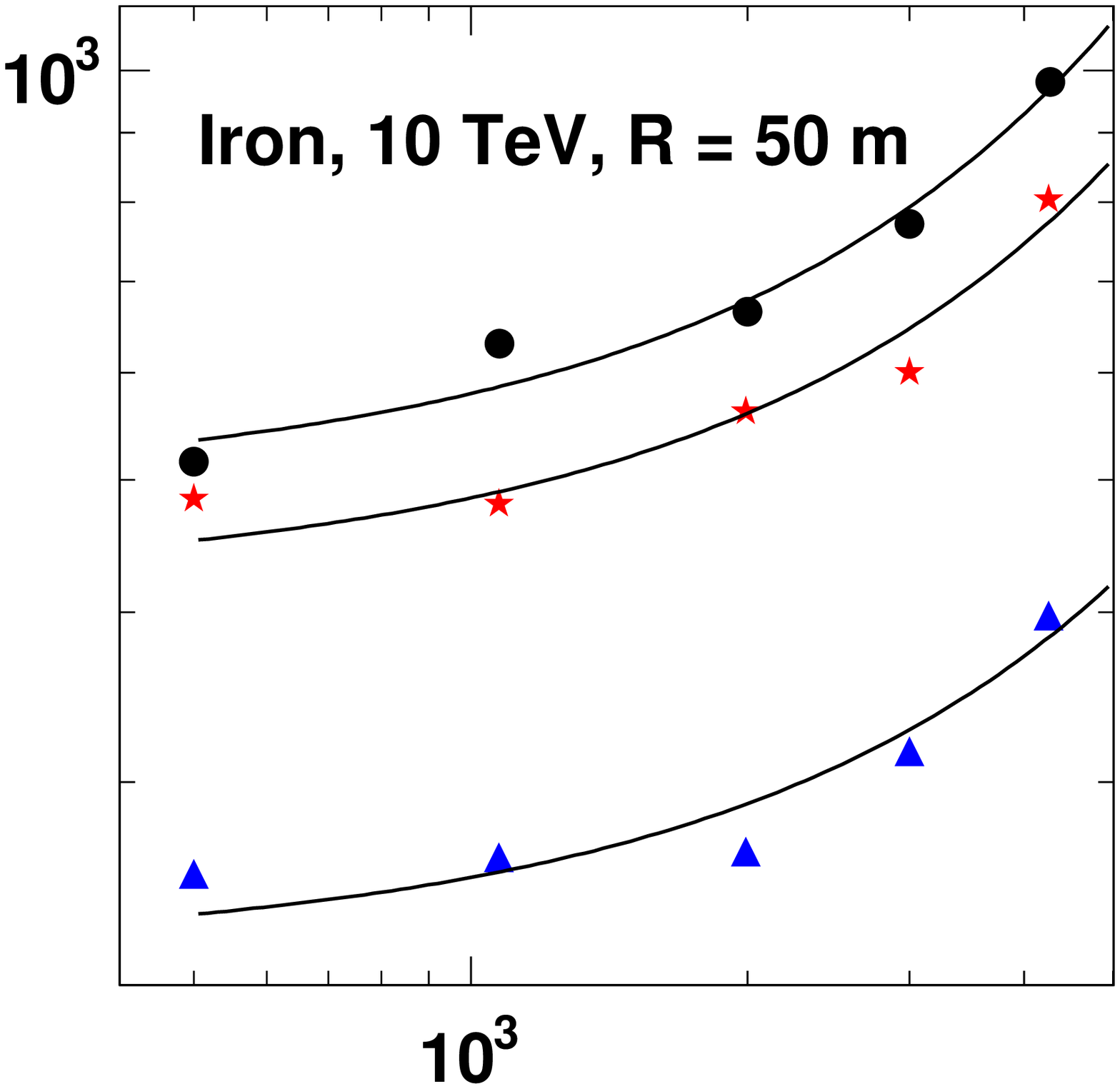}}
\centerline{\includegraphics[width=5.5cm, height=4.5cm]{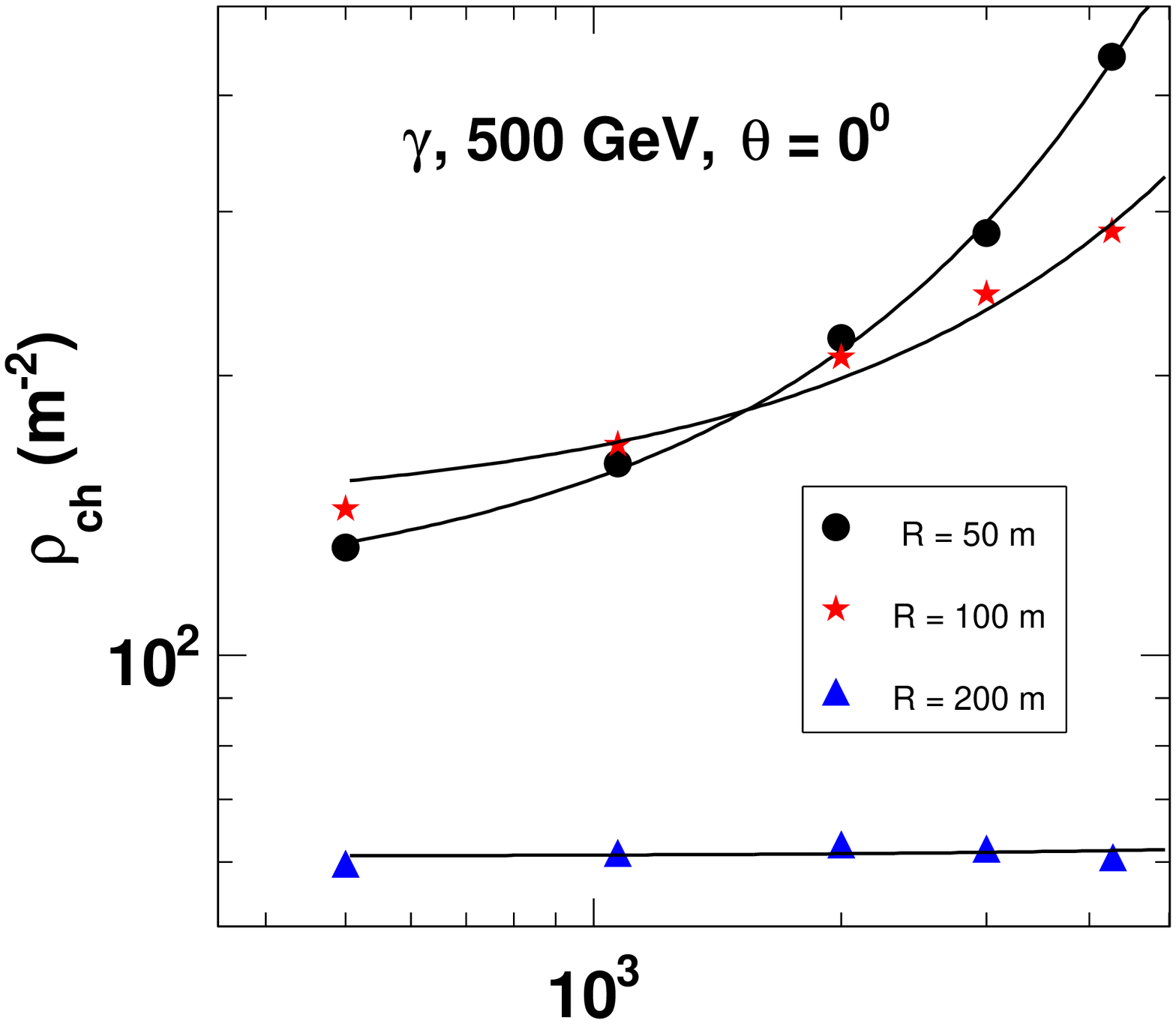}
\includegraphics[width=5.5cm, height=4.5cm]{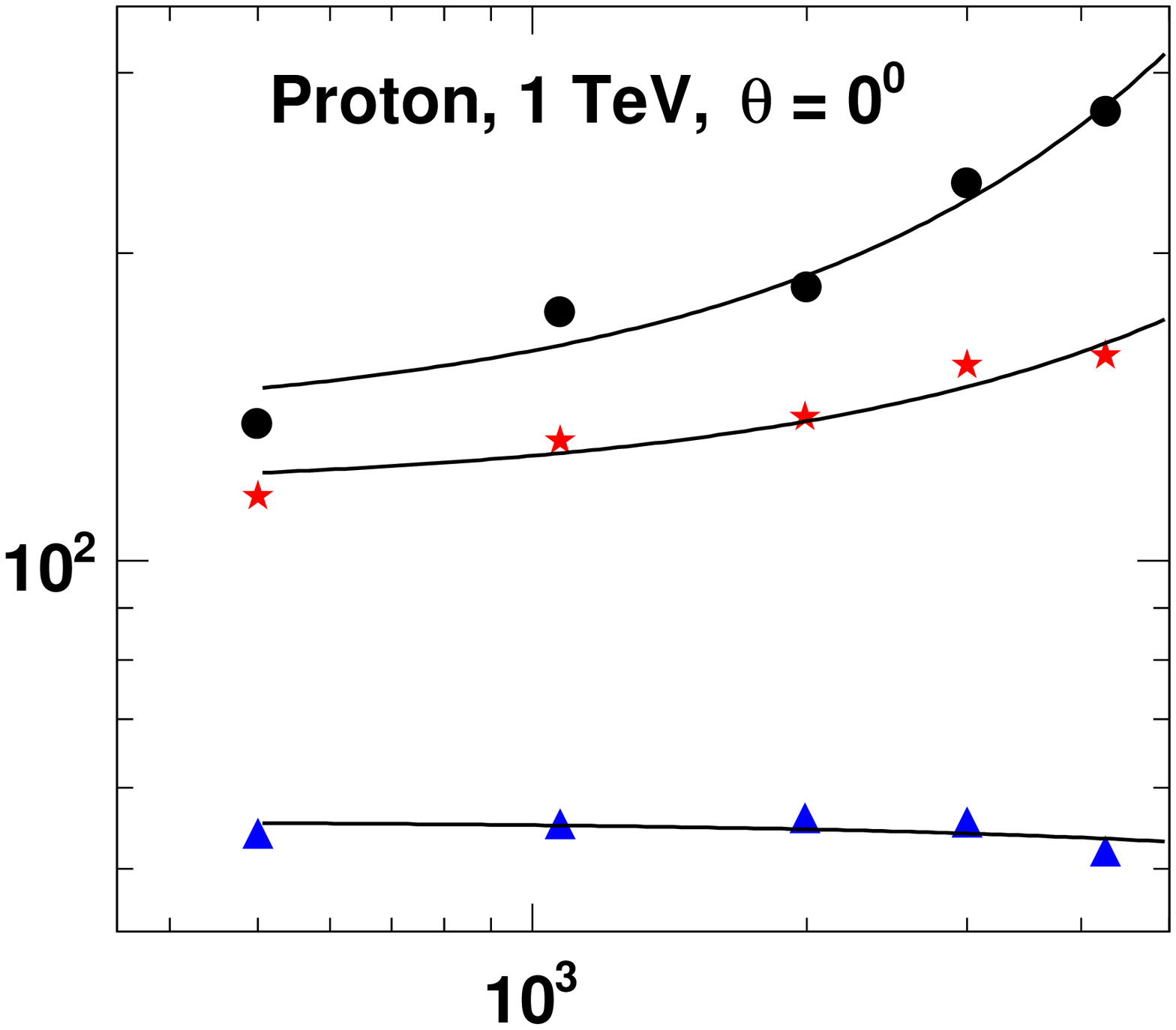}
\includegraphics[width=5.5cm, height=4.5cm]{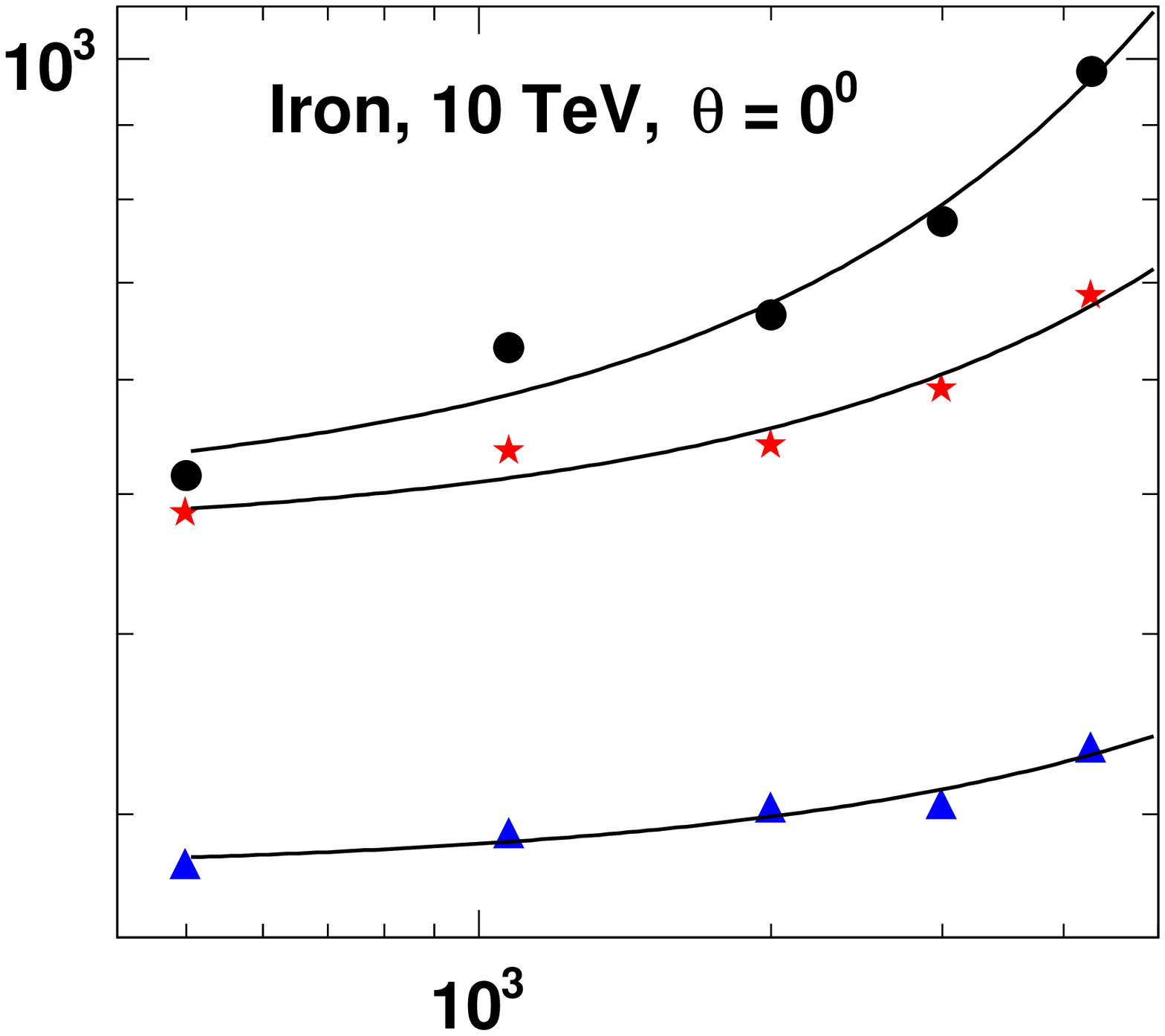}}
\centerline{\includegraphics[width=5.5cm, height=4.5cm]{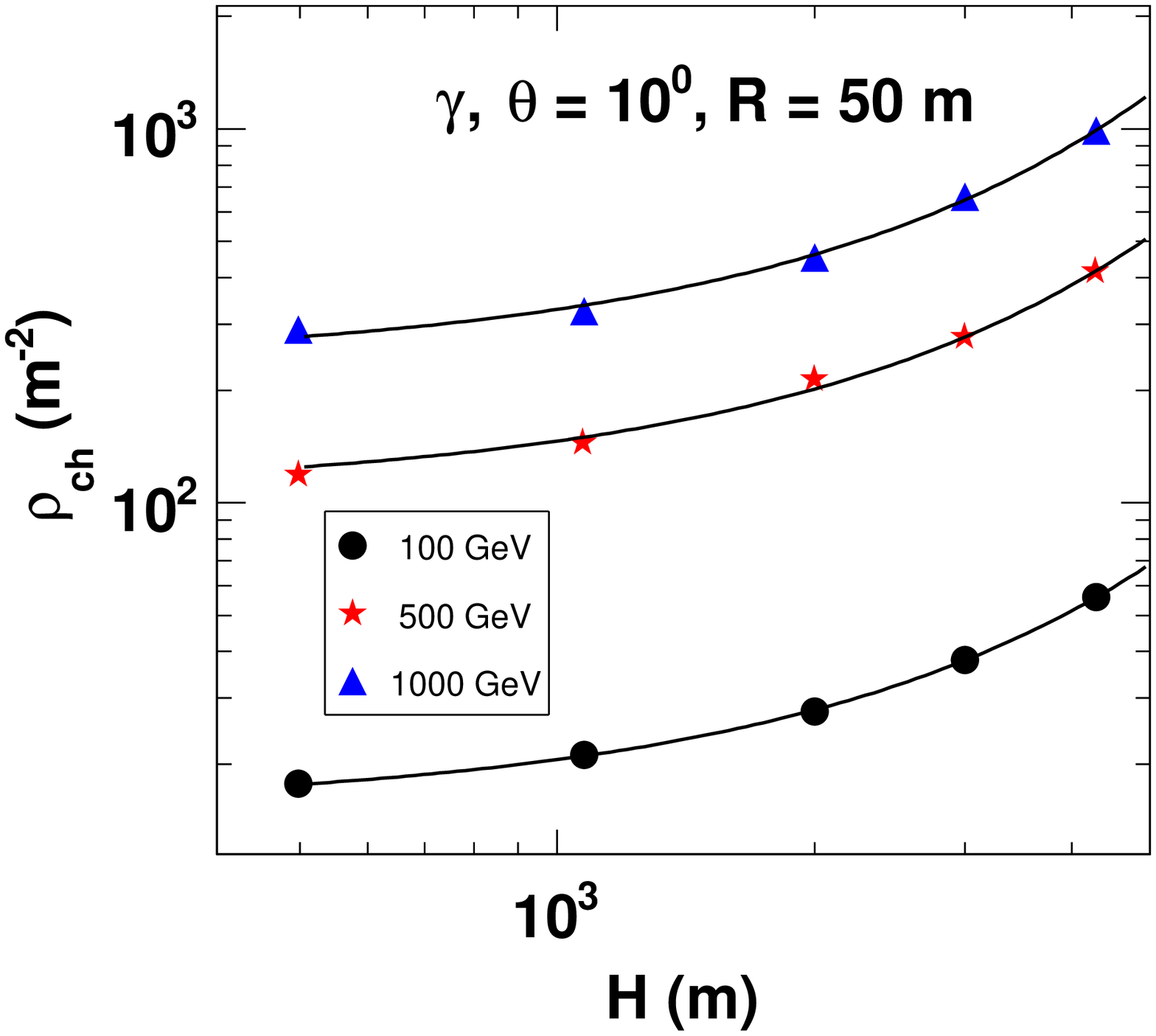}
\includegraphics[width=5.5cm, height=4.5cm]{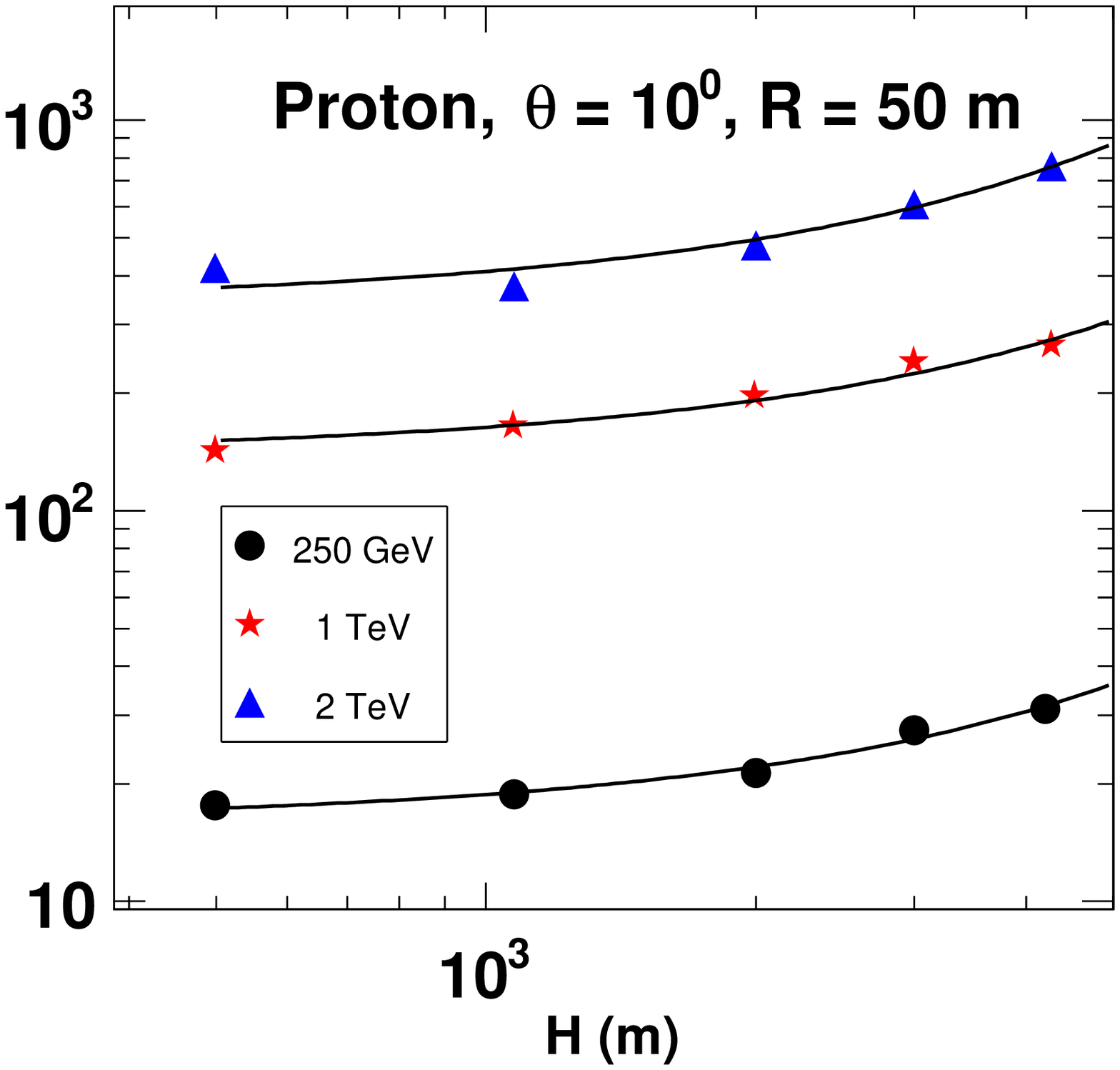}
\includegraphics[width=5.5cm, height=4.5cm]{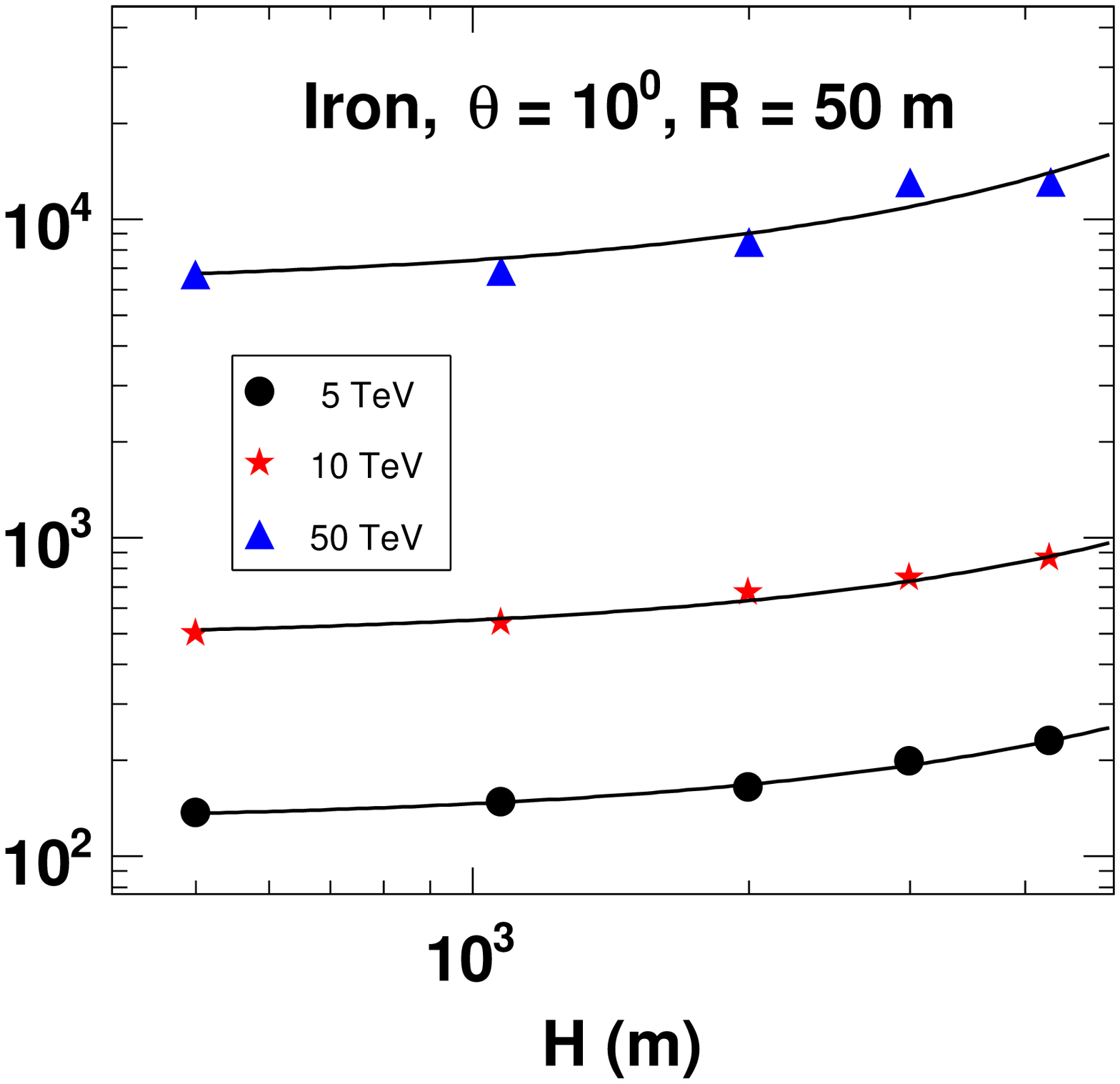}}
\caption{Average Cherenkov photon density ($\rho_{ch}$) for $\gamma$-ray, proton and iron primaries is plotted as a function of altitude of observation ($H$). The plots in the upper most panel show these variations for different zenith angles ($\theta$) keeping $E$ and $R$ fixed. The plots in the middle panel show the $\rho_{ch}$ variations 
for different values of $R$ keeping $\theta$ and $E$ fixed. Finally, the plots
in the bottom panel do the same for different values of $E$ keeping $R$ and $\theta$ fixed at a particular value. The solid lines in the respective plots shows the result 
of our parameterisation (\ref{eq4}). The fits are within the limit of 
statistical error ($< \pm 10\%)$.}
\label{fig4}
\end{figure*}
Similar to the above three subsections, here Fig.\ref{fig4} shows three different cases of the variation of average density of Cherenkov photons ($\rho_{ch}$) as a function of altitude of observation ($H$) for $\gamma$-ray, proton and iron primaries with different energies, different distances from the shower core 
($R$) and zenith angle ($\theta$) (check caption of Fig.\ref{fig4} for details). For a given primary particle, as long as the observation level is below the 
position of the shower maximum, the density of Cherenkov photons increases with increasing altitude of observation mostly near the shower core at all zenith angles and any primary energy \cite{Hazarika, Das}. This dependence is due to the fact that the Cherenkov threshold energy (via the refractive index) is a function of the altitude of the observation level \cite{Nerling}. The following equation perfectly parameterises these descriptions of altitude dependence in the form
of an exponential function
\begin{equation}
\rho(H) = a_{3} \exp(f H),
\label{eq4}
\end{equation}
where $\rho(H)$ is the altitude dependent density function of Cherenkov photons, $a_{3}$ and $f$ are parameters of the function, $H$ is the altitude of the observation level. $a_{3}$ and $f$ have different values for different primaries. 
Our parameterisation Eq.(\ref{eq4}) is in better agreement with the simulated data of the $\gamma$-ray primary compared to the proton and the iron primaries. For the proton and the iron primaries, the parameterisation slightly varies from the simulated data with increasing zenith angle, increasing energy and increasing core distance. The Table \ref{tab5} gives an idea about the values of the 
fitted parameters of the Eq.(\ref{eq4}) to the $\rho_{ch}$ distributions as a function of $H$ for $\gamma$-ray, proton and iron primary at 500 GeV, 1 TeV and 10 TeV energies respectively and at $R$ = 50 m and $\theta$ = 20$^{0}$.
\begin{table}[ht]
\caption{Values of the fitted parameters of the Eq.(\ref{eq4}) to $\rho_{ch}$ distributions as a function of $H$ for $\gamma$-ray, proton and iron primary at 500 GeV, 1 TeV and 10 TeV energies respectively and at $R$ = 50 m and $\theta$ = 20$^{0}$.} \label{tab5}
\begin{center}
\begin{tabular}{ccc}\hline
Primary & ~~$a_{3}$ & ~~~$f$ \\\hline\\[-7pt]

$\gamma$& ~~~5.08 $\pm$ 0.13 & ~~~1.891 $\pm$ 0.004\\[2pt]
Proton& ~~~5.16 $\pm$ 0.17 & ~~~1.652 $\pm$ 0.005\\[2pt]
Iron& ~~~7.32 $\pm$ 0.07 & ~~~1.355 $\pm$ 0.001\\\hline
\end{tabular}
\end{center}
\end{table}
\section{Parameterisation of Cherenkov photon's arrival time}
\subsection{As a function of radial distance ($R$)}
\begin{figure*}[ht]
\centerline
\centerline{\includegraphics[width=5.5cm, height=4.5cm]{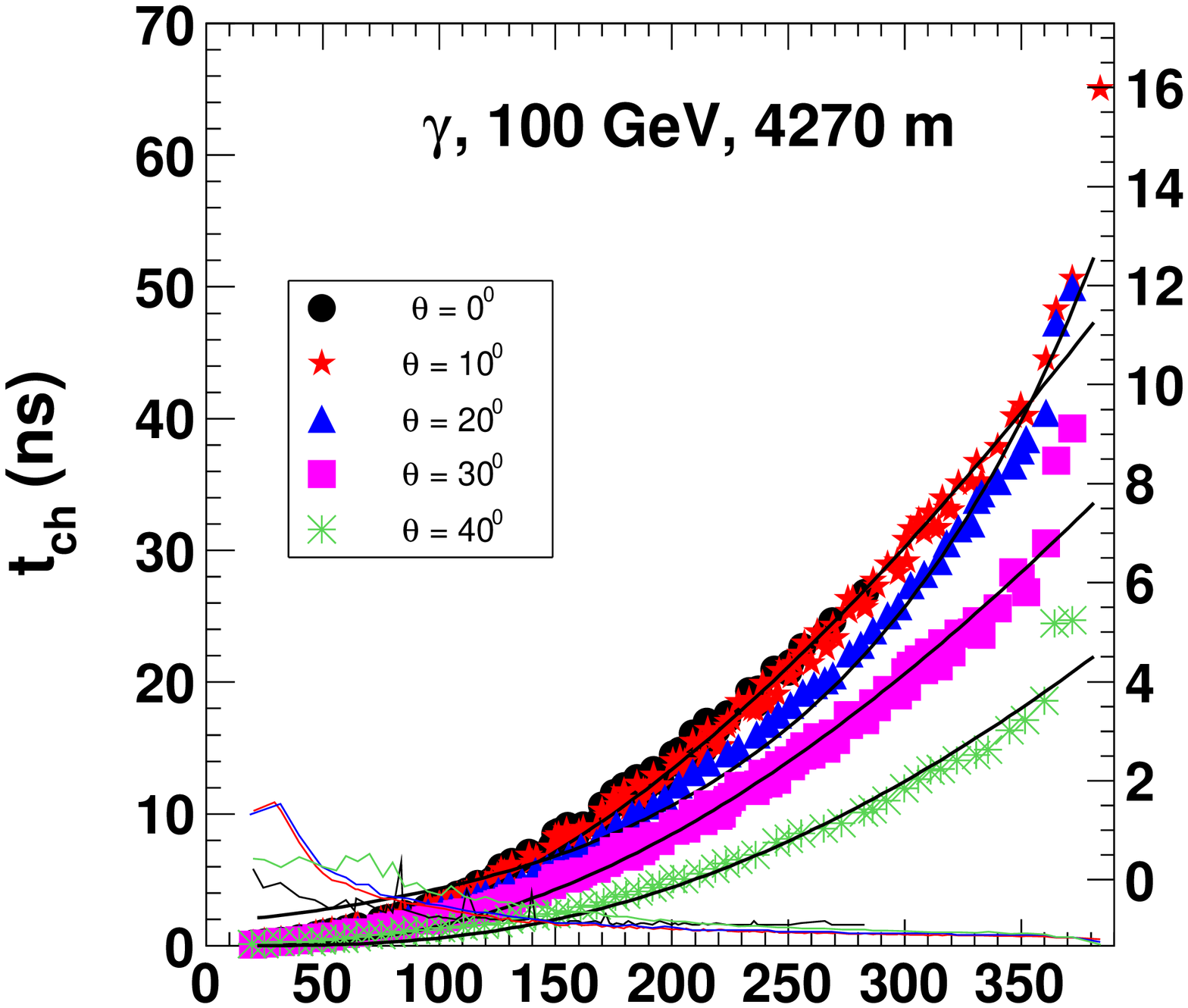}
\includegraphics[width=5.5cm, height=4.5cm]{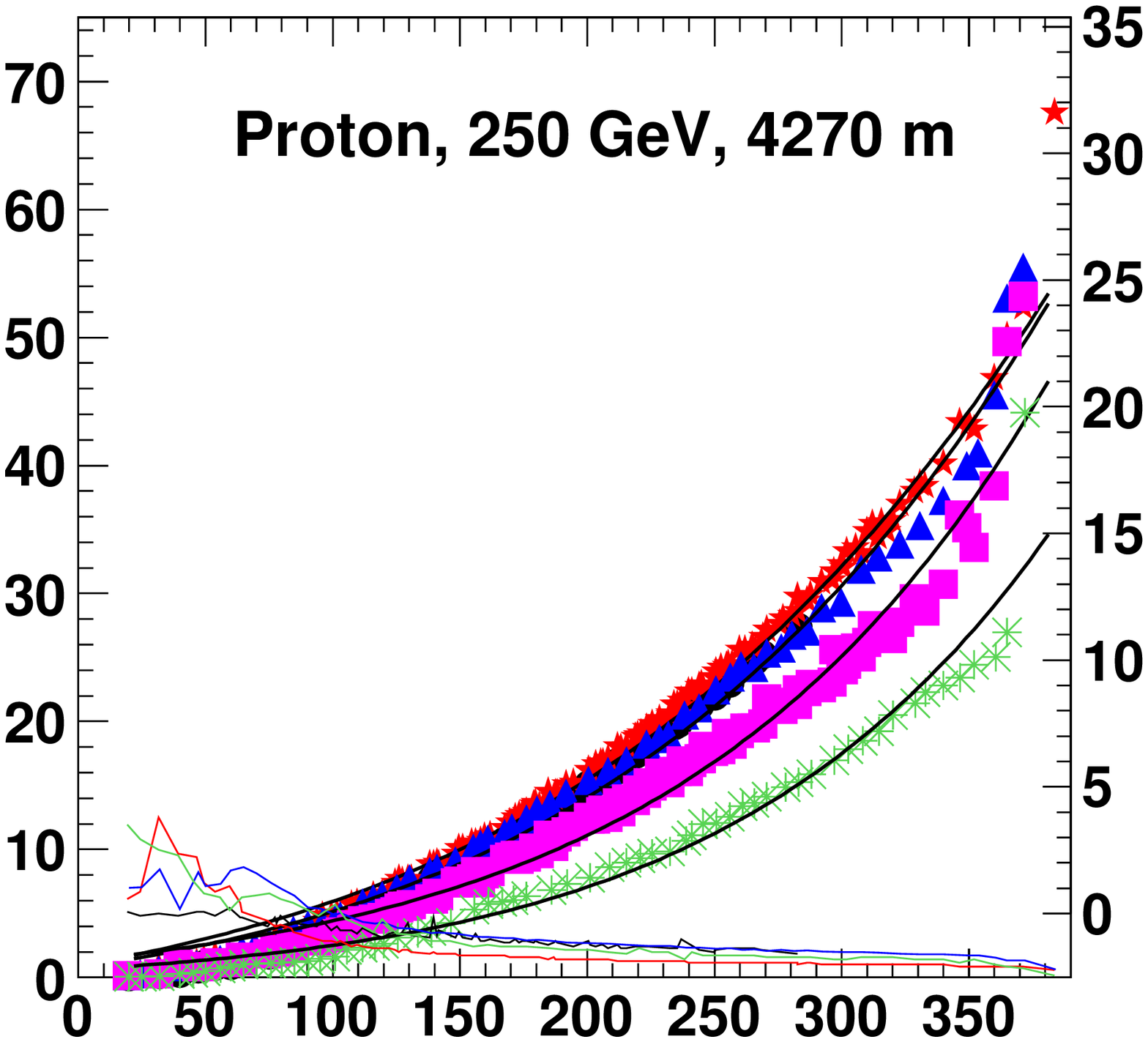}
\includegraphics[width=5.5cm, height=4.5cm]{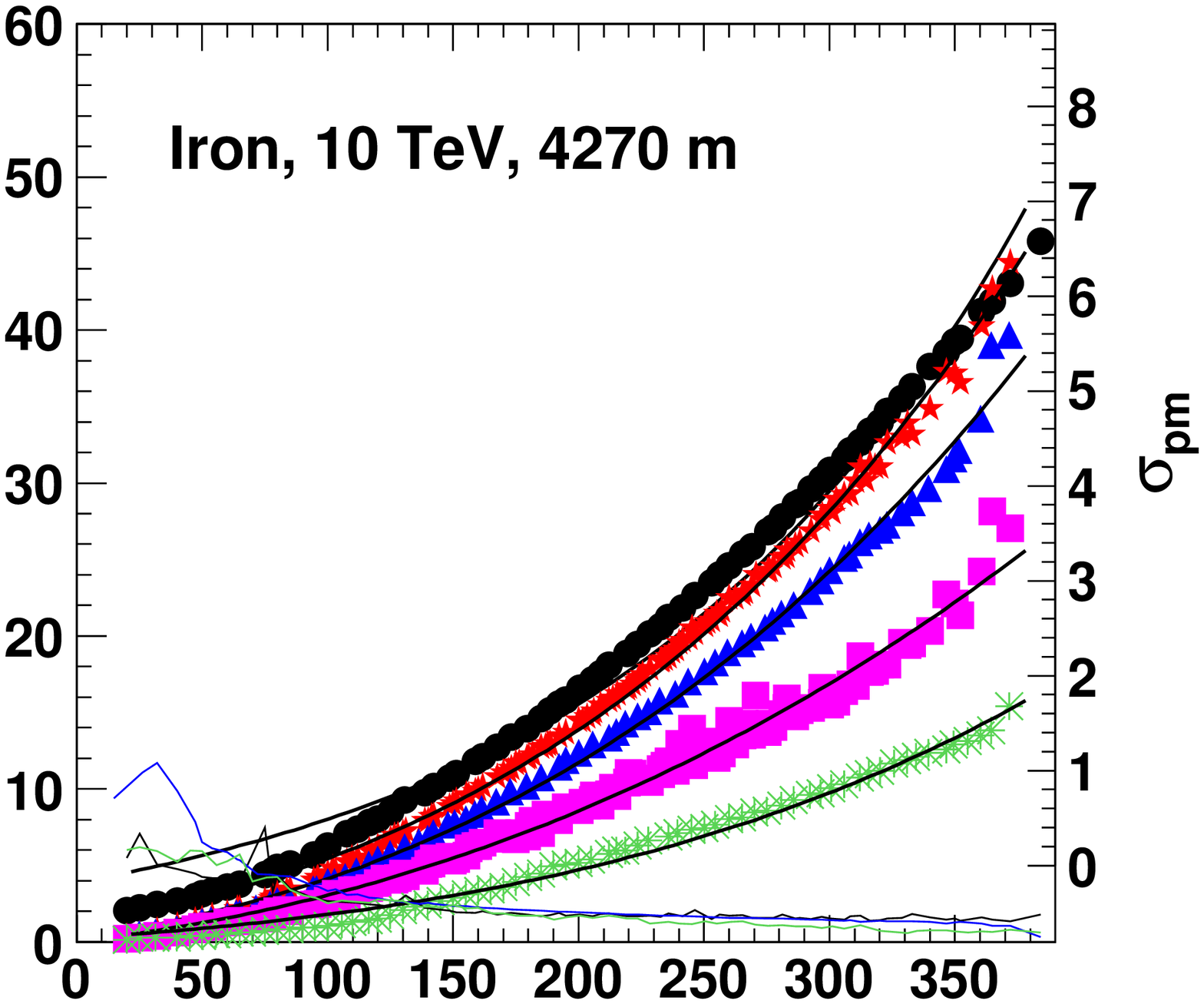}}
\centerline{\includegraphics[width=5.5cm, height=4.5cm]{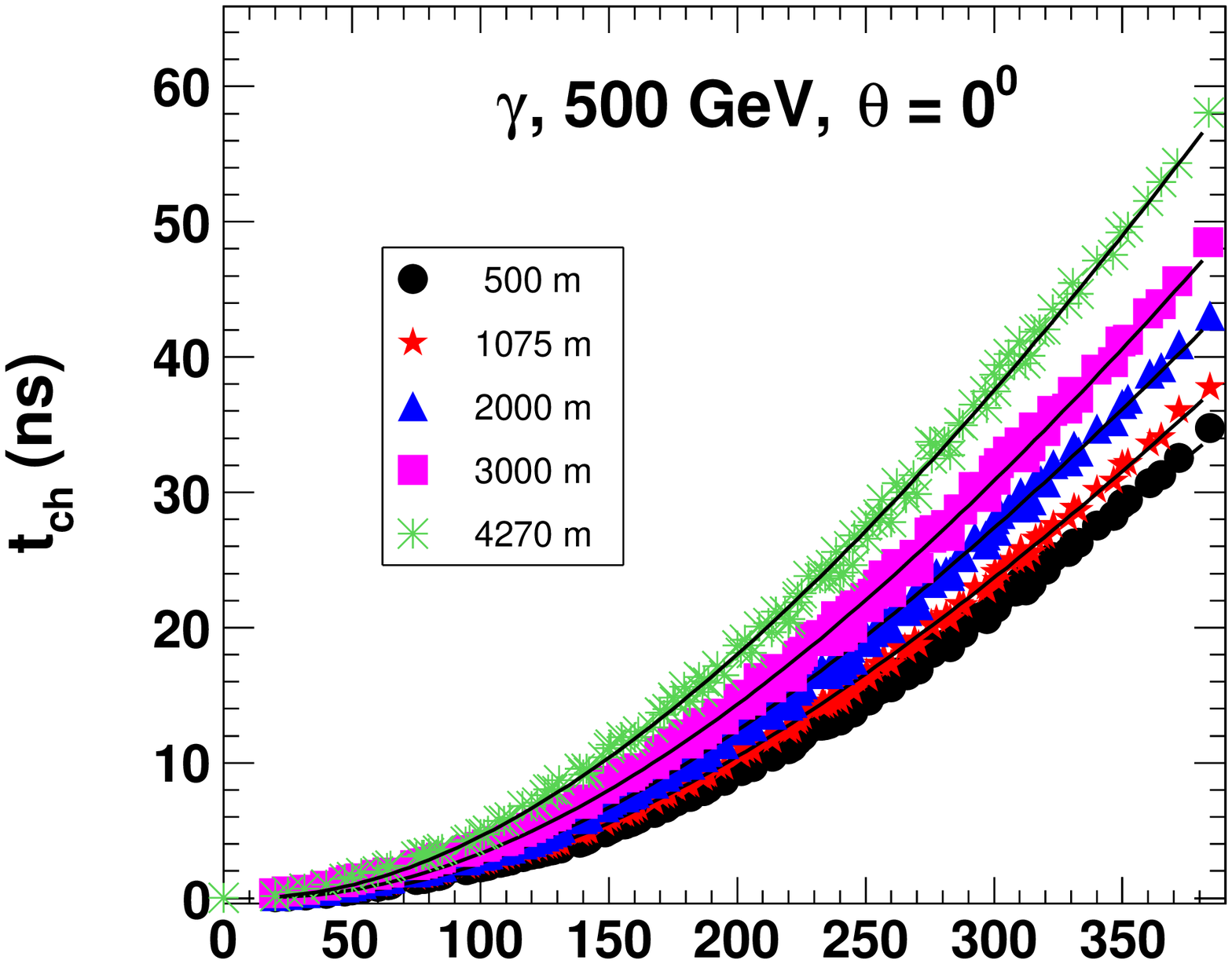}
\includegraphics[width=5.5cm, height=4.5cm]{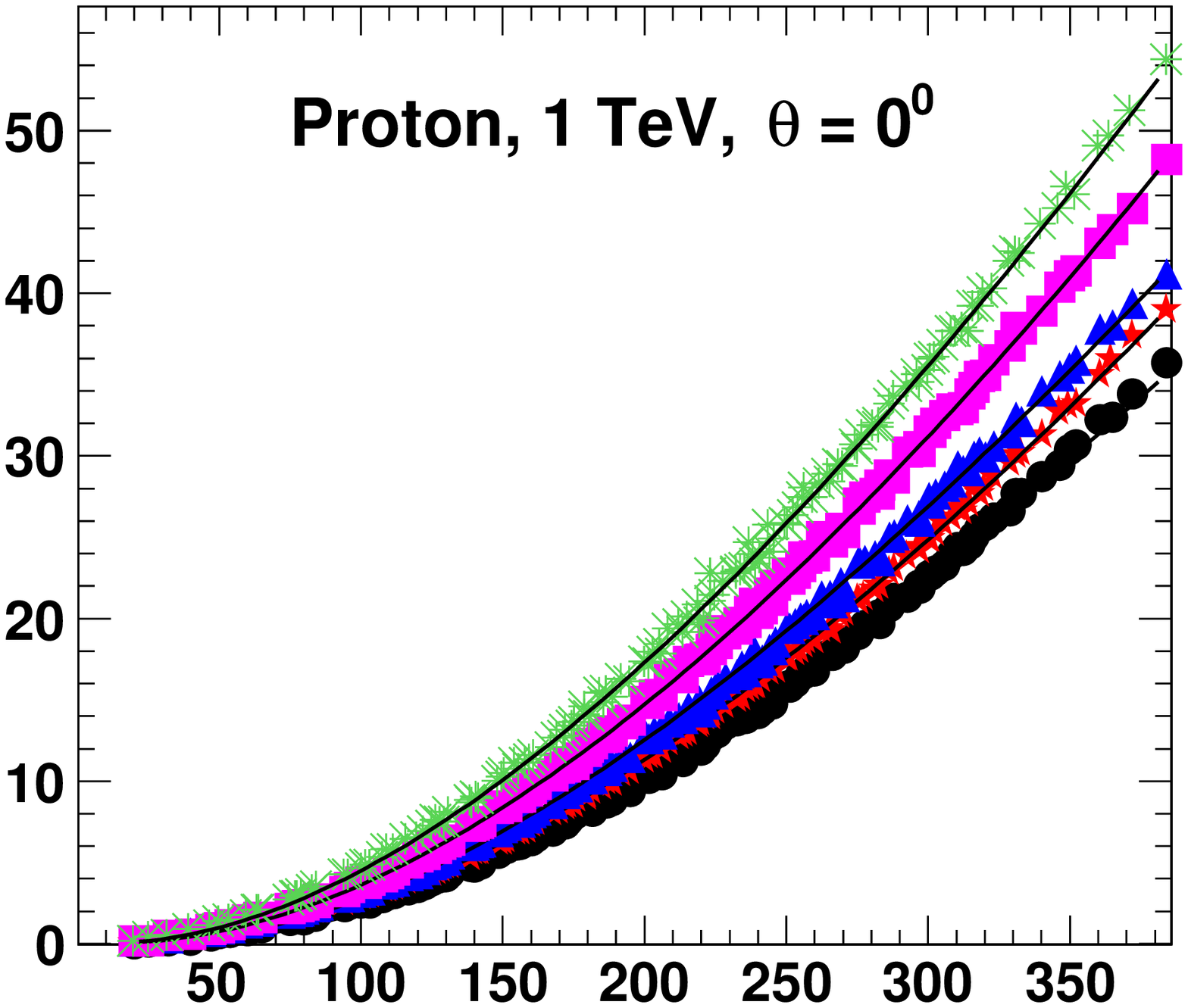}
\includegraphics[width=5.5cm, height=4.5cm]{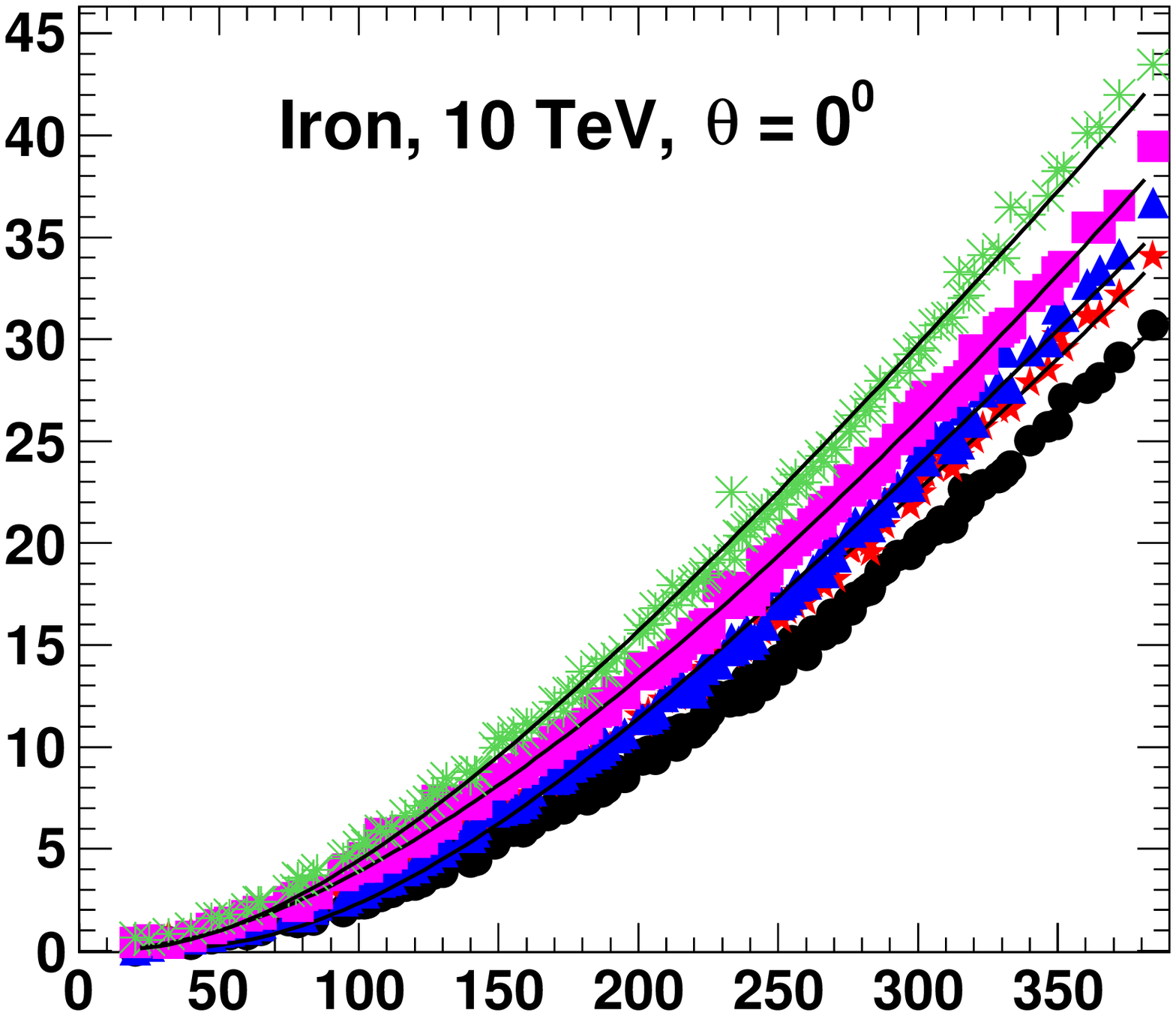}}
\centerline{\includegraphics[width=5.5cm, height=4.5cm]{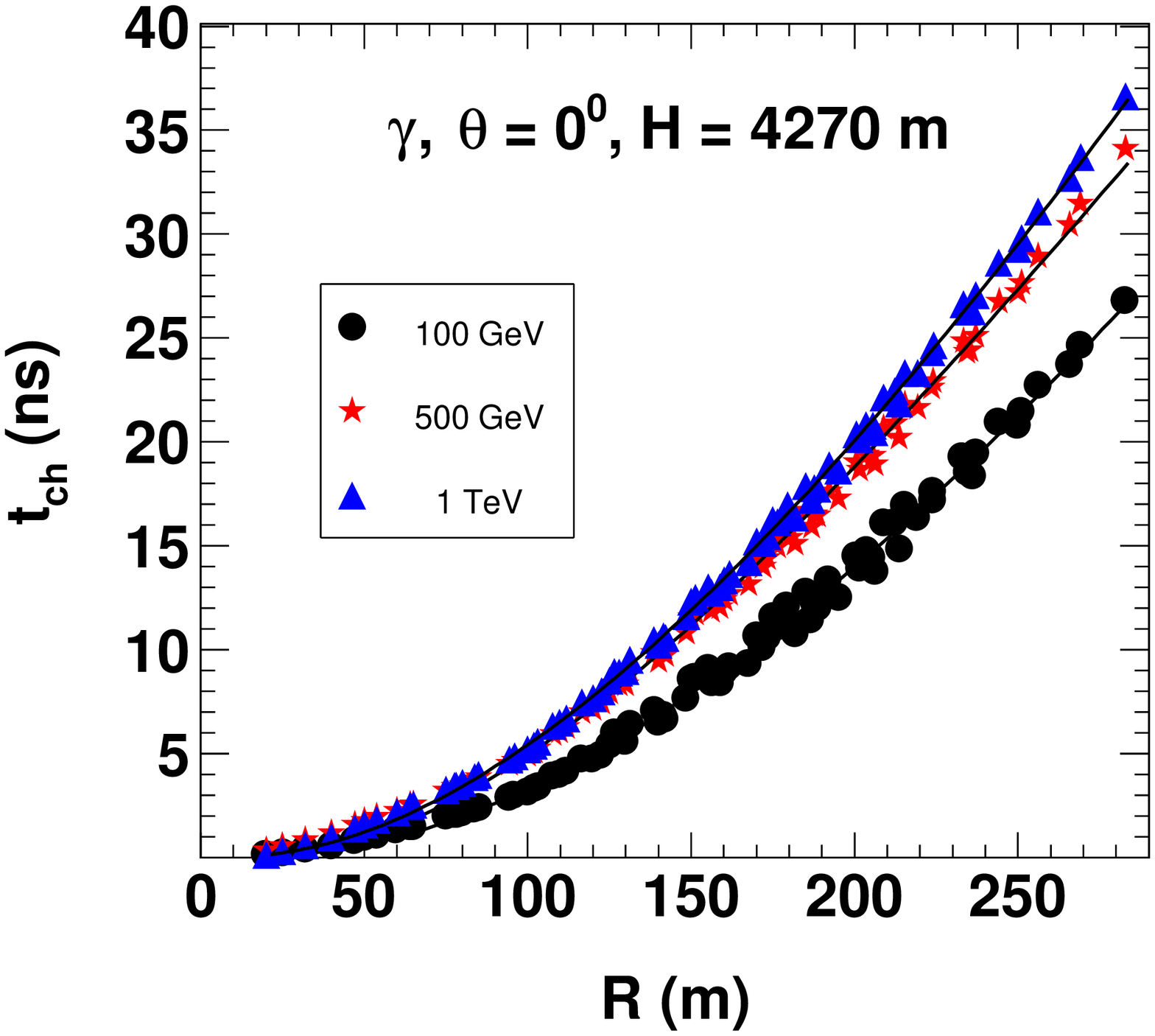}
\includegraphics[width=5.5cm, height=4.5cm]{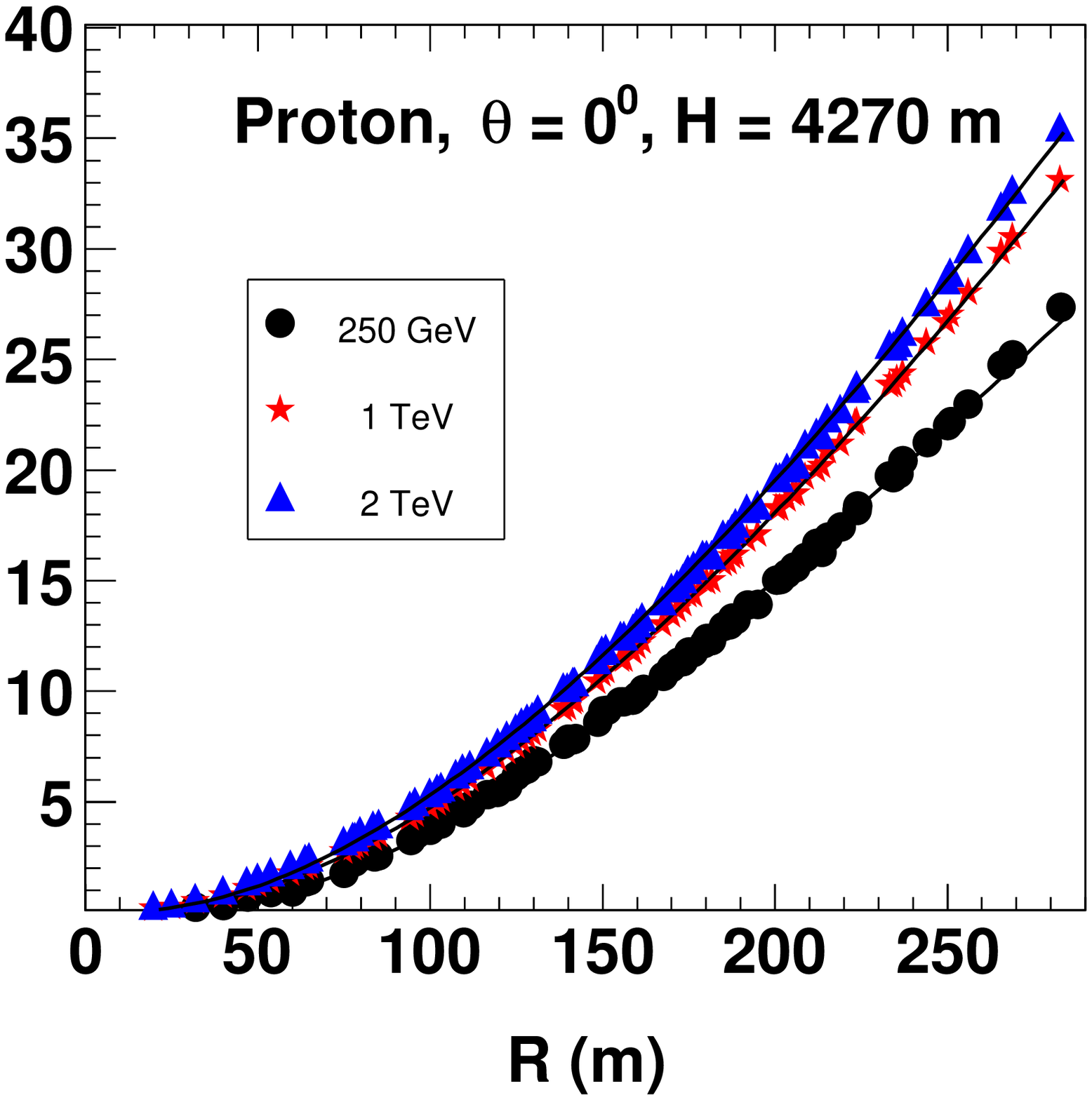}
\includegraphics[width=5.5cm, height=4.5cm]{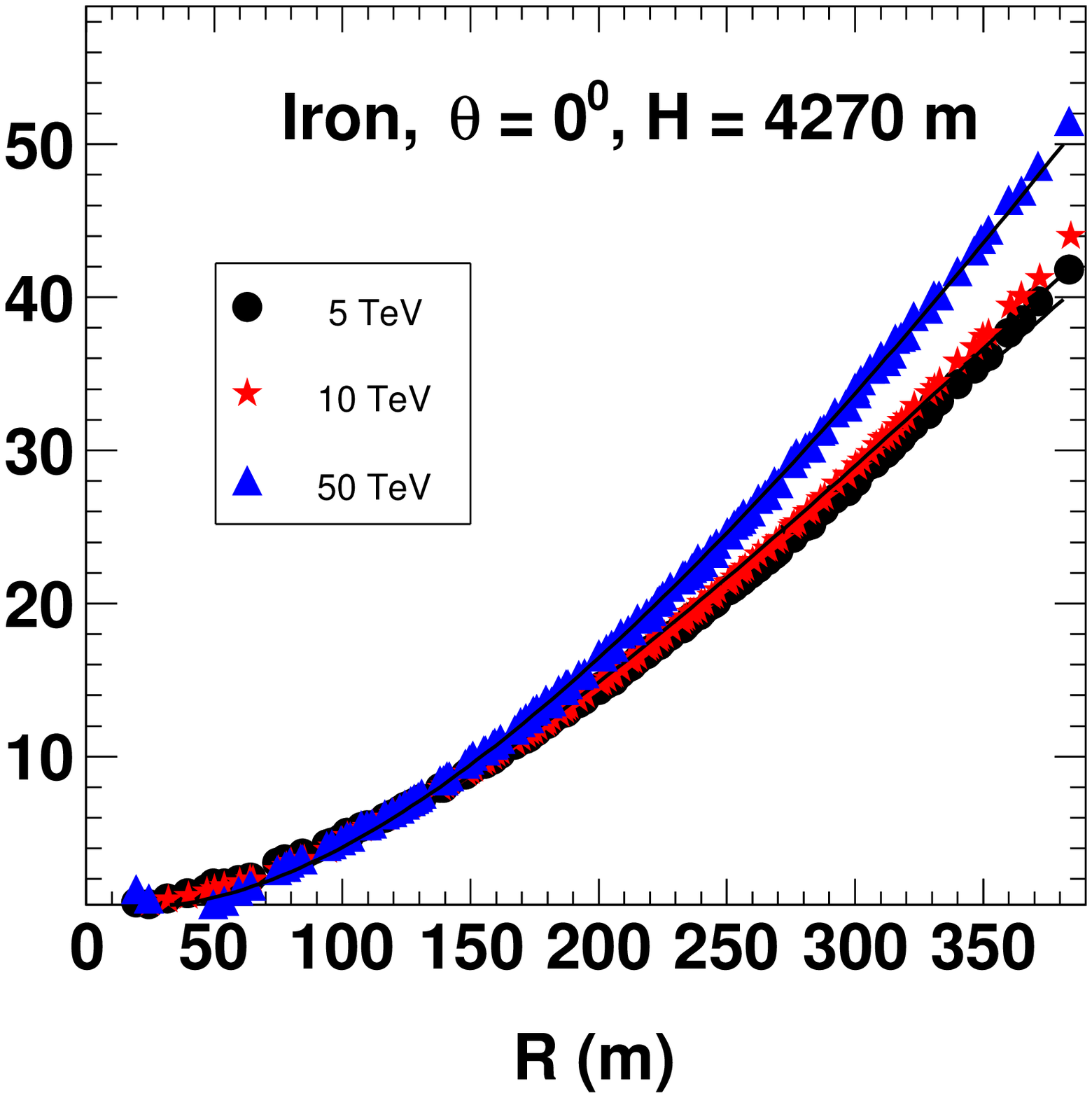}}
\caption{Average arrival time of Cherenkov photons  ($t_{ch}$) for $\gamma$-ray, proton and iron primaries is plotted as a function of distance from the shower core ($R$). The plots in the upper most panel show these variations for zenith angles 
$0^{\circ}$, $10^{\circ}$, $20^{\circ}$, $30^{\circ}$ and $40^{\circ}$ keeping 
$E$ and $H$ fixed. The r.m.s values per mean ($\sigma_{pm}$) of the Cherenkov 
photon's arrival time with respect to the distance from the shower core of 
different primaries is also shown. The plots in the middle panel show the $t_{ch}$ 
variations for different values of $H$ keeping $\theta$ and $E$ fixed. Finally,
the plots in the bottom panels do the same for different energies of the primary keeping 
$\theta$ and $H$ fixed at a particular value. The solid lines in the respective
plots shows the result of our parameterisation Eq.(\ref{eq5}). The fits are 
within the limit of statistical error ($< \pm 10\%$).}
\label{fig5}
\end{figure*}
In general, for the all primary particles, energies, zenith angles and 
altitudes of observation, the Cherenkov light front is found to be nearly 
spherical in shape. This spherical symmetry of the arrival time ($t_{ch}$) 
distribution 
is more perfect for the $\gamma$-ray primary than for the proton and iron 
primaries. This is mainly due to larger intra shower fluctuations in hadronic primaries compared to the $\gamma$-ray primary. This symmetry further deviates from sphericity for iron when compared to proton. The $t_{ch}$ distribution can be parameterised as a function of core distance ($R$) by the following equation 
\begin{equation}
t(R) = l_{0} \exp(n /R^{p}),
\label{eq5}
\end{equation}
where $t(R)$ is the mean arrival time of Cherenkov photons as a function of core distance, $l_{0}$, $n$ and $p$ are parameters of the function. $l_{0}$, $n$ and $p$ have different values for different primaries. This exponential dependence
 of $t_{ch}$ on the core distance has already been studied in our earlier work 
\cite{Hazarika, Das}. Further, similar to the case of $\rho_{ch}$ distribution,
at very low and very high values of core distance, the $t_{ch}$ distribution 
deviates from spherical symmetry. Thus, our approximation (\ref{eq5}) varies 
slightly from the CORSIKA predictions at very small and large values of core 
distance as shown in Fig.\ref{fig5}. For all other combinations of zenith 
angles, energy of the primary and altitudes of observation, parameterisation 
Eq.(\ref{eq5}) is in good agreement with the simulated data for all the three 
primaries. The best fit functions in Fig.\ref{fig5} are shown by solid lines 
and the same method as in the case of the density distributions is used here.
For instance, the Table \ref{tab6} shows the values of the fitted parameters of
the Eq.(\ref{eq5}) to the $t_{ch}$ distributions as a function of $R$ for $\gamma$-ray, proton and iron primary at 100 GeV, 250 GeV and 10 TeV energies respectively and at $H$ = 4270 m and $\theta$ = 30$^{0}$.
\begin{table}[ht]
\caption{Values of the fitted parameters of the  Eq.(\ref{eq5}) to the $t_{ch}$ distributions as a function of $R$ for $\gamma$-ray, proton and iron primary at 100 GeV, 250 GeV and 10 TeV energies respectively and at $H$ = 4270 m and $\theta$ = 30$^{0}$.} \label{tab6}
\begin{center}
\begin{tabular}{cccc}\hline
Primary & ~~$l_{0}$ & ~~~$n$ & ~~~$p$ \\\hline\\[-7pt]

$\gamma$&~6.67$\pm$0.48&~-1.303$\pm$0.107&~-31.88$\pm$0.30\\[2pt]
Proton&~5.06$\pm$0.19&~-1.323$\pm$0.047&~-27.19$\pm$0.23\\[2pt]
Iron&~8.68$\pm$0.99&~-0.203$\pm$0.005&~-27.95$\pm$0.34\\\hline
\end{tabular}
\end{center}
\end{table}
\subsection{As a function of energy ($E$)}
\begin{figure*}[hbt]
\centerline
\centerline{\includegraphics[width=5.5cm, height=4.5cm]{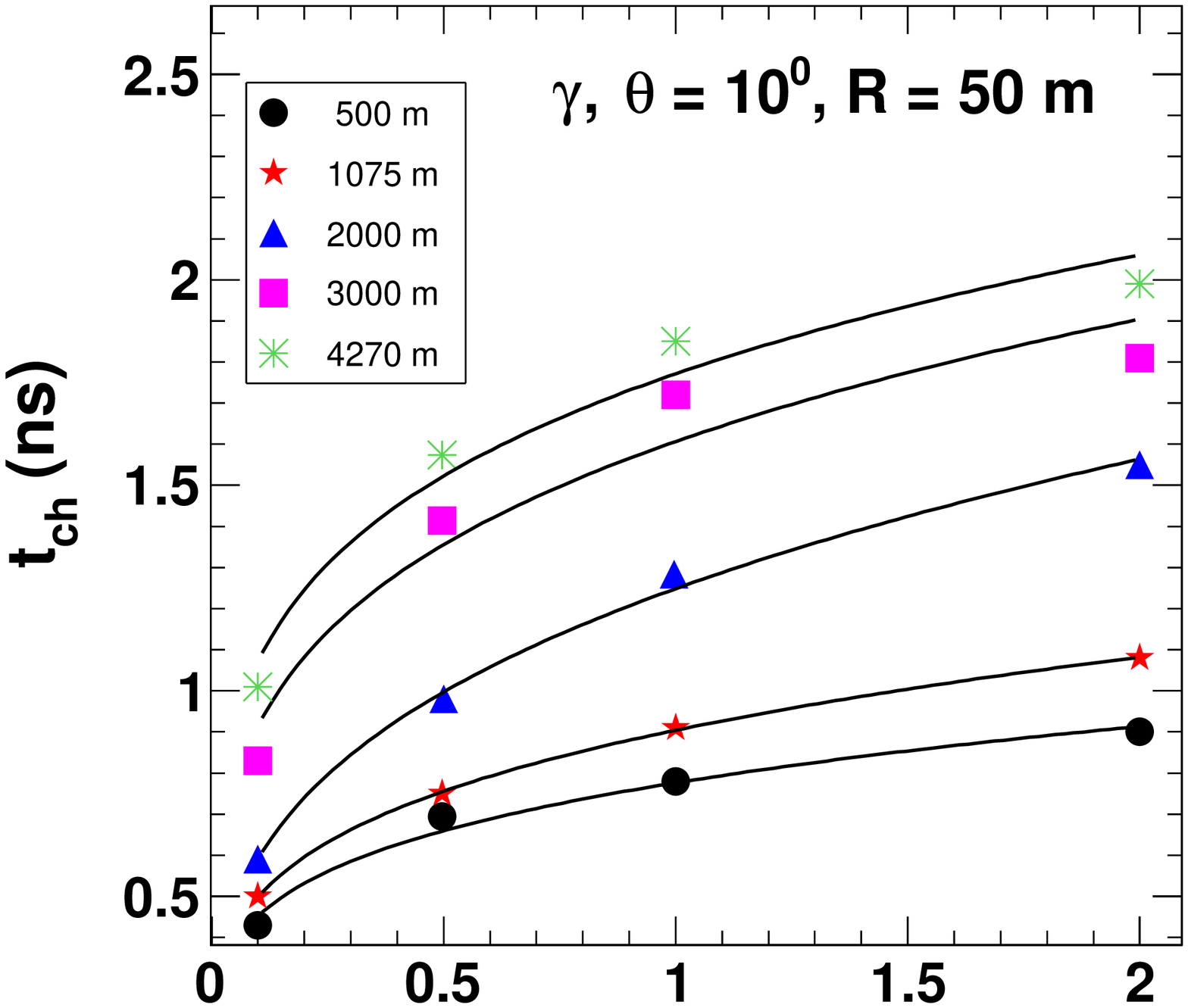}
\includegraphics[width=5.5cm, height=4.5cm]{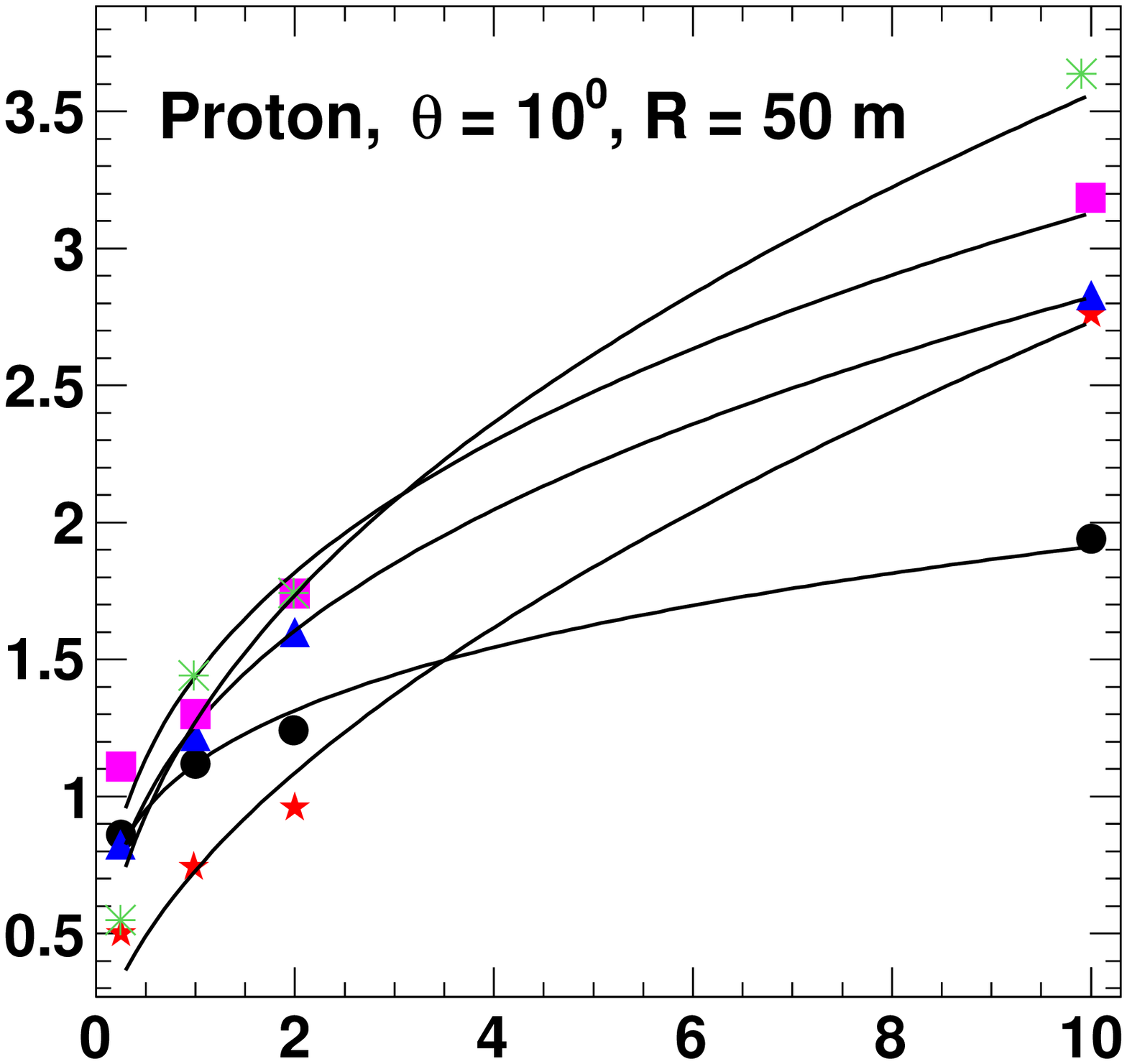}
\includegraphics[width=5.5cm, height=4.5cm]{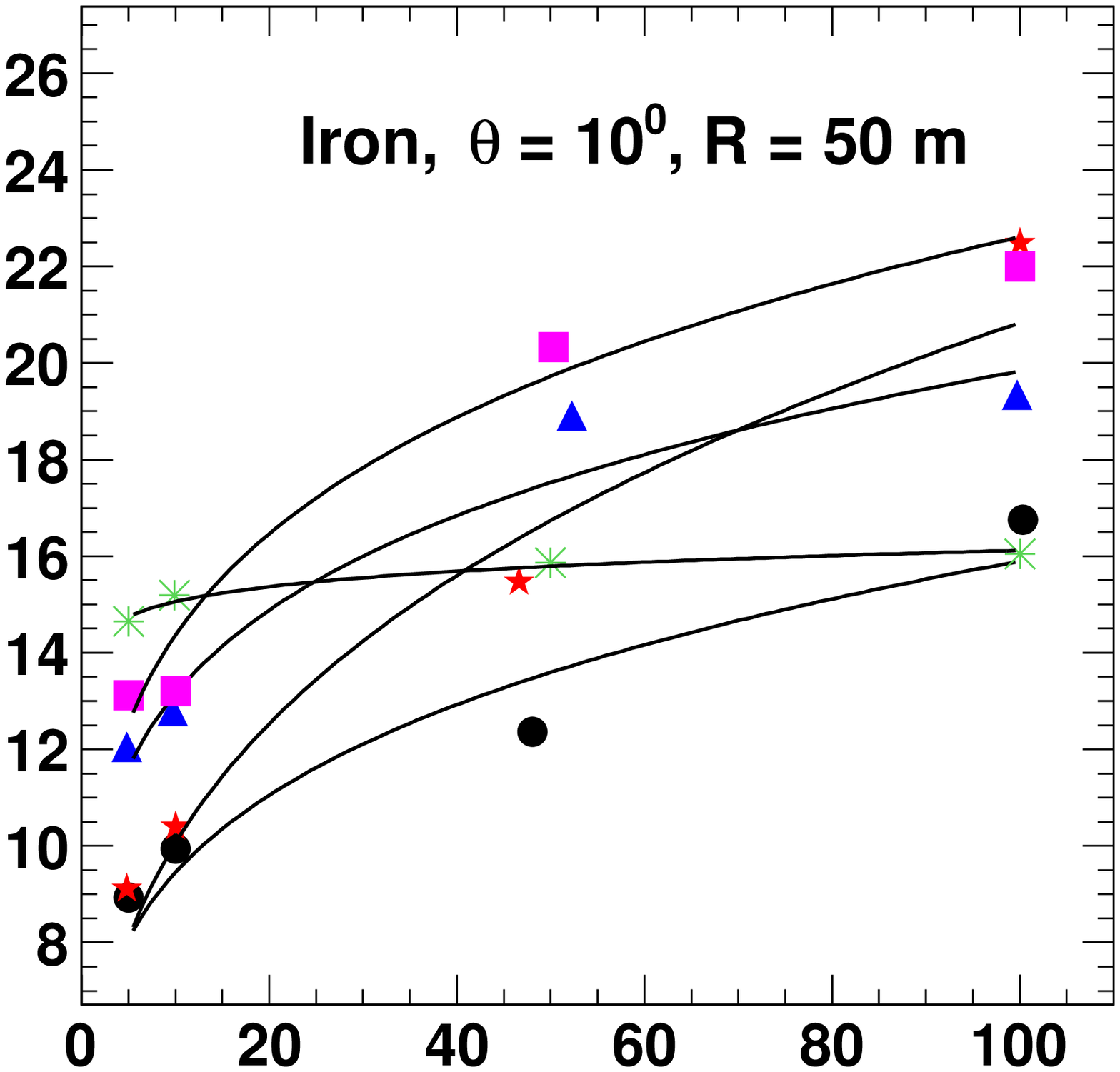}}
\centerline{\includegraphics[width=5.5cm, height=4.5cm]{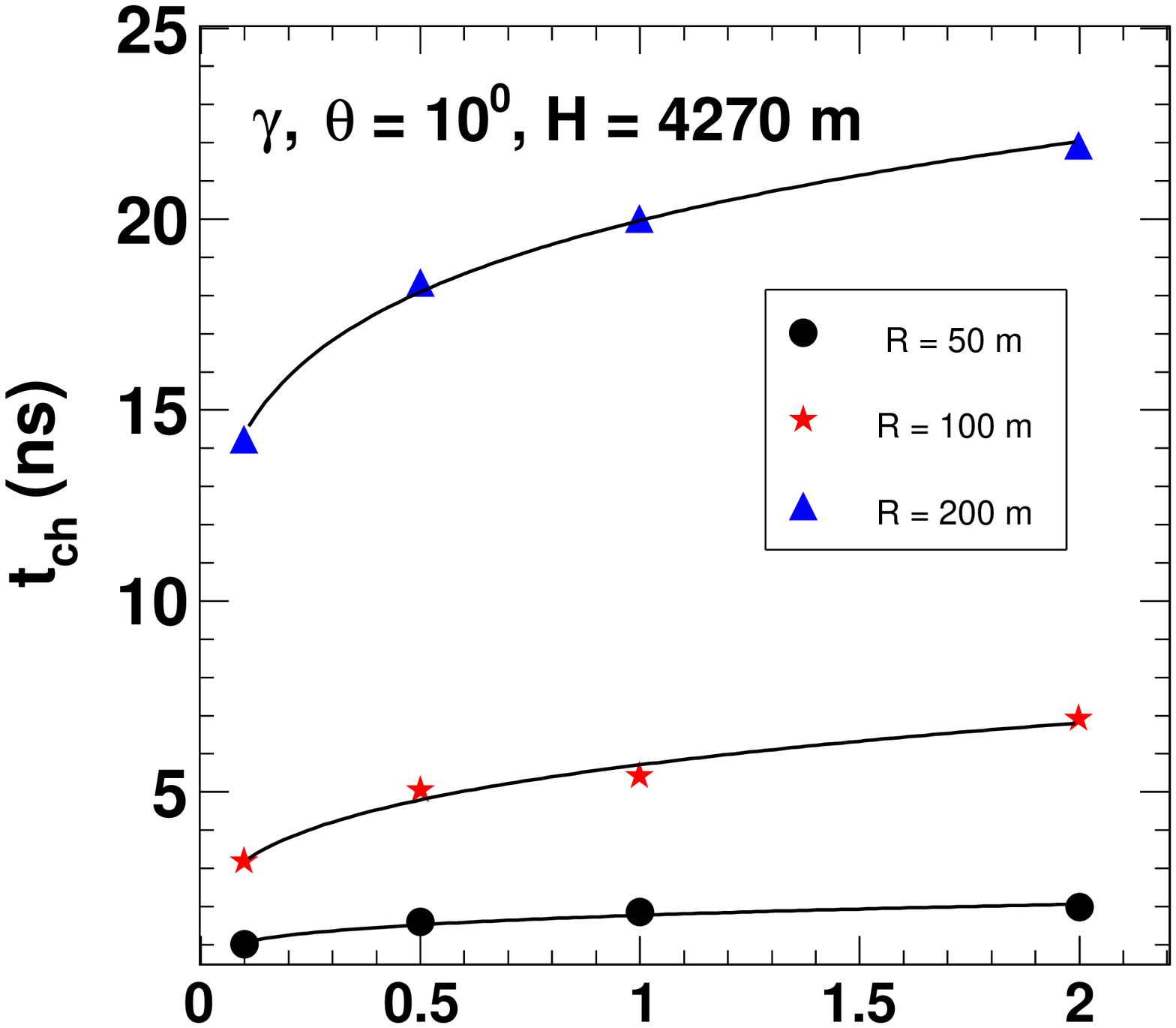}
\includegraphics[width=5.5cm, height=4.5cm]{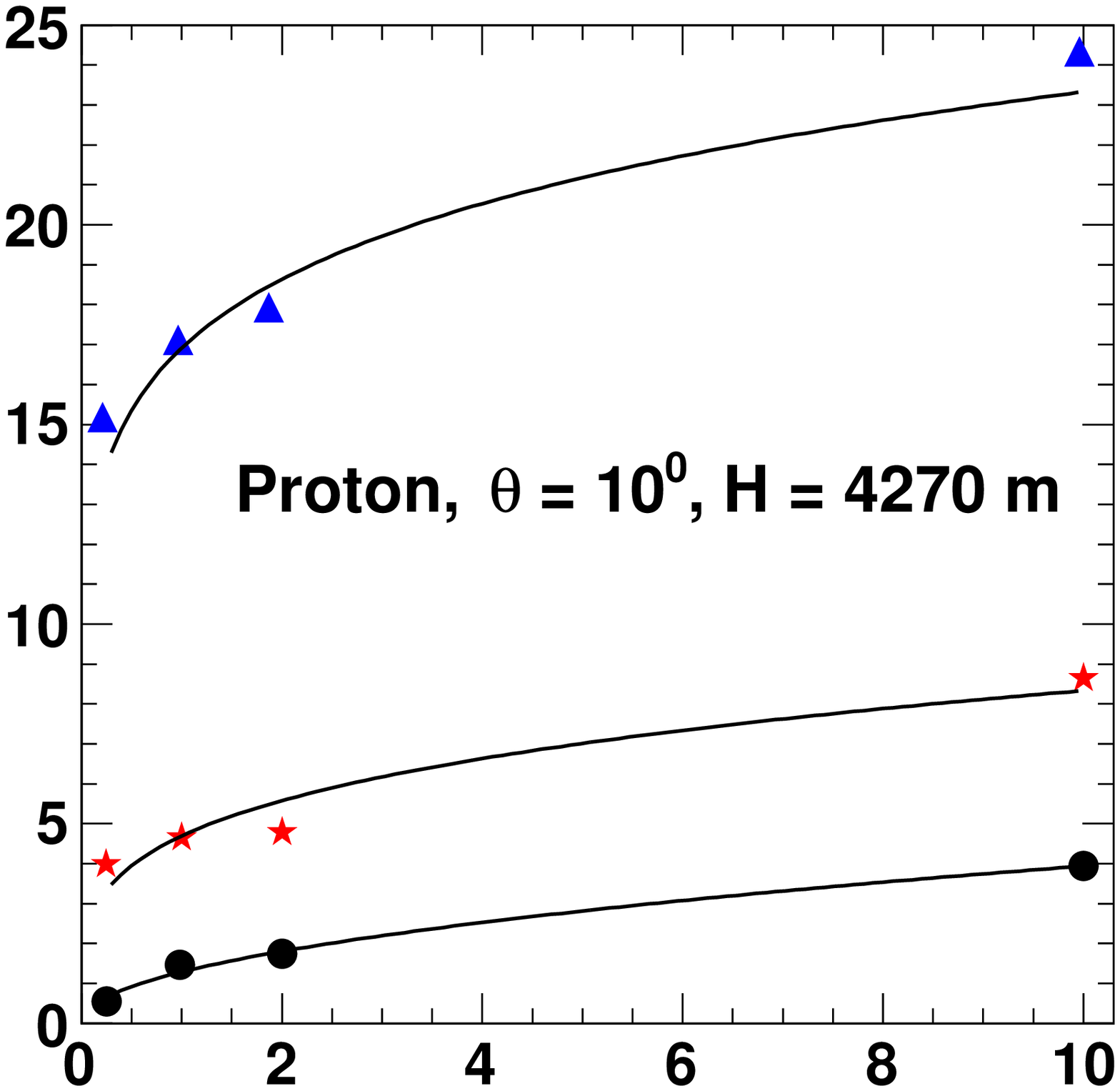}
\includegraphics[width=5.7cm, height=4.5cm]{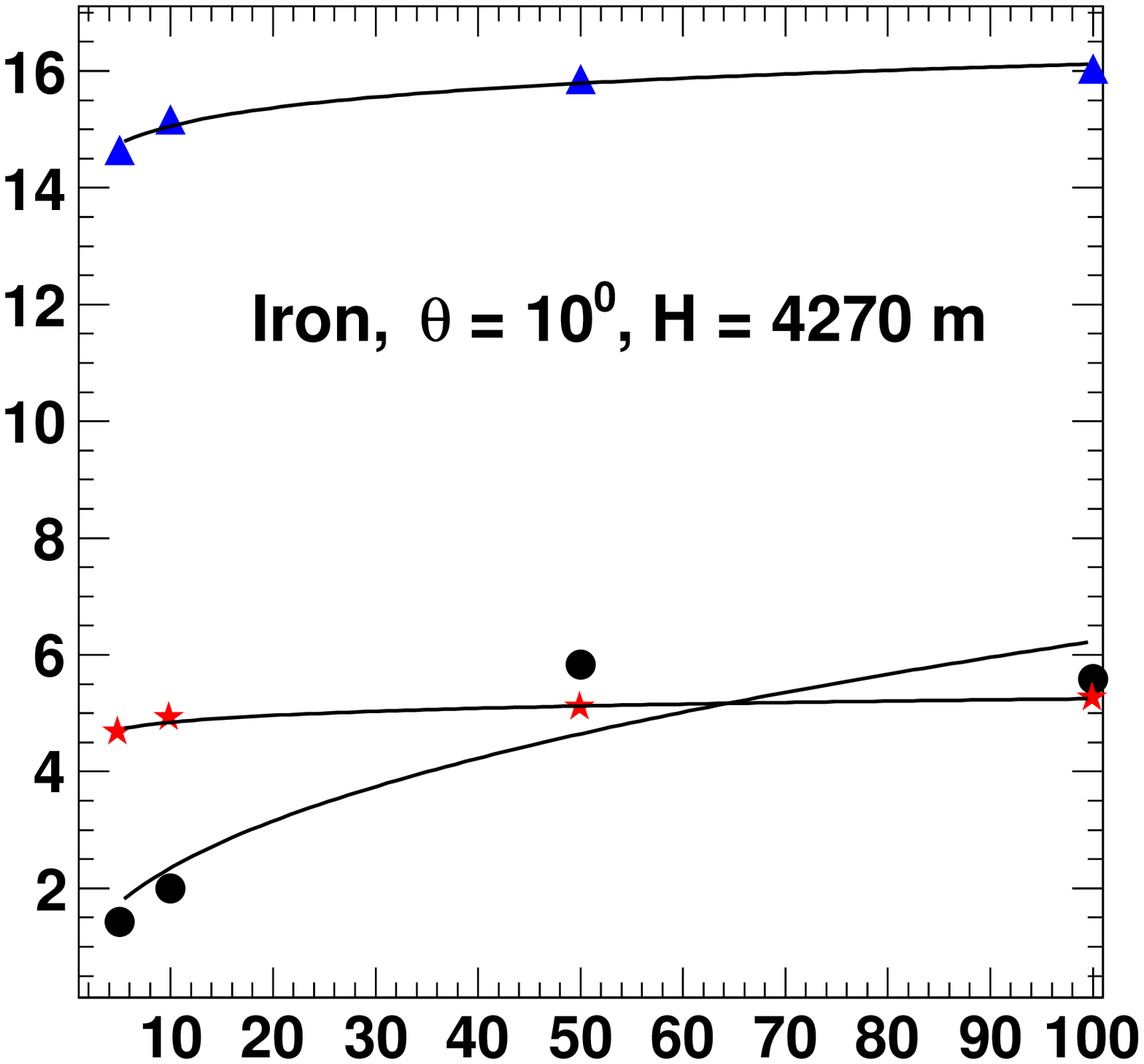}}
\centerline{\includegraphics[width=5.6cm, height=4.5cm]{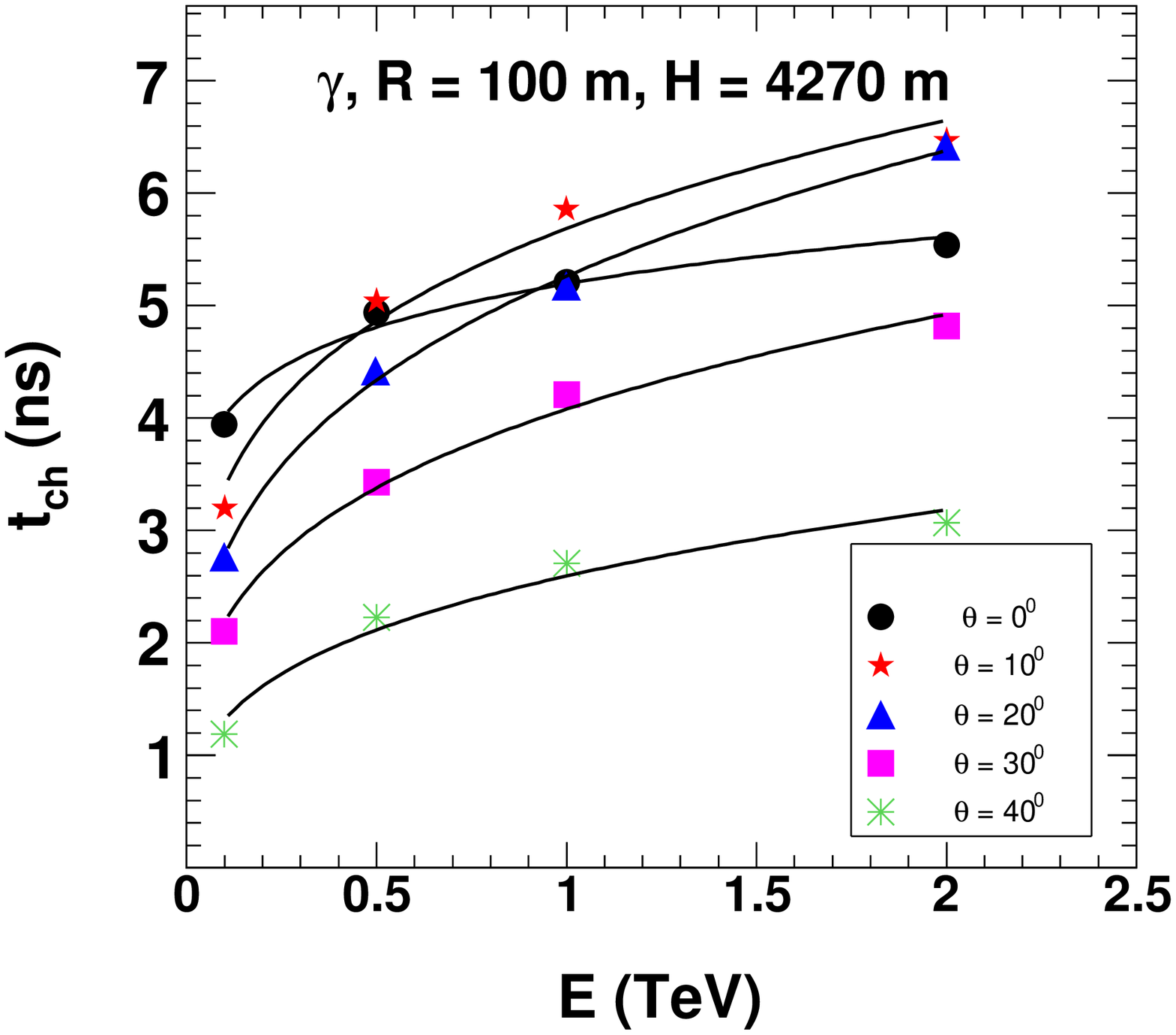}
\includegraphics[width=5.6cm, height=4.5cm]{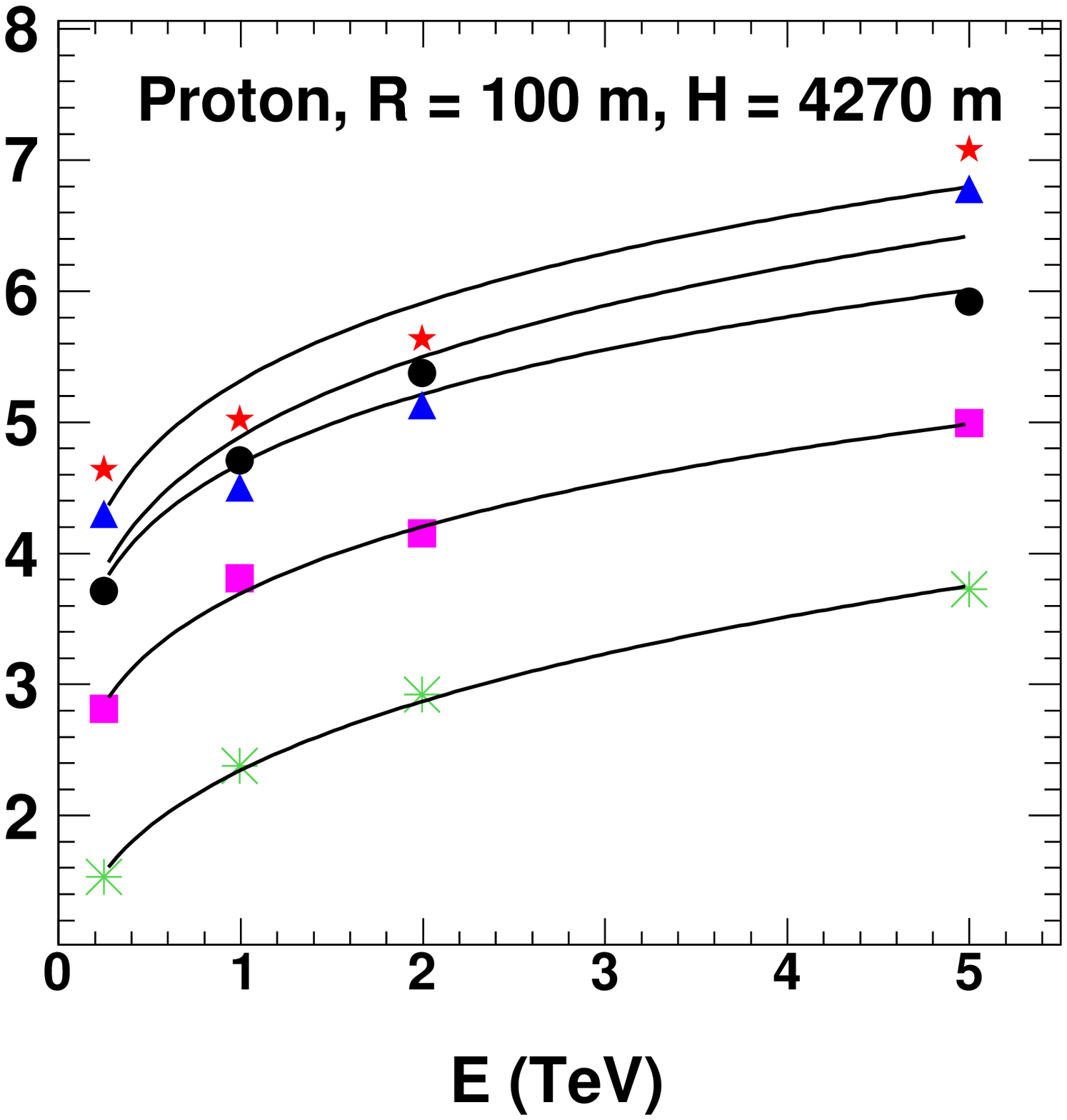}
\includegraphics[width=5.6cm, height=4.5cm]{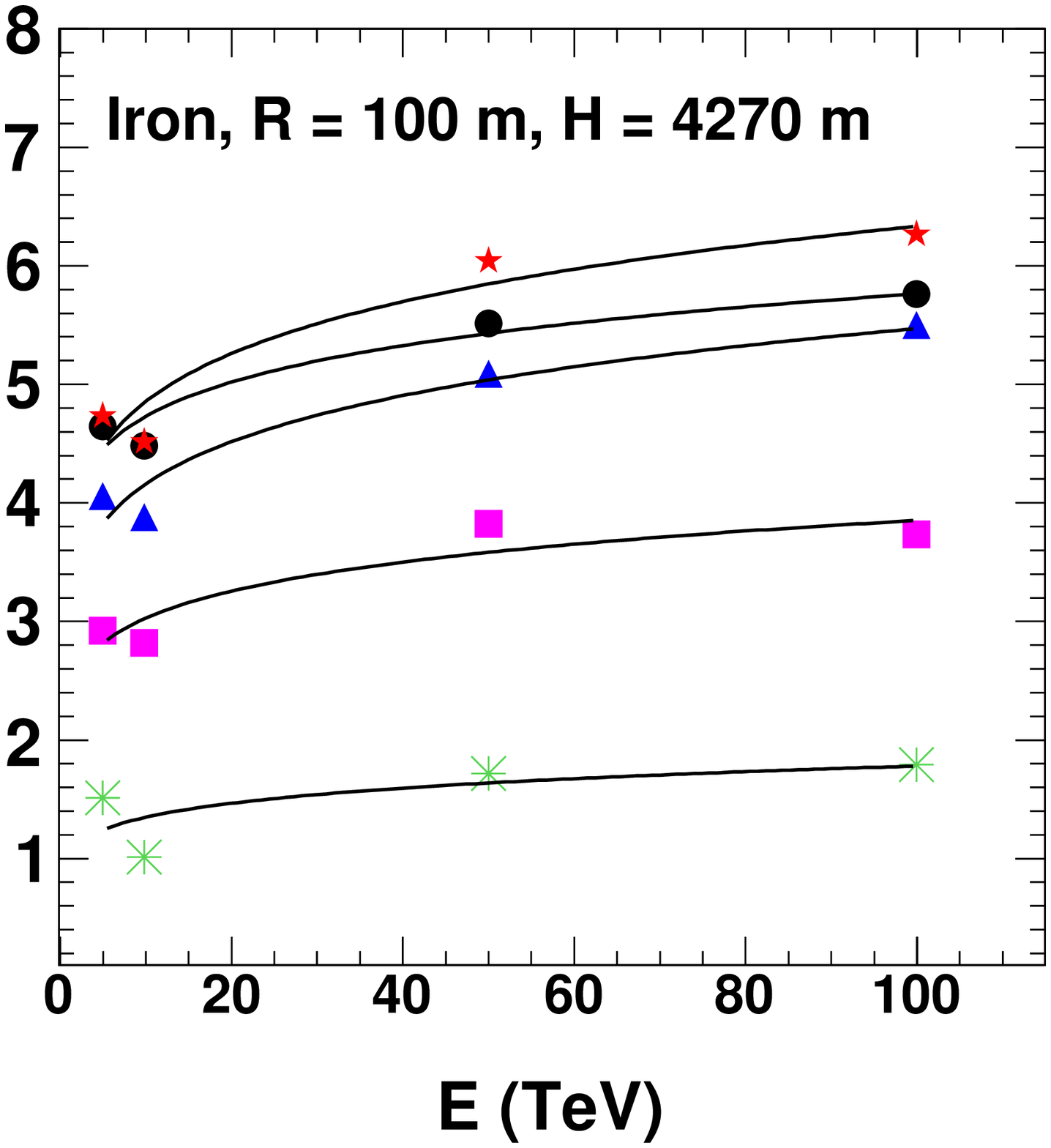}}
\caption{Average arrival time of Cherenkov photons ($t_{ch}$) for $\gamma$-ray, proton and iron primaries is plotted as a function of energy ($E$). The plots in
the upper most panel show these variations for different altitudes of observation $(H)$ keeping $\theta$ and $R$ fixed. The plots in the middle panel show the $t_{ch}$ variations for different values of $R$ keeping $\theta$ and $H$ fixed. The plots in the bottom panel do the same for different values of $\theta$ keeping $R$ and $H$ fixed at a particular value. The solid lines in the respective 
plots show the result of our parameterisation Eq.(\ref{eq6}). The fits are 
within the limit of statistical error ($< \pm 10\%$).}
\label{fig6}
\end{figure*}

The dependence of $t_{ch}$ distribution on the primary energy ($E$) can be 
parameterised as a power law in energy as follows 
\begin{equation}
t(E) = l_{1} E^{-m},
\label{eq6}
\end{equation}
where $t(E)$ is the mean arrival time of Cherenkov photons as a function of 
primary energy, $l_{1}$ and $m$ are parameters of the function. $l_{1}$ and $m$ take different values for different primaries. With increasing energy of the 
primary, the slant depth of the shower maximum increases and hence the distance of the detector array from the shower maximum of the primary particle decreases \cite{Das, Chitnis}. Thus with increasing energy of the primary the average arrival time of the Cherenkov photons will increase. However, it can be seen that the iron initiated photons have a flatter average arrival time. This may be because of the larger muon content in the shower of the iron primary. As such, the $\gamma$-ray and proton initiated showers have a non-linear dependence of $t_{ch}$ on primary energy but for iron initiated showers its almost a linear dependence. This trend is followed for all the combinations of core distances, zenith angles and altitudes of observation, refer Fig.\ref{fig6}. Hence, our approximation (\ref{eq6}) is best applicable to the $\gamma$-ray primary. The proton and iron primaries have large shower to shower fluctuations which affects the arrival time of the Cherenkov photons particulary at the tail region. Lower statistics further accentuates this affect. So to improve the systematics we must increase the total number of showers produced for proton and iron primaries. The Table \ref{tab7} shows values of the fitted parameters of the Eq.(\ref{eq6}) to the $t_{ch}$ distributions as a function of $E$ for $\gamma$-ray, proton and iron primary at $H$ = 4270 m, $\theta$ = 10$^{0}$ and $R$ = 50 m, to give an idea about the
possible values of parameters in the equation. 
\begin{table}[ht]
\caption{Values of the fitted parameters of the Eq.(\ref{eq6}) to the $t_{ch}$ distributions as a function of $E$ for $\gamma$-ray, proton and iron primary at $H$ = 4270 m, $\theta$ = 10$^{0}$ and $R$ = 50 m.} \label{tab7}
\begin{center}
\begin{tabular}{ccc}\hline
Primary & ~~$l_{1}$ & ~~~$m$ \\\hline\\[-7pt]

$\gamma$& ~~~0.0046 $\pm$ 0.0004 & ~~~-0.226 $\pm$ 0.102\\[2pt]
Proton& ~~~0.0028 $\pm$ 0.0025 & ~~~-0.152 $\pm$ 0.089\\[2pt]
Iron& ~~~1.6113 $\pm$ 1.4282 & ~~~-0.115 $\pm$ 0.076\\\hline
\end{tabular}
\end{center}
\end{table}
\subsection{As a function of zenith angle ($\theta$)}
\begin{figure*}[hbt]
\centerline
\centerline{\includegraphics[width=5.5cm, height=4.5cm]{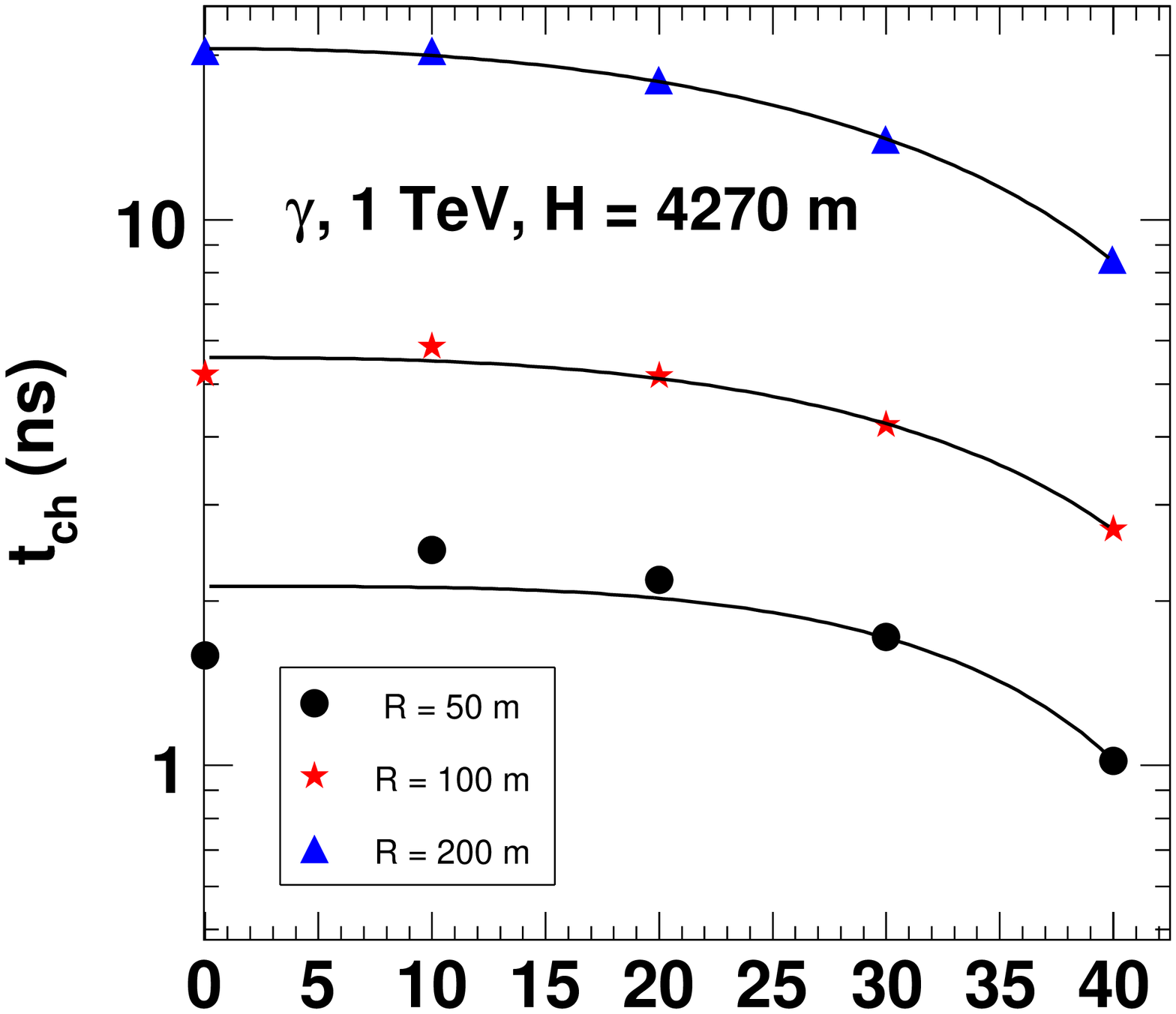}
\includegraphics[width=5.5cm, height=4.6cm]{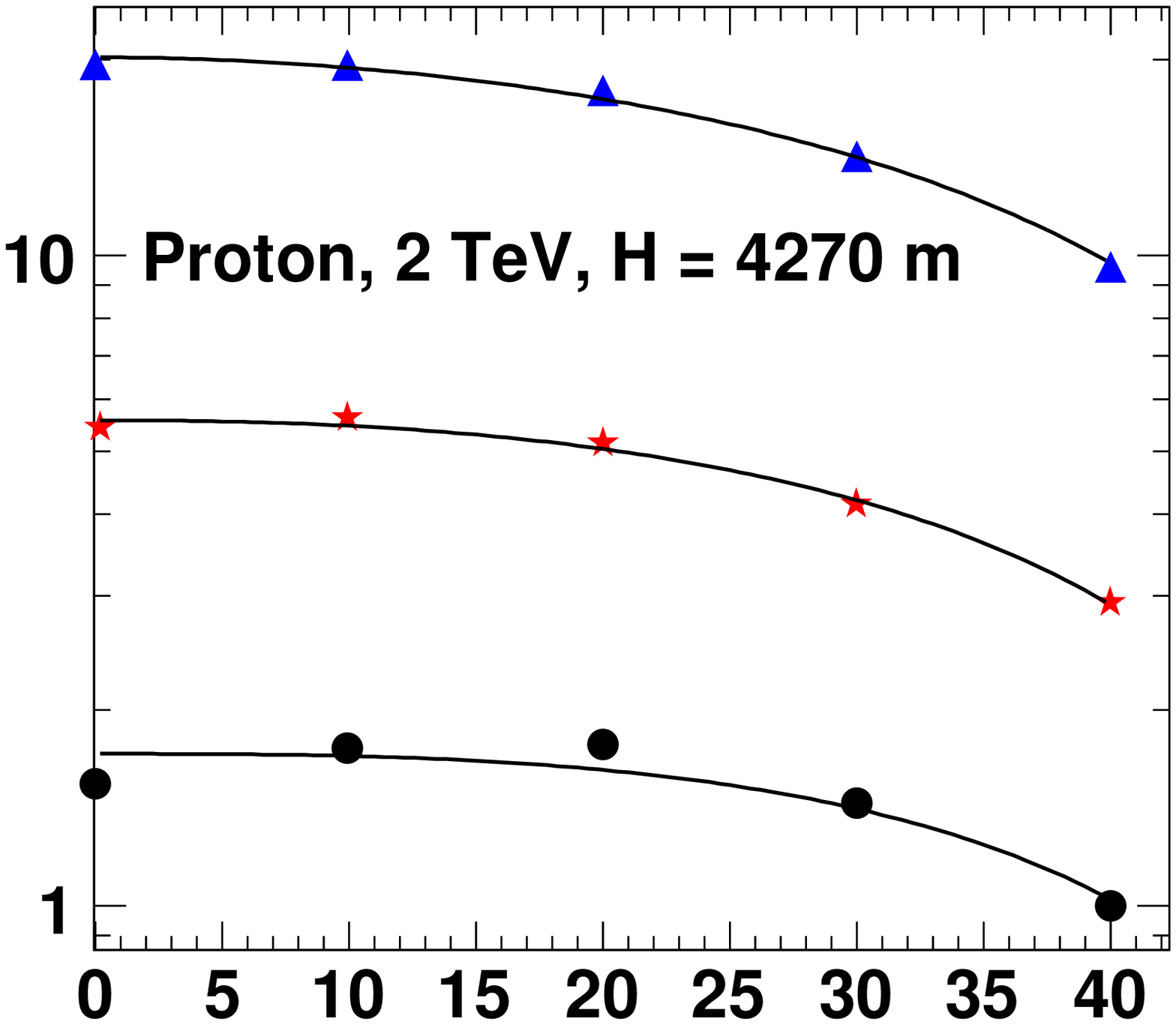}
\includegraphics[width=5.5cm, height=4.4cm]{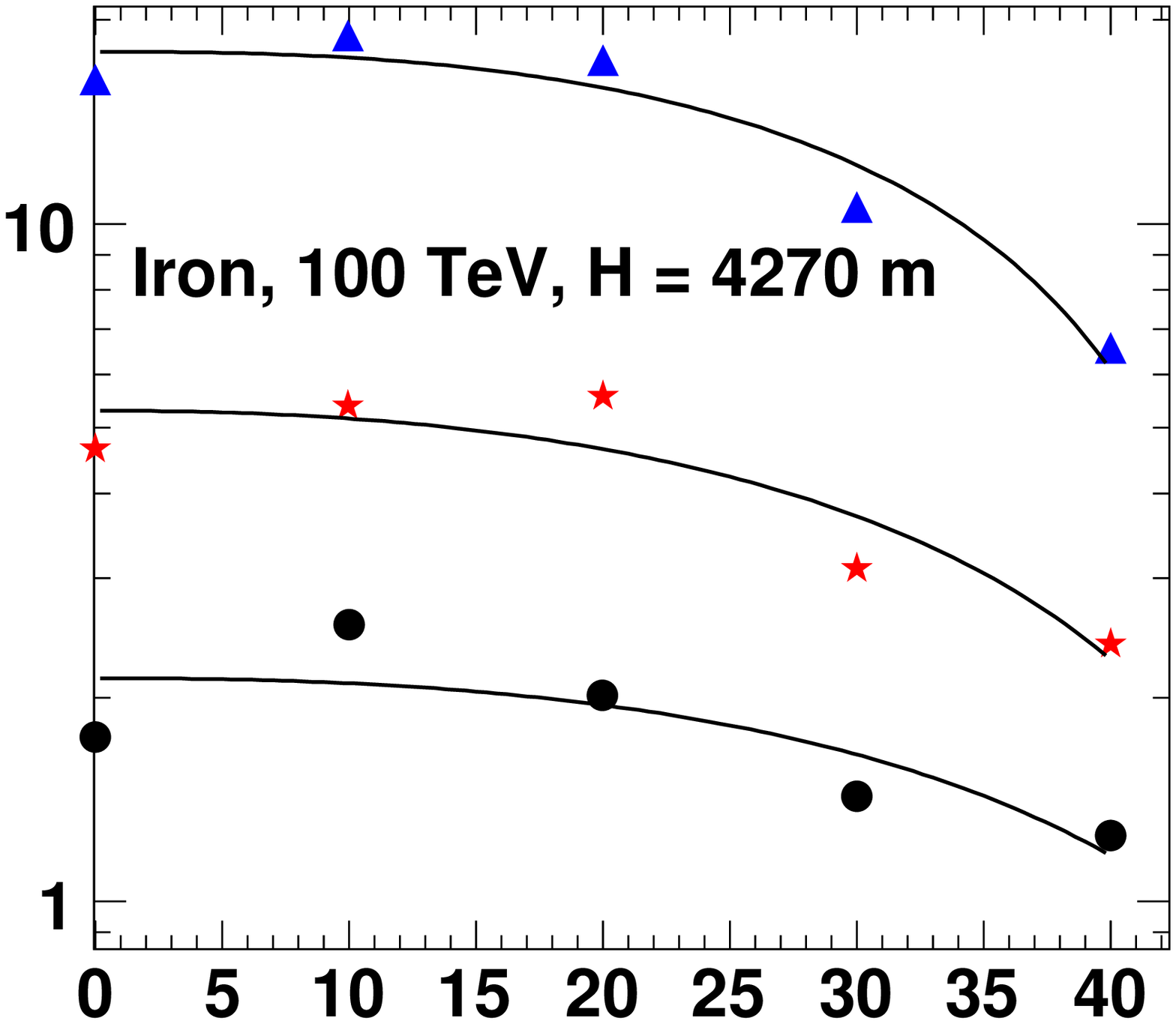}}
\centerline{\includegraphics[width=5.5cm, height=4.5cm]{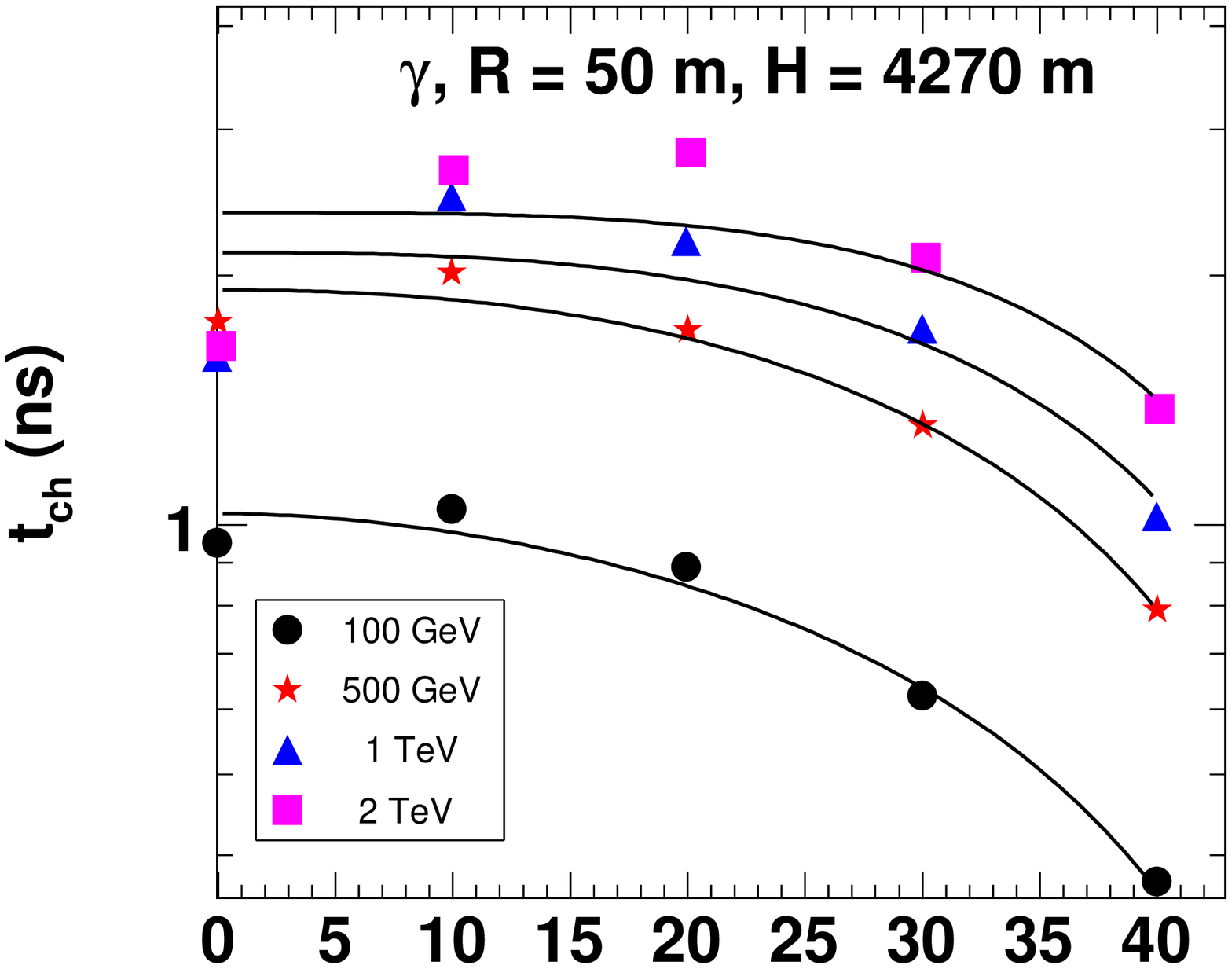}
\includegraphics[width=5.5cm, height=4.4cm]{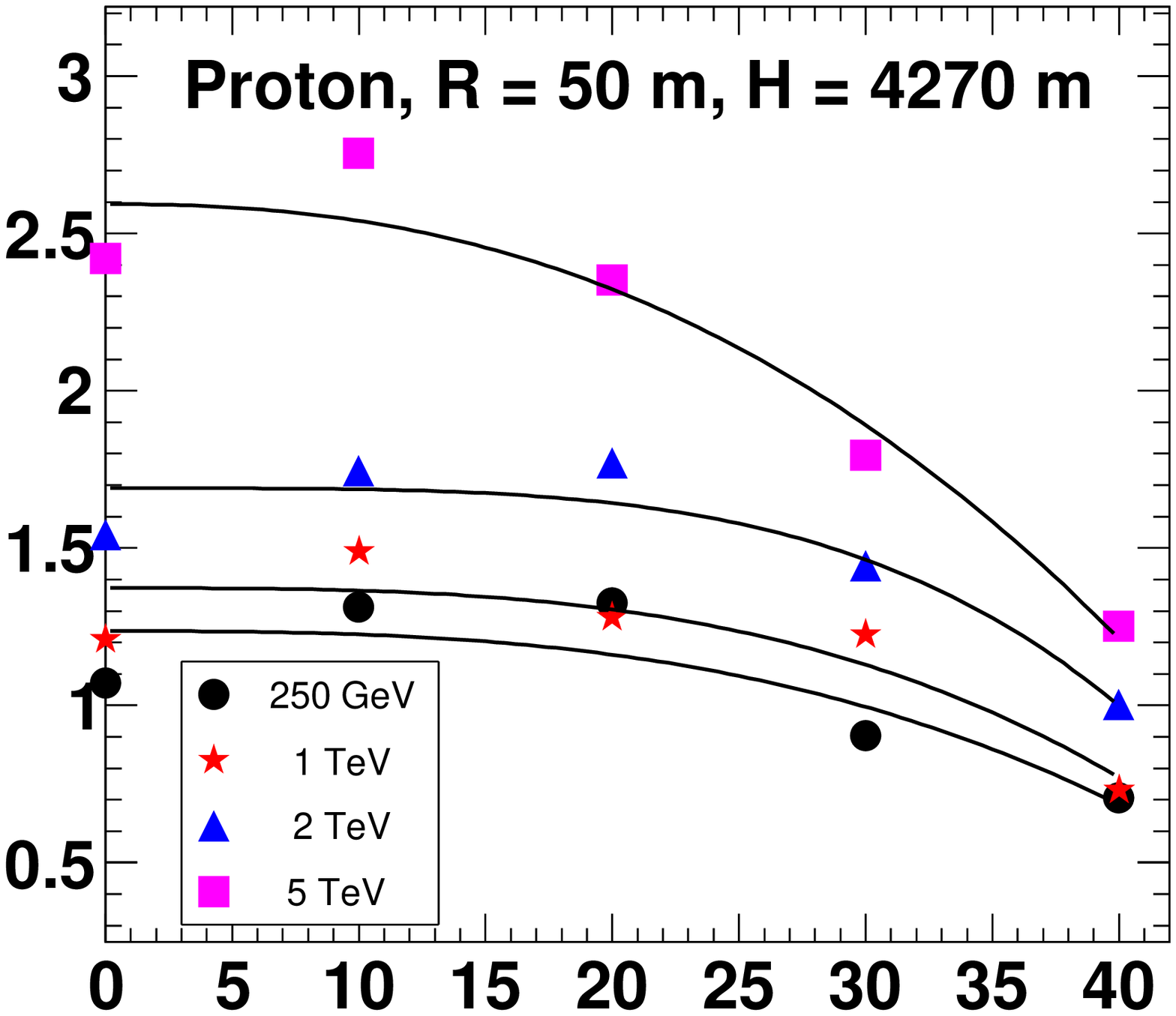}
\includegraphics[width=5.4cm, height=4.4cm]{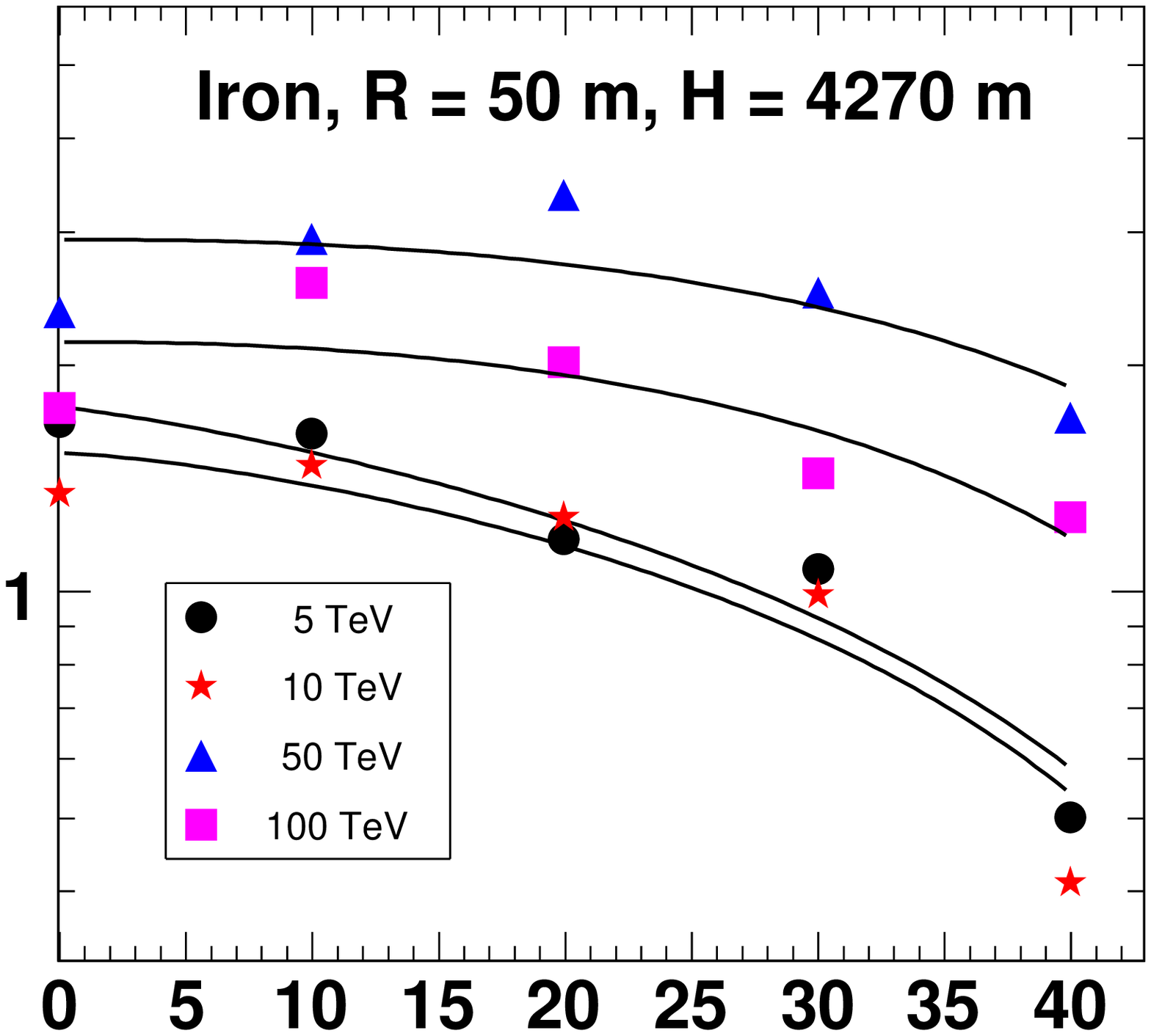}}
\centerline{\includegraphics[width=5.5cm, height=4.5cm]{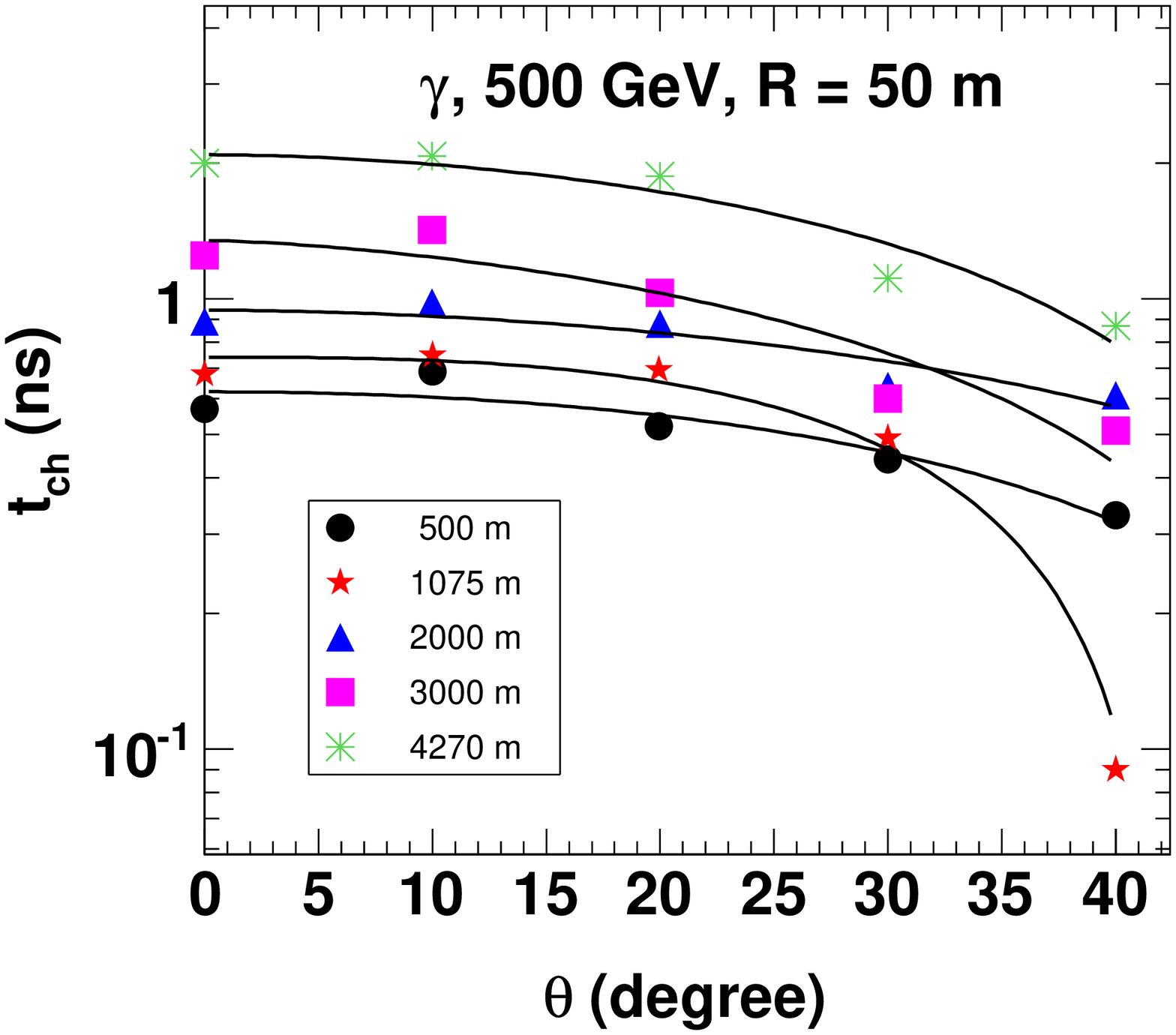}
\includegraphics[width=5.6cm, height=4.6cm]{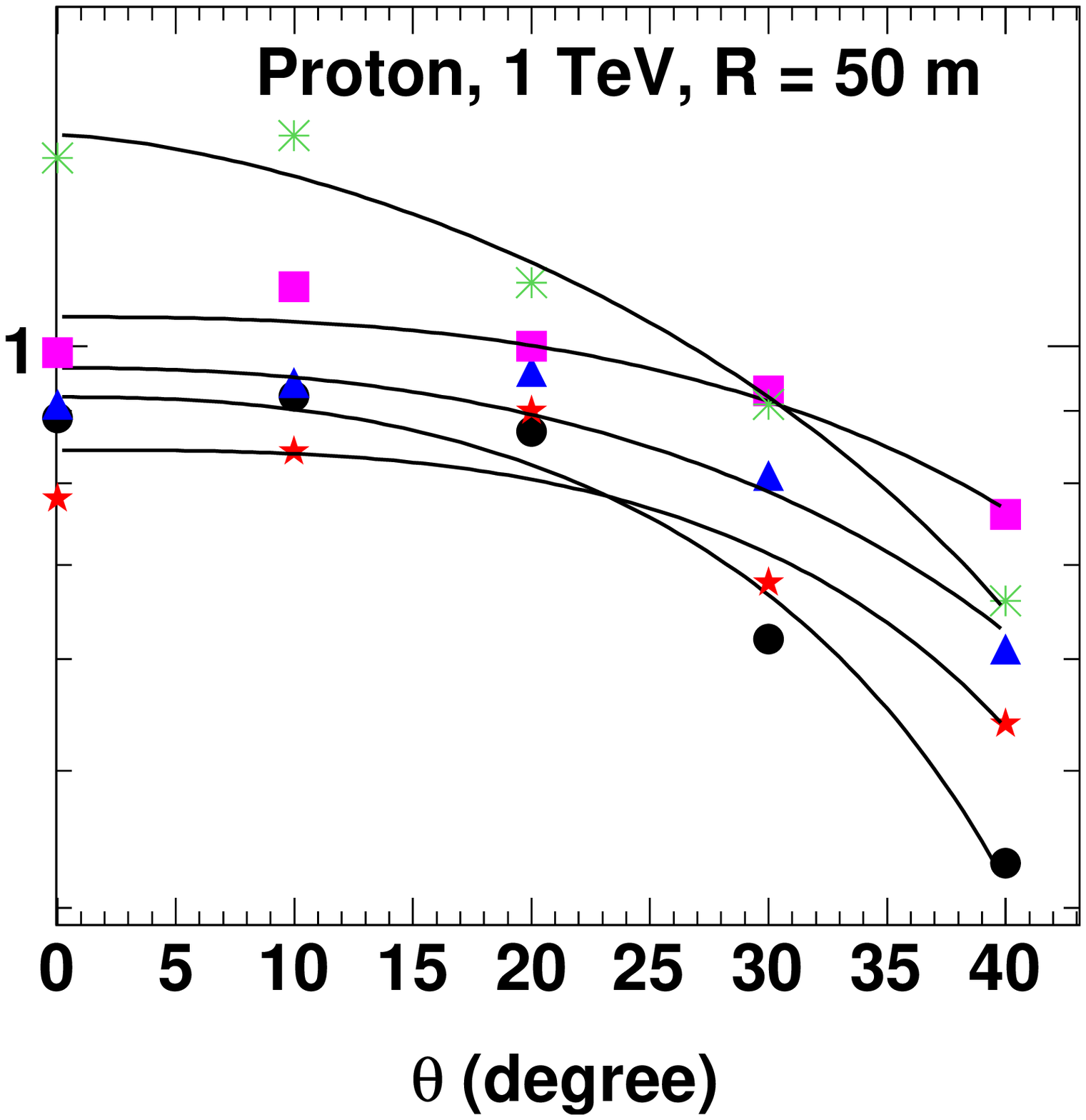}
\includegraphics[width=5.5cm, height=4.5cm]{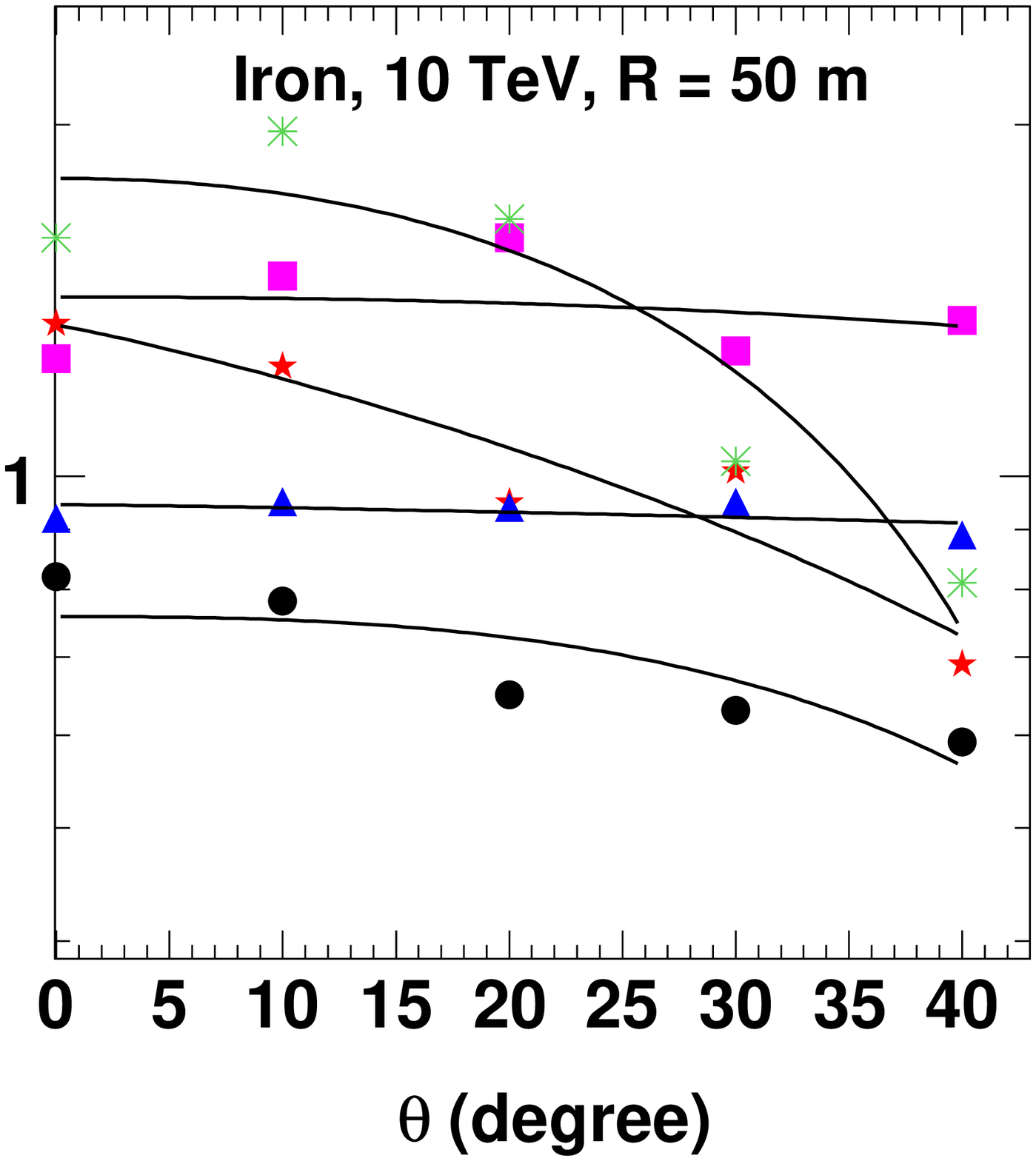}}
\caption{Average arrival time of Cherenkov photons ($t_{ch}$) for $\gamma$-ray, proton and iron primaries is plotted as a function of zenith angle ($\theta$). The plots in the upper most panel show these variations for different core distances $(R)$ keeping $E$ and $H$ fixed. The plots in the middle panel show the $t_{ch}$ variations for different values of $E$ keeping $H$ and $R$ fixed and the plots in the bottom panel do the same for different values of $H$ keeping $R$ and $E$ fixed at a particular value. The solid lines in the respective plots shows the result of our parameterisation 
Eq.(\ref{eq7}). The fits are within the limit of statistical error ($< \pm 10\%$).}
\label{fig7}
\end{figure*}

Because of the decrease number of Cherenkov photons over a detector array as a consequence of decrease in slant depth of the shower maximum of the primary particle with increasing zenith angle, the variation of $t_{ch}$ as a function of zenith angle follows a falling trend for all the primary particles and for all combinations of energies, core distances and altitudes of observation, refer Fig.\ref{fig7}. However, this variation is steeper in case of $\gamma$-ray and proton primaries in comparison to the iron primary. As mentioned earlier, the larger 
muon content and lower statistics can contribute to this observation. The variation of $t_{ch}$ with zenith angle can be parameterised by an equation of the form
\begin{equation}
t(\theta) = l_{2} ({\theta}^q+s),
\label{eq7}
\end{equation}
where $t(\theta)$ is mean arrival time of Cherenkov photons as a function of zenith angle, $l_{2}$, $q$ and $s$ are parameters of the function, $\theta$ is the zenith angle. $l_{2}$, $q$ and $s$ have different values for different primaries. For reasons mentioned above the fit is smoothest for the $\gamma$-ray 
primary. The Table \ref{tab8} shows, as an example, the values of the fitted parameters of the Eq.(\ref{eq7}) to the $t_{ch}$ distributions as a function of $\theta$ for $\gamma$-ray, proton and iron primary at 1 TeV, 2 TeV and 100 TeV energies respectively and at $H$ = 4270 m and $R$ = 100 m.
\begin{table}[ht]
\caption{Values of the fitted parameters of the Eq.(\ref{eq7}) to the $t_{ch}$ distributions as a function of $\theta$ for $\gamma$-ray, proton and iron primary at 1 TeV, 2 TeV and 100 TeV energies respectively and at $H$ = 4270 m and $R$ = 100 m.} \label{tab8}
\begin{center}
\begin{tabular}{cccc}\hline
Primary & ~$l_{2}$ & ~$q$ & ~$s$  \\\hline\\[-7pt]

$\gamma$&~0.067$\pm$0.019&~1.820$\pm$0.285&~-39.69$\pm$1.14\\[2pt]
Proton&~0.071$\pm$0.022&~1.299$\pm$0.003&~-39.25$\pm$0.08\\[2pt]
Iron&~3.757$\pm$1.657&~1.264$\pm$0.011&~-38.85$\pm$0.33\\\hline
\end{tabular}
\end{center}
\end{table}
\subsection{As a function of altitude of observation ($H$)}
\begin{figure*}[hbt]
\centerline
\centerline{\includegraphics[width=5.3cm, height=4.5cm]{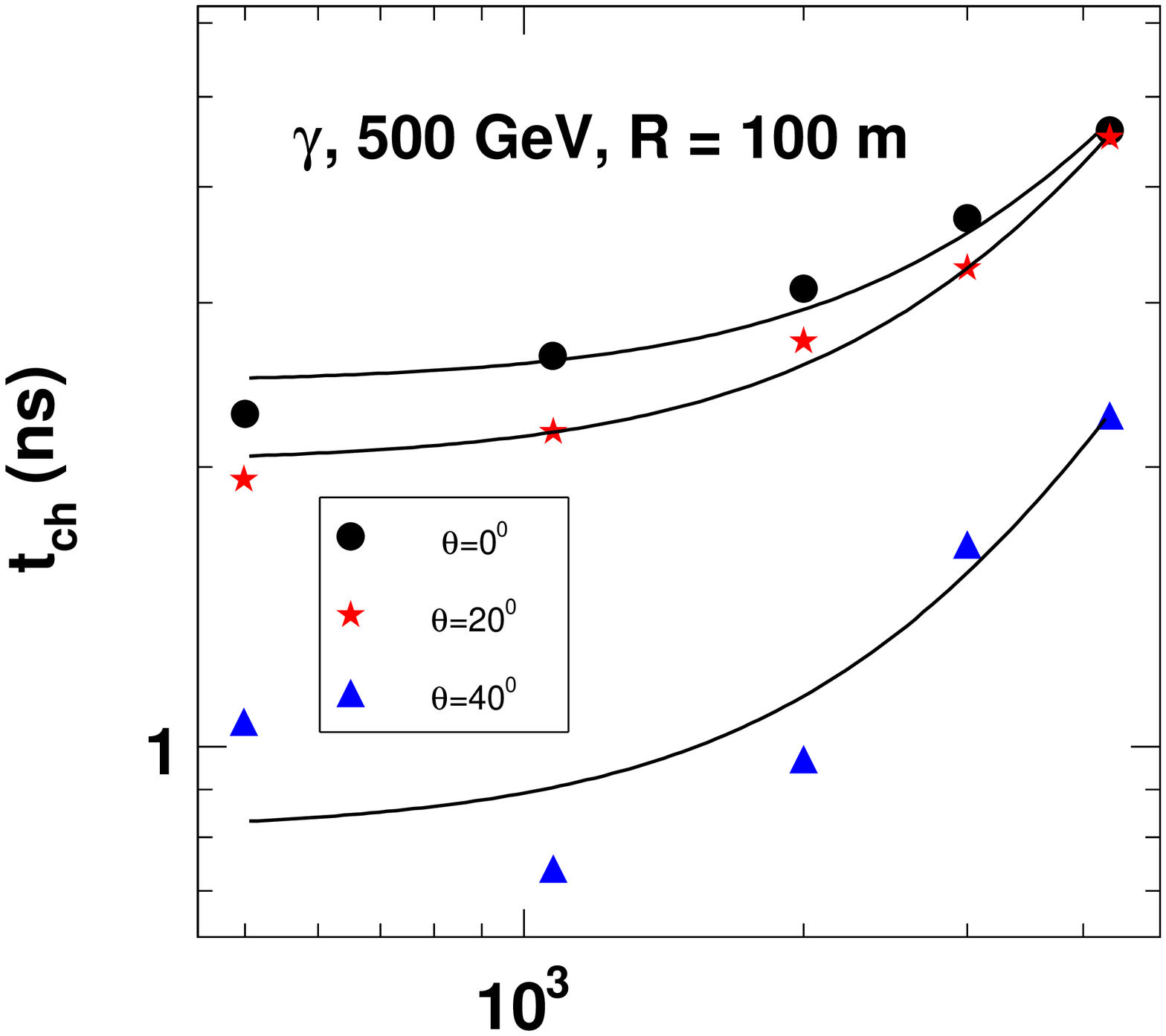}
\includegraphics[width=5.3cm, height=4.5cm]{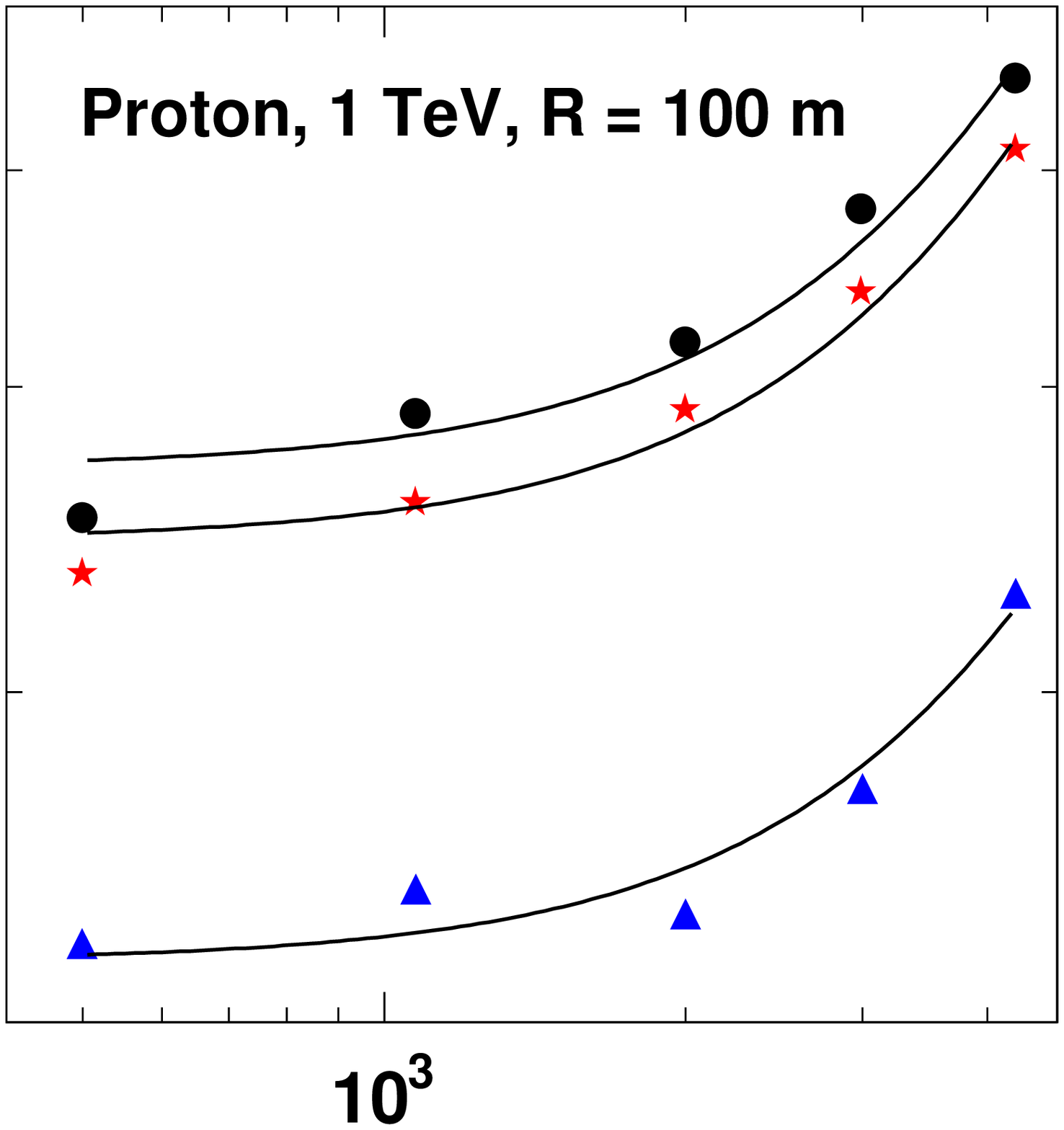}
\includegraphics[width=5.3cm, height=4.5cm]{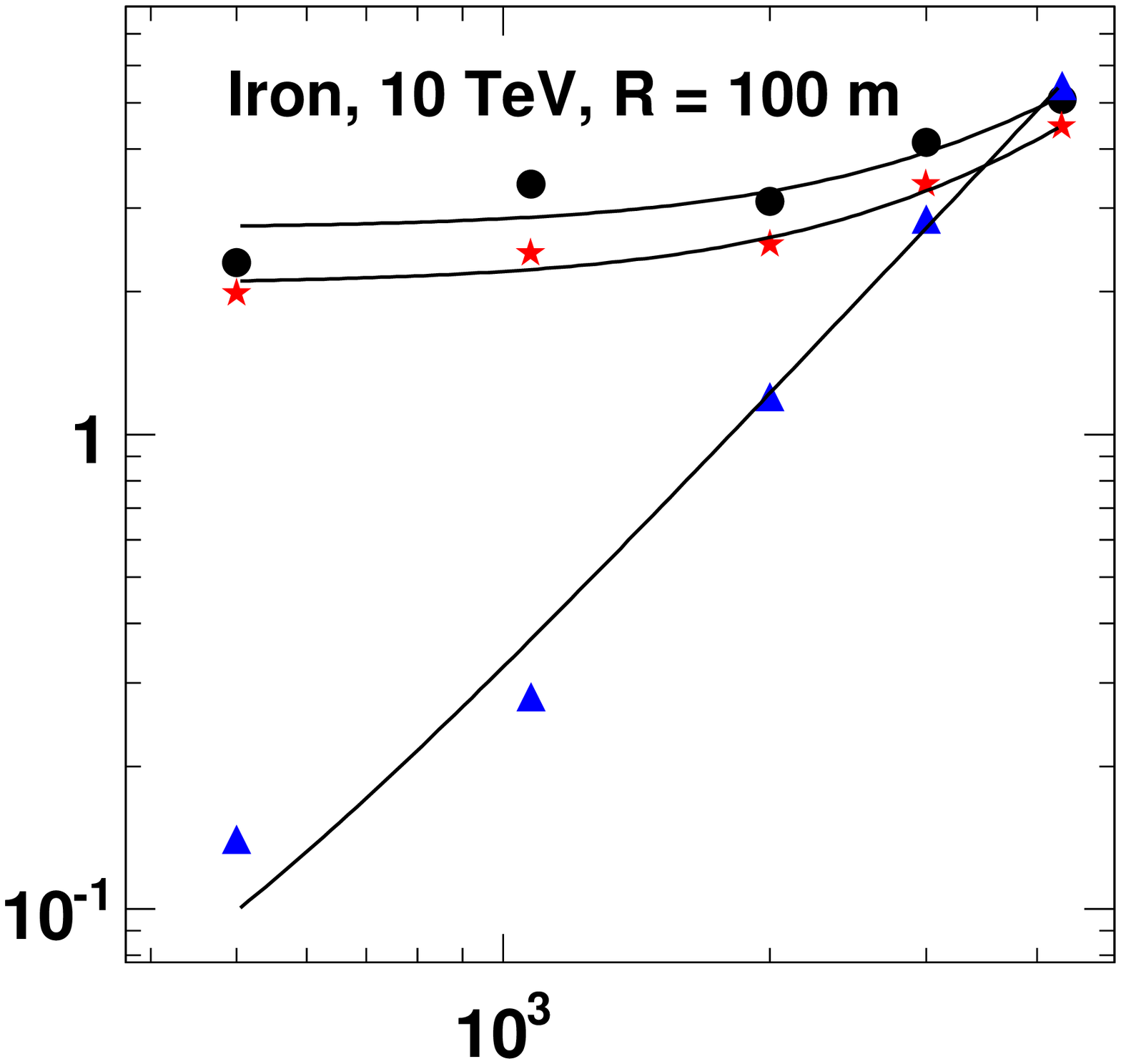}}
\centerline{\includegraphics[width=5.2cm, height=4.5cm]{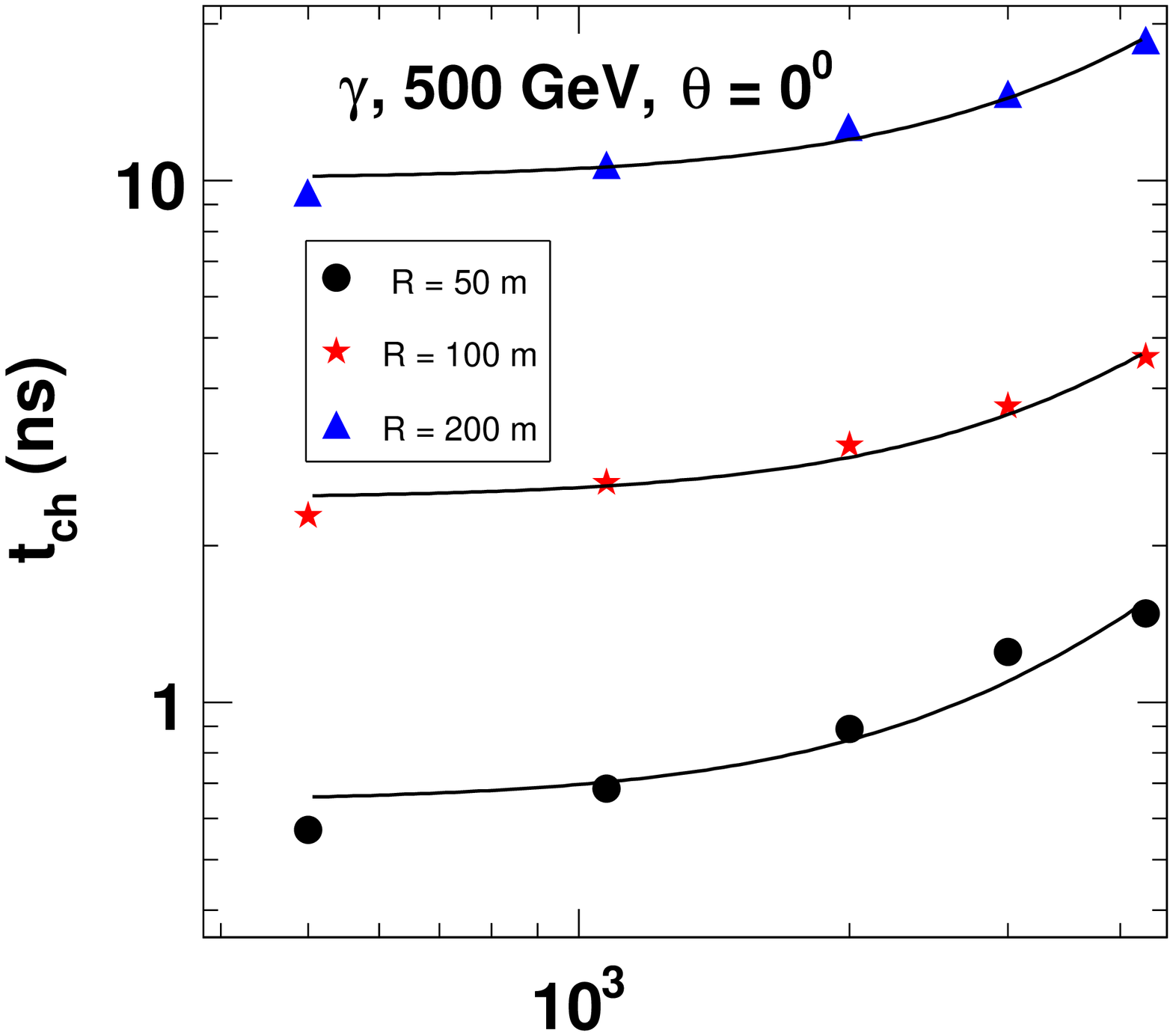}
\includegraphics[width=5.3cm, height=4.5cm]{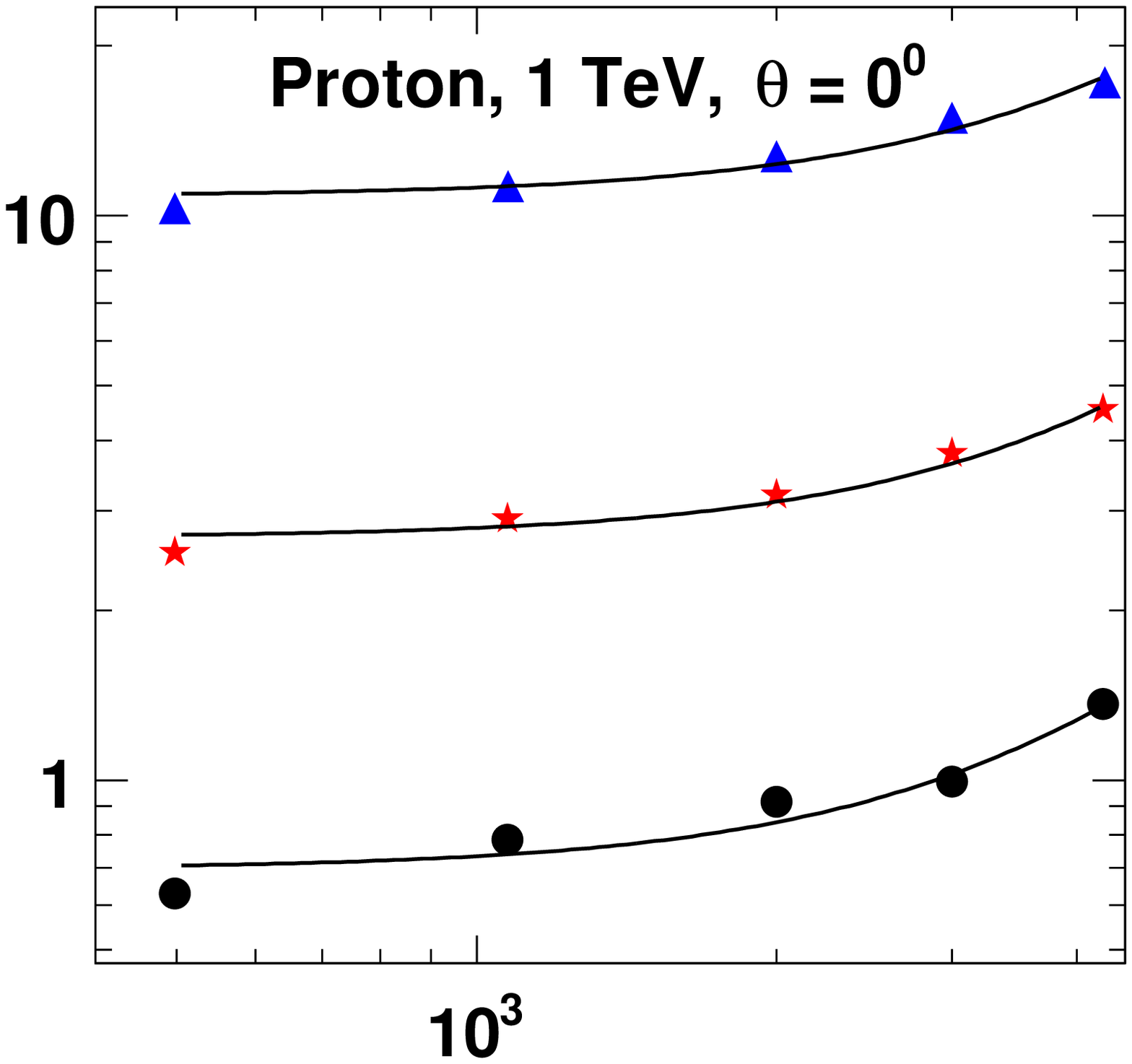}
\includegraphics[width=5.3cm, height=4.3cm]{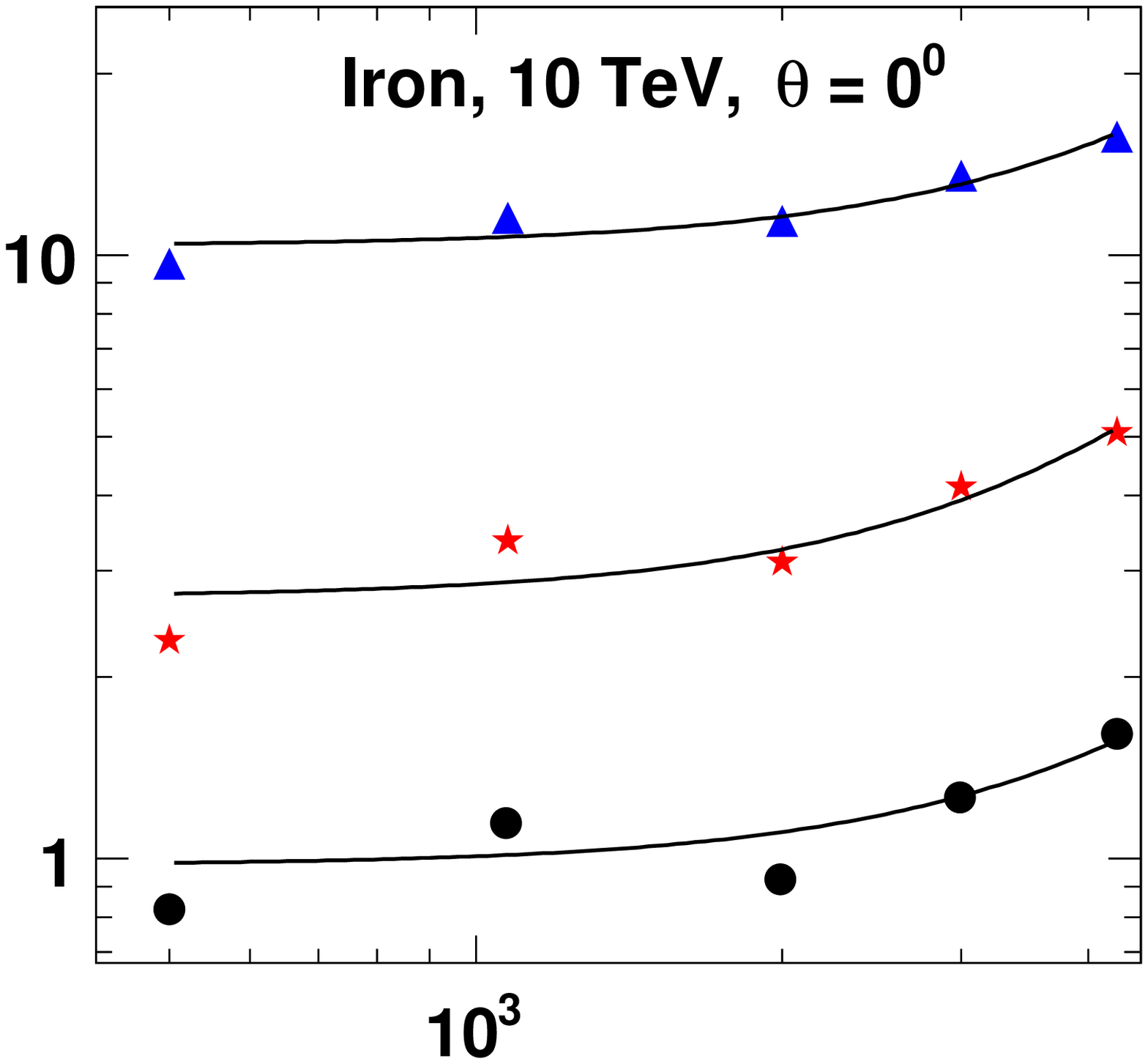}}
\centerline{\includegraphics[width=5.3cm, height=4.5cm]{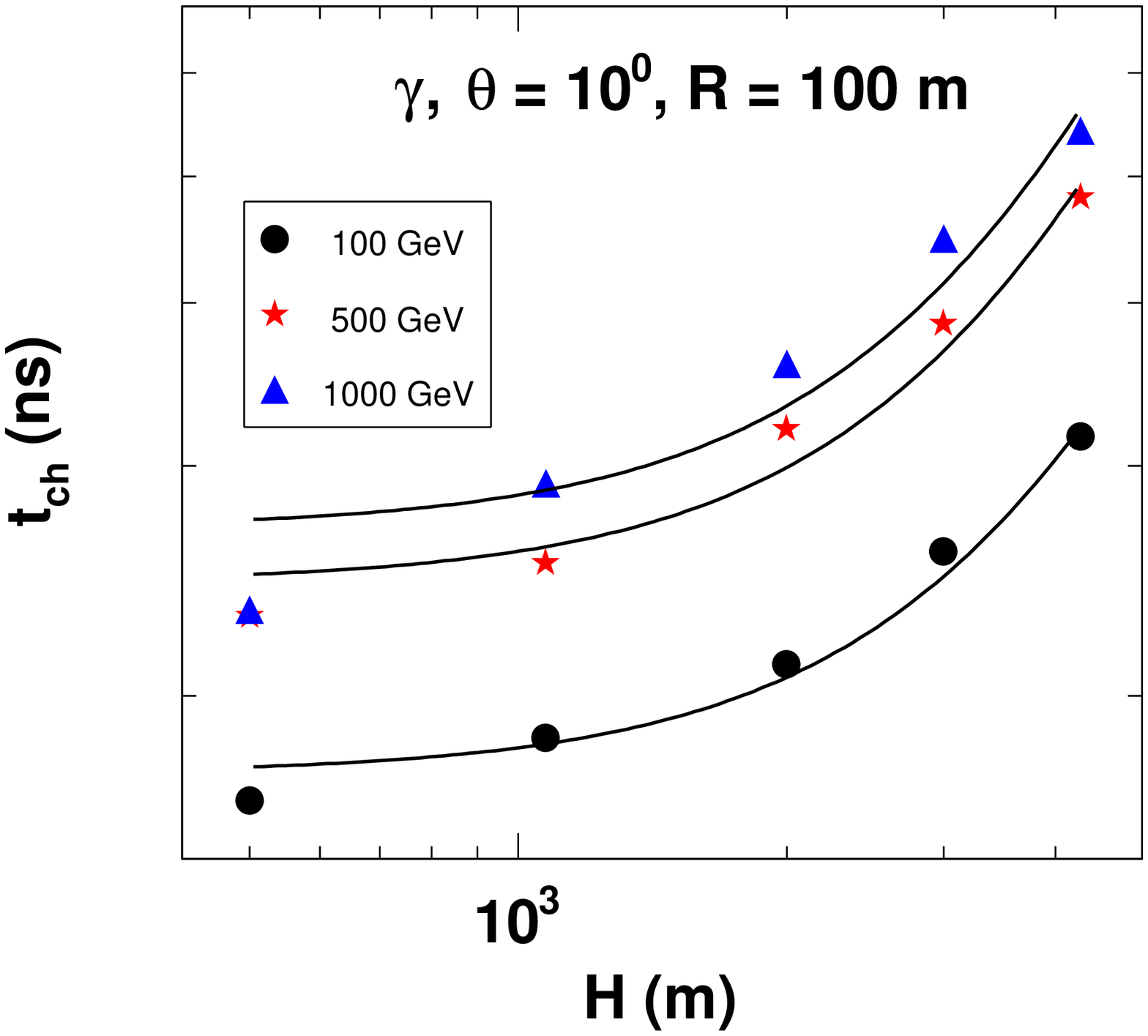}
\includegraphics[width=5.3cm, height=4.5cm]{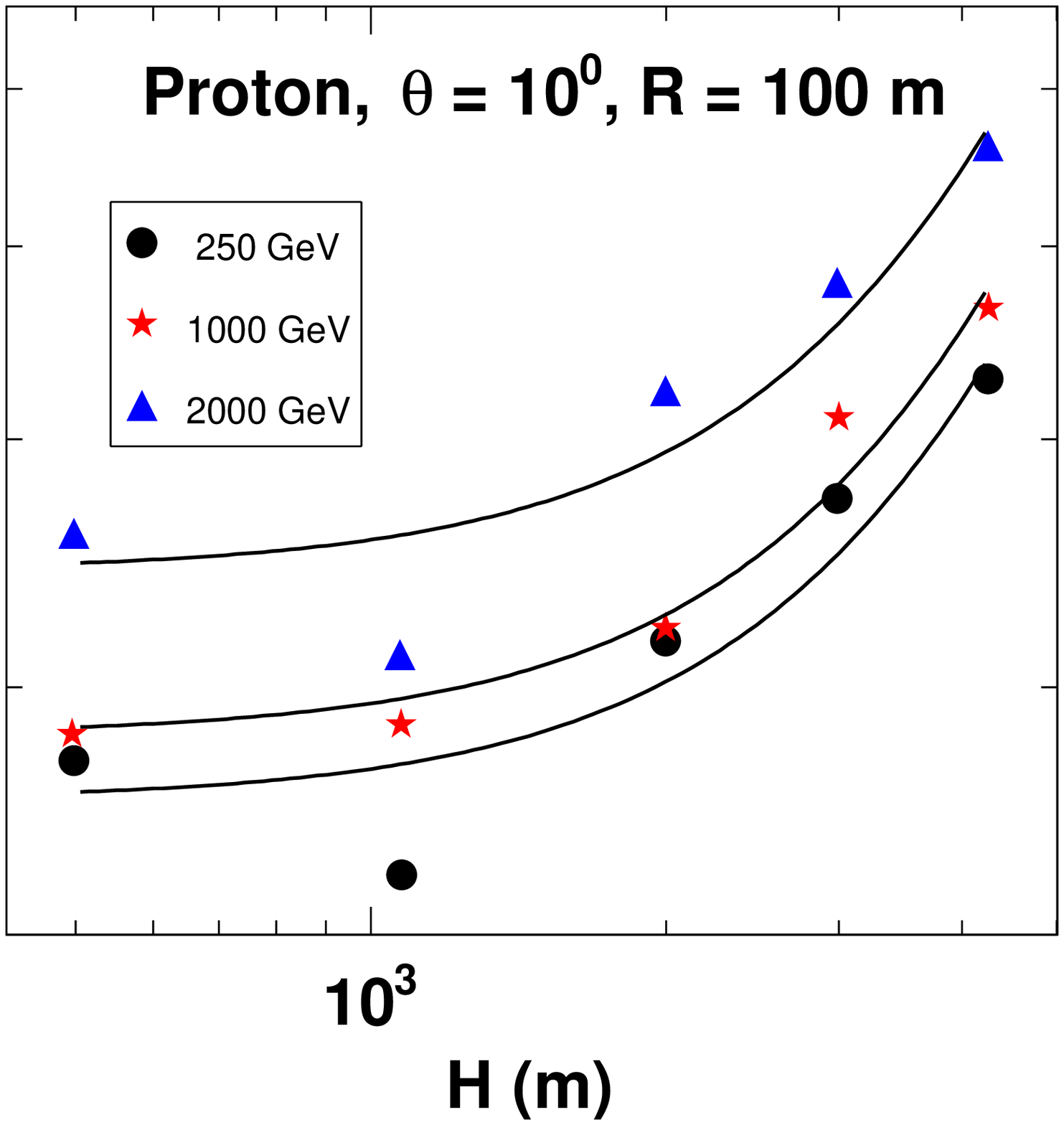}
\includegraphics[width=5.3cm, height=4.3cm]{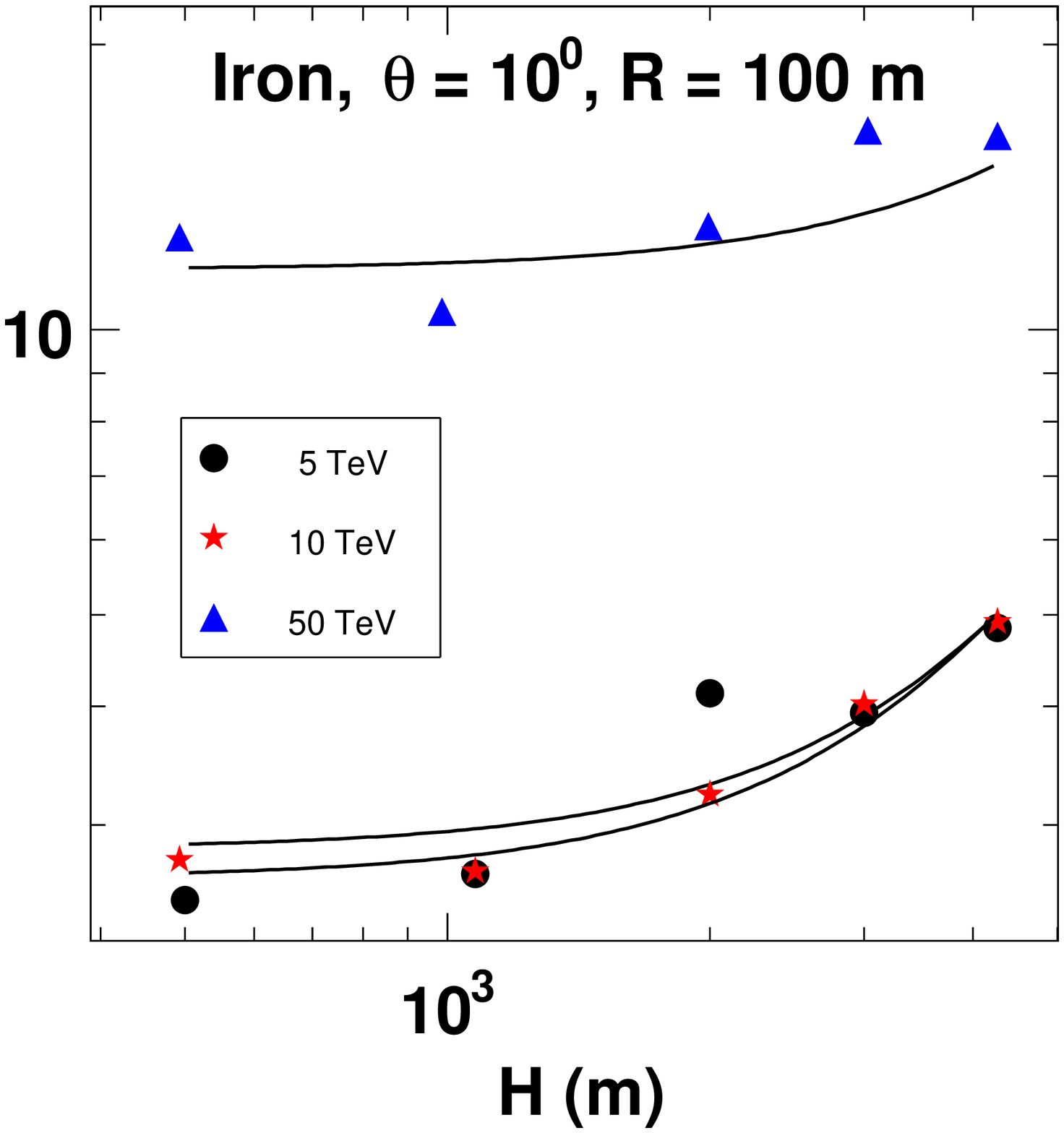}}
\caption{Average arrival time of Cherenkov photons ($t_{ch}$) for $\gamma$-ray, proton and iron primaries is plotted as a function of altitude of observation ($H$). The plots in the upper most panel show these variations for different zenith angles ($\theta$) keeping $E$ and $R$ fixed. The plots in the middle panel show the $t_{ch}$ variations for different values of $R$ keeping $\theta$ and $E$ fixed. The plots in the bottom panels do the same for different values of $E$ keeping $R$ and $\theta$ fixed at a particular value. The solid lines in the respective plots shows the result of our parameterisation (\ref{eq8}). The fits are within the limit of statistical error 
($< \pm 10\%$).}
\label{fig8}
\end{figure*}

In general the Cherenkov light front becomes steeper for increasing altitude of observation as the observation level comes closer to the shower maximum, whereas it gets flatter for decreasing altitudes of observation. This trend is followed for both vertical as well as inclined showers and at all core distances and at all energies of the three primary. For reasons as mentioned above in the earlier two cases, this trend is most prominent for the $\gamma$-ray primary and least for the iron primary, refer Fig.\ref{fig8}. The solid lines in Fig.\ref{fig8} are the result of our parameterisation of $t_{ch}$ distribution as a function of altitude of observation as follows:
\begin{equation}
t(H) = l_{3}(u{H}^2+v),
\label{eq8}
\end{equation}
where $t(H)$ is mean arrival time of Cherenkov photons as a function of altitude of observation, $l_{3}$, $u$ and $v$ are parameters of the function, $H$ is the altitude of observation. $l_{3}$, $u$ and $v$ have different values for different primaries. For the reason of lower statistics the parameterisation (\ref{eq8}) vary slightly from the predicted behaviour for proton and iron primaries compared to the $\gamma$-ray primary. As an illustration of the possible values of 
the paramters in the Eq.(\ref{eq8}), the Table \ref{tab9} shows the values of 
the fitted parameters of the Eq.(\ref{eq8}) to the $t_{ch}$ distributions as a function of $H$ for $\gamma$-ray, proton and iron primary at 500 GeV, 1 TeV and 10 TeV energies respectively and at $R$ = 50 m and $\theta$ = 20$^{0}$.
\begin{table}[ht]
\caption{Values of the fitted parameters of the Eq.(\ref{eq8}) to the $t_{ch}$ distribution as a function of $H$ for $\gamma$-ray, proton and iron primary at 500 GeV, 1 TeV and 10 TeV energies respectively and at $R$ = 50 m and $\theta$ = 20$^{0}$.} \label{tab9}
\begin{center}
\begin{tabular}{cccc}\hline
Primary & ~$l_{3}$ & ~$u$ & ~$v$  \\\hline\\[-7pt]

$\gamma$&~0.90$\pm$2.10&~0.413$\pm$0.374&~-10.47$\pm$1.35\\[2pt]
Proton&~0.58$\pm$1.42&~0.414$\pm$1.159&~~~-9.85$\pm$1.41\\[2pt]
Iron&~2.08$\pm$1.46&~0.355$\pm$0.386&~~~-1.69$\pm$1.38\\\hline
\end{tabular}
\end{center}
\end{table}
\section{Summary and Conclusions}
\begin{figure*}[hbt]
\centerline
\centerline{\includegraphics[width=6.5cm, height=5.4cm]{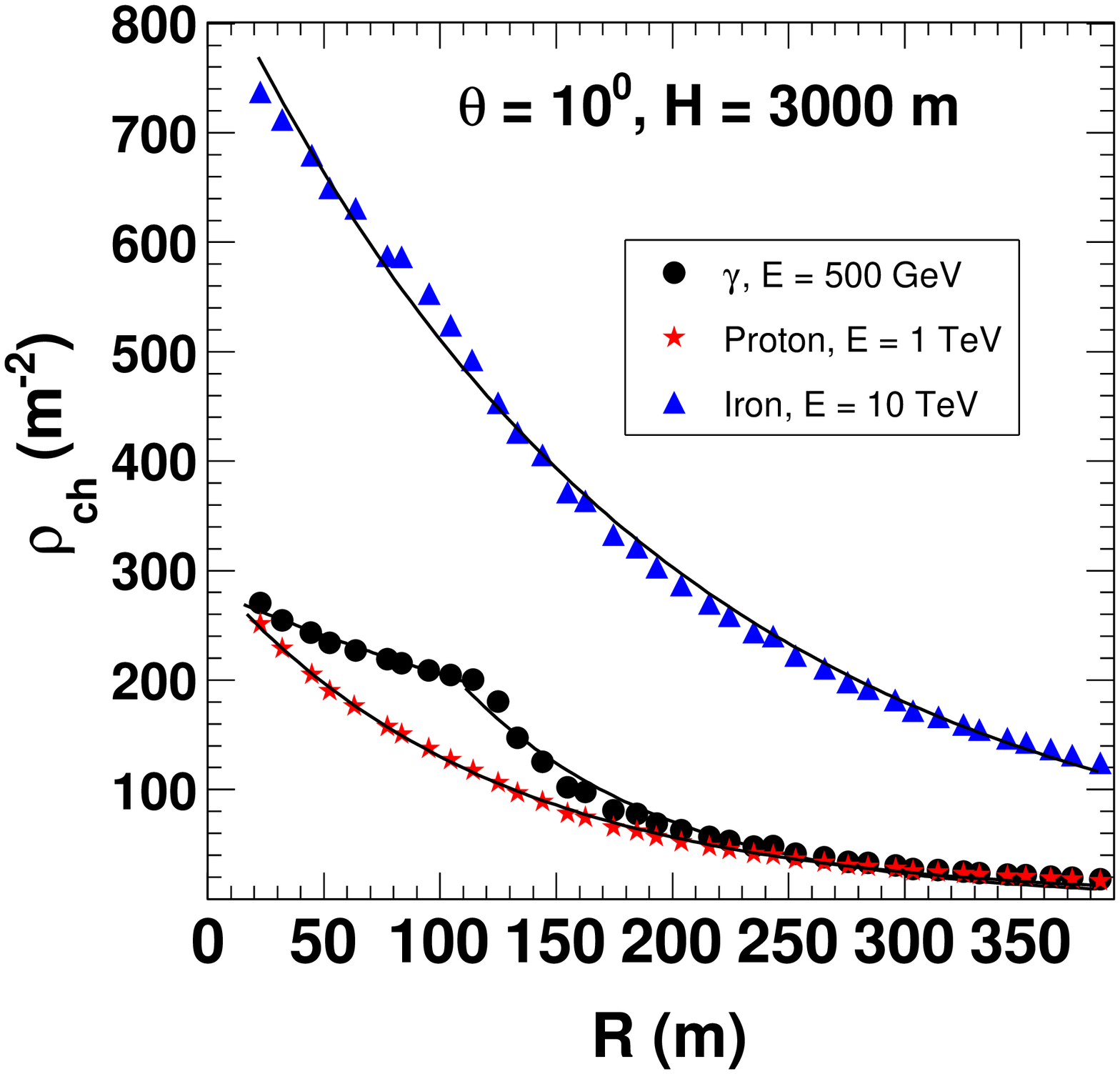}
\includegraphics[width=6.5cm, height=5.4cm]{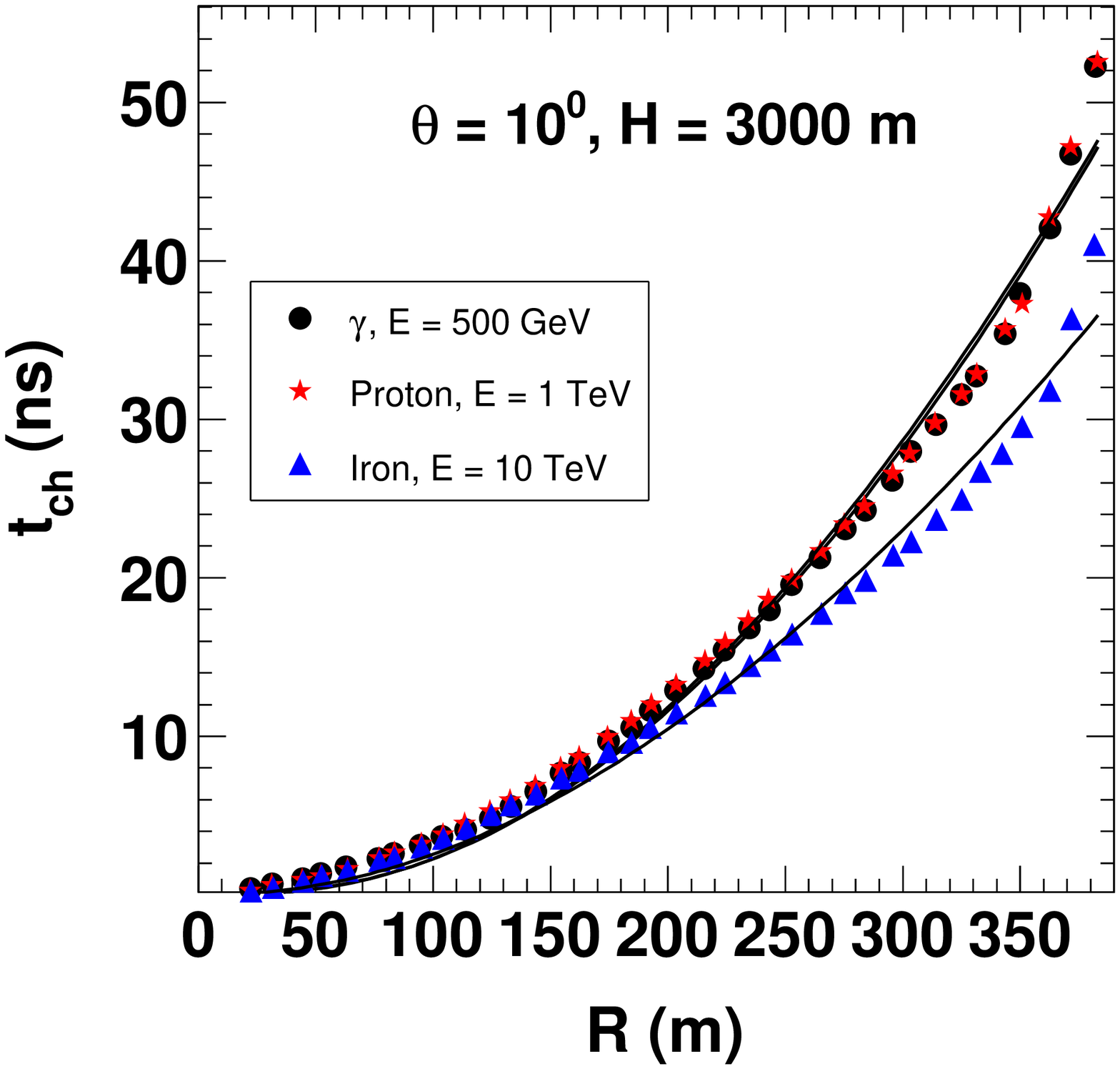}}
\centerline{\includegraphics[width=6.5cm, height=5.4cm]{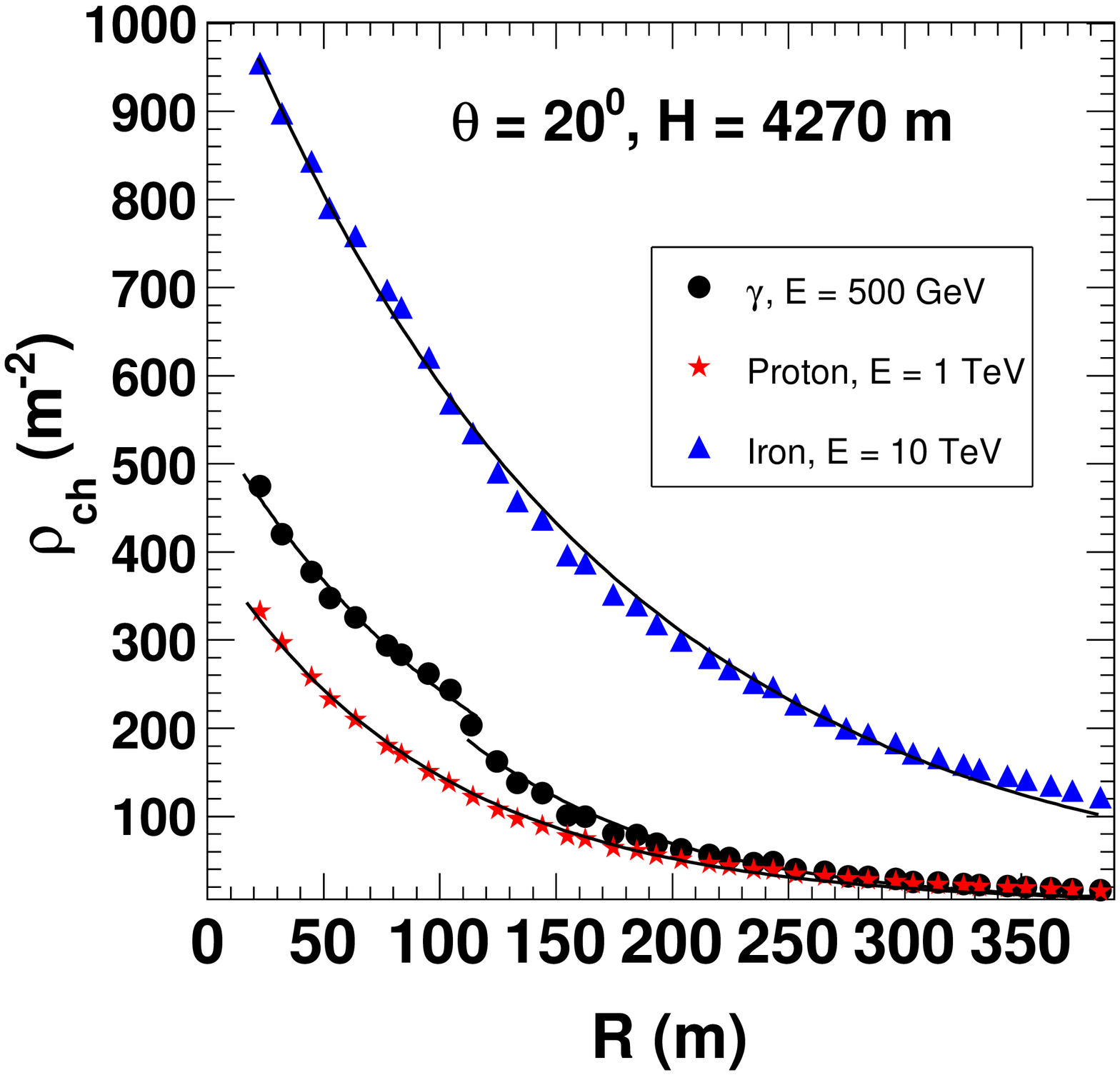}
\includegraphics[width=6.5cm, height=5.4cm]{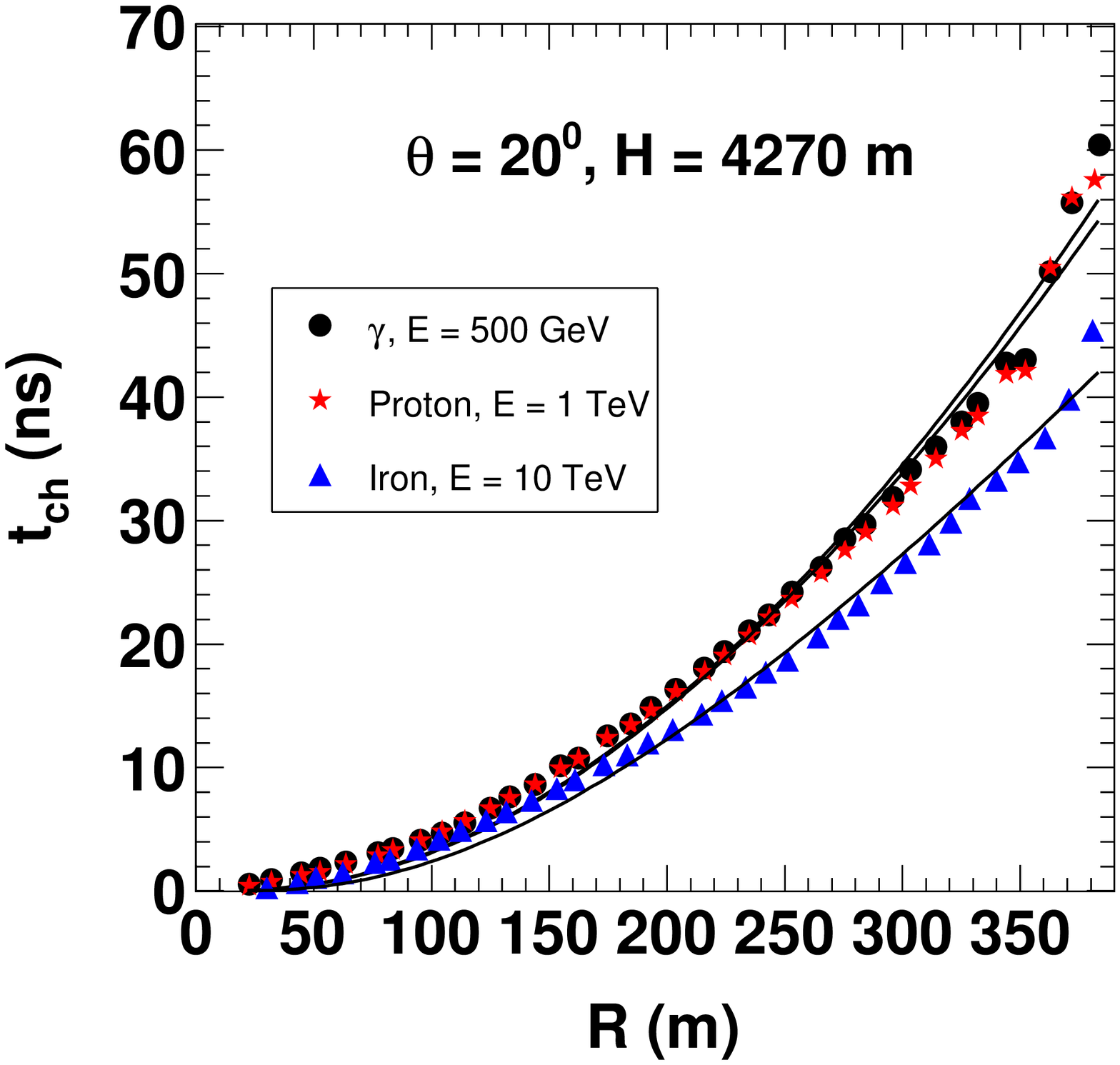}}
\caption{Average density ($\rho_{ch}$) and arrival time ($t_{ch}$) of Cherenkov photons for $\gamma$-ray, proton and iron primaries at 500 GeV, 1 TeV and 10 TeV energies respectively are plotted as a function of radial distance ($R$) keeping zenith angle ($\theta$) fixed at $10^{0}$ and $20^{0}$ and the altitude of observation ($H$) at 4270 m and 3000 m. The solid lines in the respective plots show the results of our parameterisation functions (\ref{eq9}) and (\ref{eq10}). 
The fits are within the limit of statistical error ($< \pm 10\%$).}
\label{fig9}
\end{figure*}

For the purpose of developing an analytical description for the production of the Cherenkov photons in EAS, we parameterise the distributions of average density and arrival time of Cherenkov photons as a function of distance from the shower core ($R$), zenith angle ($\theta$), altitude of observation ($H$) and the energy of the primary ($E$). The proposed parameterisations are as follows: 
\begin{equation}
\rho(E,R,\theta,H) = a E^{b}\exp[-\{c R + (\theta /d)^{2}-f H\}],
\label{eq9}
\end{equation}
\begin{equation}
t(E,R,\theta,H) = l E^{-m}\exp(n /R^{p})({\theta}^q+s)(u {H}^2+v),
\label{eq10}
\end{equation}
where $\rho(E,R,\theta,H)$ and $t(E,R,\theta,H)$ are the average density and arrival time as a function of core distance ($R$), zenith angle ($\theta$), altitude of observation ($H$) and the energy of the primary ($E$). $a$, $b$, $c$, $d$, $f$, $l$, $m$, $n$, $p$, $q$, $s$, $u$ and $v$ are function parameters whose values are different for different primaries. The estimated results of these parameterisations (\ref{eq9}) and (\ref{eq10}) are then compared with the Monte Carlo predictions using the CORSIKA 6.990 simulation package \cite{Knapp} for the $\gamma$-ray, proton and iron primaries at different values of $R$, $\theta$, $H$ and $E$. As an example, we have shown the compared results of the simulated 
CORSIKA data with parameterisations (\ref{eq9}) and (\ref{eq10}) for $\gamma$, 
proton and iron primary at energies 500 GeV, 1 TeV and 10 TeV respectively 
keeping $H$ fixed at 3000 m and 4270 m, and $\theta$ at $10^{0}$ and $20^{0}$ 
in the Fig.\ref{fig9}. The related fitting parameters of the functions 
(\ref{eq9}) and (\ref{eq10}) are given in Table \ref{tab10} and Table 
\ref{tab11} respectively.
  
In case of lateral density of Cherenkov photons, the parameterisation (\ref{eq9}), in general is in good agreement with the simulated CORSIKA data. However for all the three primaries, at very small (< 50 m) and very large core distances, 
and large values of zenith angles the parameterisation do show a slight variation from the predicted data. This may be due to lesser number of Cherenkov photons produced near the shower axis and tail of the distribution, and also for large zenith angles. In addition to this, for $\gamma$-ray primary, the presence of the significant characteristic hump, affects the quality of the parameterisation. However as the hump becomes smaller as energy of the primary and the altitude of observation increases, the parameterisation becomes better even for the $\gamma$-ray primary.

The parameterisation (\ref{eq10}) can very well describe the general features of $t_{ch}$ distributions of Cherenkov photons for the three primaries. However similar to the case of $\rho_{ch}$ distributions, approximation (\ref{eq10}) varies slightly from the CORSIKA predictions at very small and large values of core distances. These deviations are higher for the iron primaries compared to the $\gamma$-ray and proton primaries. Due to greater muon content of the iron primary, larger shower to shower fluctuations affects the arrival time of the Cherenkov photons particulary at the tail region. Lower statistics further accentuates this affect.

Thus the parameterisations (\ref{eq9}) and (\ref{eq10}) can in general describe the Cherenkov photon production in an EAS irrespective of the primary particle, primary energy, core distance, zenith angle and altitude of observation (within shower-to-shower fluctuations). To explain the deviations between the simulated data and the parameterisations in the region of very low and very high core distances one have to introduce further degrees of freedom in the proposed parameterisations. As the mean arrival time of Cherenkov photons at each detector is calculated with respect to the first photon hitting the detector array, so the $t_{ch}$ distributions are largely affected by lower statistics compared to the $\rho_{ch}$ distributions. Also higher muon content of the iron primary also contributes to this fact. So to improve our systematics we must increase the number of showers produced. The parameteristions can also be tested over a wider range of energy. With this concerns taken care of the proposed parameterisations can help for the reconstruction of the shower variables and hence in estimation of the primary energy and primary particle identification. In future we plan to further extend our study to include the effect of geomagnetic field. 
\\
\begin{table*}[ht]
\caption{Values of the fitted parameters of the Eq.(\ref{eq9}) to the $\rho_{ch}$ distributions as a function of $R$ for $\gamma$-ray, proton and iron primary at 500 GeV, 1 TeV and 10 TeV energies respectively and at $H$ = 4270 m and 3000 m, and $\theta$ = 20$^{0}$ and 10$^{0}$.} \label{tab10}
\begin{center}
\begin{tabular}{cccccccc}\hline
$\theta$ (deg) & ~$H$ (m) & Primary & ~$a$ & ~$b$ & ~$c$ & ~$d$ & ~$f$ \\\hline\\[-7pt]

&&$\gamma$&2.16$\pm$0.297&0.6032$\pm$0.0221&0.0082$\pm$0.0004&7.838$\pm$0.083&0.0019$\pm$0.0003\\[2pt]
$20$&4270&Proton&1.35$\pm$0.122&1.1999$\pm$0.0131&0.0103$\pm$0.0002&7.801$\pm$0.053&0.0009$\pm$0.0002\\[2pt]
&&Iron&1.96$\pm$0.016&0.9441$\pm$0.0006&0.0062$\pm$0.0001&7.873$\pm$0.003&0.0010$\pm$0.0001\\\\[2pt]
&&$\gamma$&1.36$\pm$0.117&0.5927$\pm$0.0137&0.0031$\pm$0.0002&7.787$\pm$0.051&0.0019$\pm$0.0001\\[2pt]
$10$&3000&Proton&1.09$\pm$0.059&1.1957$\pm$0.0078&0.0083$\pm$0.0001&7.779$\pm$0.032&0.0009$\pm$0.0001\\[2pt]
&&Iron&2.19$\pm$0.012&0.8973$\pm$0.0004&0.0052$\pm$0.0001&7.900$\pm$0.003&0.0009$\pm$0.0001\\\hline
\end{tabular}
\end{center}
\end{table*}
\begin{table*}[ht]
\caption{Values of the fitted parameters of the Eq.(\ref{eq10}) to the $t_{ch}$ distributions as a function of $R$ for $\gamma$-ray, proton and iron primary at 500 GeV, 1 TeV and 10 TeV energies respectively and at $H$ = 4270 m and 3000 m, and $\theta$ = 20$^{0}$ and 10$^{0}$.} \label{tab11}
\begin{center}
\begin{tabular}{ccccccccccc}\hline
$\theta$ (deg) & ~$H$ (m) & Primary & ~$l$ & ~$m$ & ~$n$ & ~$p$ & ~$q$ & ~$s$ & ~$u$ & ~$v$\\\hline\\[-7pt]

&&$\gamma$&5.42$\pm$0.71&-1.85$\pm$0.31&-31.92$\pm$1.41&0.14$\pm$0.007&0.002$\pm$0.0005&0.01$\pm$0.012&0.0001$\pm$0.00002&~421.46$\pm$169.64\\[2pt]
$20$&4270&Proton&6.61$\pm$0.33&-1.28$\pm$0.12&-30.59$\pm$1.08&0.17$\pm$0.002&0.002$\pm$0.0003&0.01$\pm$0.006&0.0001$\pm$0.00001&-249.53$\pm$186.83\\[2pt]
&&Iron&1.42$\pm$0.64&-0.48$\pm$0.04&-36.78$\pm$2.79&0.33$\pm$0.009&0.002$\pm$0.0002&0.01$\pm$0.005&0.0001$\pm$0.00001&-327.16$\pm$368.34\\\\[2pt]
&&$\gamma$&5.87$\pm$0.65&-2.01$\pm$0.23&-34.41$\pm$1.37&0.13$\pm$0.005&0.003$\pm$0.0006&0.02$\pm$0.011&0.0001$\pm$0.00002&~486.75$\pm$149.04\\[2pt]
$10$&3000&Proton&6.14$\pm$0.72&-1.55$\pm$0.19&-32.96$\pm$1.58&0.16$\pm$0.004&0.002$\pm$0.0001&0.01$\pm$0.001&0.0002$\pm$0.00001&-151.60$\pm$101.81\\[2pt]
&&Iron&6.34$\pm$0.72&-1.26$\pm$0.20&-31.36$\pm$1.48&0.12$\pm$0.006&0.003$\pm$0.0006&0.02$\pm$0.012&0.0001$\pm$0.00003&-431.15$\pm$113.26\\\hline
\end{tabular}
\end{center}
\end{table*}

\end{document}